\documentclass[12pt, a4paper]{article}

\usepackage{geometry}
\usepackage{verbatim}
\usepackage{graphicx}
\usepackage{subfigure}
\usepackage{overpic}
\usepackage[nofiglist]{endfloat}  % figures are at end of document
\usepackage{amsmath}
\usepackage{bm}
\usepackage{txfonts}
\usepackage{array}
\usepackage{cite}
\usepackage{color}

\usepackage{epstopdf}

\newcommand{\mbB}{\bm{\mathrm{B}}}
\newcommand{\mbJ}{\bm{\mathrm{J}}}
\newcommand{\mbE}{\bm{\mathrm{E}}}
\newcommand{\mbV}{\bm{\mathrm{V}}}

\newcommand{\mm}{\mathrm{m}}

\newcommand{\ms}{\mathrm{s}}
\newcommand{\mT}{\mathrm{T}}
\newcommand{\mkeV}{\mathrm{keV}}

\newcommand{\mkHz}{\mathrm{kHz}}

\newcommand{\efit}{\textsc{EFIT }}

\usepackage{lineno}

\begin{document}

%\begin{CJK*}{GBK}
%\begin{xeCJK*}{GBK}{song}
%\linenumbers
\begin{center}
  \Large\textbf{Frequency multiplication with toroidal mode
    number of kink/fishbone modes on a static HL-2A-like
    tokamak}
\end{center}
%Zhihui ZOU (邹志慧)$^{1}$, Ping ZHU (朱平)$^{2, 3\ast}$, Charlson C. KIM$^{4}$,
%Wei DENG (邓玮)$^{5}$, Xianqu WANG (王先驱)$^{6}$,
%Yawei HOU (侯雅巍)$^{1\ast\ast}$ % ...... authors
%\\
Zhihui ZOU$^{1}$, Ping ZHU$^{2, 3\ast}$, Charlson C. KIM$^{4}$,
 Wei DENG$^{5}$, Xianqu WANG$^{6}$,
Yawei HOU$^{1\ast \ast}$ % ...... authors
\\

{\small
  $^{1}$ CAS Key Laboratory of Geospace Environment and Department of
  Plasma Physics and Fusion Engineering,
  University of Science and Technology of China, Hefei,
  230026, Peoples's Republic of China \\
  $^{2}$ International Joint Research Laboratory of Magnetic Confinement
  Fusion and Plasma Physics, State Key Laboratory of Advanced
  Electromagnetic Engineering and Technology, School of Electrical
  and Electronic Engineering, Huazhong University of Science and
  Technology, Wuhan, 430074, Peoples's Republic of  China\\
  $^{3}$ Department of Engineering Physics, University of
  Wisconsin-Madison, Madison, Wisconsin 53706,USA \\
  $^{4}$ SLS2 Consulting, San Diego, California 92107, USA \\
  $^{5}$ Southwestern Institute of Physics, PO Box 432,
  Chengdu 610041, People's Republic of China \\
  $^{6}$ Institute of Fusion Science, School of Physical Science and
  Technology, Southwest Jiaotong University, Chengdu,
  610031, Peoples's Republic China \\
}

\date{\today}
\textbf{Abstract}\\
In the presence of energetic particles (EPs), the Long-Lived Mode (LLM)
frequency multiplication with $n=1, 2, 3$ or higher
is often observed on HL-2A, where $n$ is the toroidal mode number.
Hybrid kinetic-MHD model simulations of the
energetic particle (EP) driven kink/fishbone
modes on a static HL-2A-like tokamak using NIMROD code find that,
when the background plasma pressure is relatively high,
and the EP pressure and the beam energy are relatively low,
the mode frequency increases almost linearly with EP pressure,
and the frequency is proportional to $n$ (``frequency multiplication''),
even in absence of any equilibrium plasma rotation.
In addition, the frequency multiplication
persists as the safety factor at magnetic axis $q_0$ varies.
In absence of EPs, the growth rate of the $1/1$ mode is the largest;
however, as the EP pressure increases,
the growth rate of $2/2$ modes or $3/3$ modes becomes
dominant, suggesting that the higher-$n$ modes are
more vulnerable to EPs. These results may shed light on the understanding
about the toroidal mode number dependence
of kink/fishbone modes in the advanced scenarios of tokamaks with weak
or reversed central magnetic shear.

(Some figures may appear in colour only in the online journal)
%\end{xeCJK*}
%\end{CJK*}

\section{\label{Introduction}Introduction}
Confinement improvement and higher beta limit have been
achieved in advanced tokamak (AT) operation scenarios, in which
the $q$-profile is reversely or weakly sheared in the
core region\cite{Taylor97, Gormezano04, Sips05}, and
the AT operation scenarios have been proposed for the
steady-state operation of ITER\cite{Gormezano07, Kikuchi12}.
With weak shear and  plasma pressure
gradient exceeding a critical value in the core region,
besides $1/1$ internal kink mode\cite{Shafranov70, Rosenbluth73, Bussac75},
higher-harmonics modes,  which are usually called
``infernal modes''\cite{Holties96, Manickam87},
also become unstable. Fishbone modes,
which are believed to be driven by the resonant interaction
between internal kink mode and the energetic particles (EPs)
\cite{ChenL84, Coppi86, Betti93, Wang01},
have been observed on many tokamaks with auxiliary heating since
1980s\cite{McGuire83, Heidbrink90, Nave91, Chen10, Xu15}. During
  the fishbone burst events, which usually last about 1
  millisecond, significant loss of EPs has been observed, and
  they degrade the heating efficiency and limit the
  beta achieved in the experiments. Besides, the lost EPs
  can cause the damage of the first wall.
  The time interval between two adjacent fishbone bursts
  is usually several milliseconds, and the $q$-profile is monotonic
  with finite shear in the core region. Similar to the energetic particle (EP)
  driven fishbone modes, another kind of
  EP driven modes are observed in auxiliary heating experiments with
  flat $q$ in the core plasma. The modes
  last 100 milliseconds or longer after the saturation, which are
  called long-lived modes (LLMs) for the long-lasting feature. The LLMs
  often cause loss of EPs and braking of the plasma toroidal rotation.
  Unlike the bursty fishbone modes, the LLMs may serve as an option
  for the AT operation scenarios because of the gentle and continuous energy
  loss in the core plasma of tokamaks. \par
  Because the potential for the AT operation scenarios,
  many experiments have been performed to study LLMs.
  During the NBI heating of plasmas with flat central safety
factor profile, LLMs are observed on HL-2A
that can last for several hundred milliseconds, and the mode frequency is
approximately proportional to the toroidal mode number $n$, a phenomenon
we refer to as frequency multiplication (FM), as shown in figure 1 of
reference \cite{Deng14}, where all $1/1$, $2/2$, $3/3$ and $4/4$ modes
are displayed in the frequency range of $10\,\mkHz\sim 60\,\mkHz$.
The LLMs can be suppressed by ECRH or
SMBI (supersonic molecular beam injection)
which is related to changing of $q$ profile and
pressure gradient\cite{Deng14}. On MAST, the FM
for LLMs is also observed during NBI heating, not only for weakly sheared,
but also for slightly and reversely sheared $q$-profiles,
where the critical plasma pressure, above which the mode becomes
unstable, increases with $n$ \cite{Chapman10}.
For KSTAR plasma with ECRH and NBI, the pressure-driven LLMs lasts
up to 40 seconds along with the FM,
when $q$-profile is above $2$ with broad weak shear in the core
region\cite{Lee16}. As a characteristic feature of LLMs, the FM
has not been well understood.\par
Several analytical or numerical studies have investigated the
properties of LLMs.
Solving an analytical dispersion relation developed for LLMs based on
the HL-2A equilibrium, Zhang \emph{et al} find the growth rate of
the $2/2$ mode greater than that of the $1/1$ mode,
and the frequency is proportional to $n$ ($n=1,2,3$) \cite{Zhang14}.
However, they have not considered the FM
in detail, and the $3/3$ mode is stable in their calculations,
different from the experimental observations.
Including the equilibrium rotation in the dispersion relation,
Xie reports that fishbone mode can turn into LLM with low magnetic shear,
when the EP pressure and the plasma rotation frequency both exceed certain
critical value \cite{Xie22}, however, neither the $n>1$ mode or the
FM is discussed. Using the kinetic-MHD hybrid code M3D-K
for simulations of a circular cross-section tokamak with central
weakly sheared $q$-profile, Ren \emph{et al} obtain the
FM of LLMs in the presence of EPs,
where the growth rates of higher harmonics ($n>1$) can be greater
than $1/1$ mode\cite{Ren17}. Their study is based on the analytical
equilibrium profile for a model tokamak, and
the conditions in which the FM is broken remains unclear. \par
In this work, we study the FM of LLMs, with a focus on its
onset conditions. Based on the equilibrium generated from the
LLM experiments on HL-2A, using the hybrid kinetic-MHD model implemented
in the NIMROD code\cite{Sovinec04, Kim08, Brenna12}, our simulations show
that the EP-driven fishbone mode frequency is almost proportional to $n$
in absence of any equilibrium toroidal flow, and the FM
appears when the background plasma pressure gradient
is above certain threshold, but neither the beam energy or
the  EP $\beta$ fraction
(the ratio between EP pressure and total plasma pressure)
is too high. The $1/1$ mode is
the most unstable when the EP pressure is not too high\cite{Zou21}.
However, higher-$n$ kink/fishbone modes are found more
dominant when EP pressure becomes relatively high. \par
The rest of paper is organized as follows. The
simulation model is reviewed briefly in
section 2. The simulation set-up is described
in section 3, which is followed by the report
on the main results in section 4. First, the
$q_0$ effect on kink/fishbone mode is studied
in both absence or presence of EPs,
and then the effects of EP $\beta$ fraction, beam energy, and
EP pressure gradient are also investigated.
The summary and discussion are provided in section 5. \par
\section{Simulation model}
The simulation model is based on the following hybrid kinetic-MHD
equations implemented in the NIMROD code\cite{Sovinec04, Kim08}.
\begin{equation}
  \frac{\partial\rho}{\partial t} + \nabla\cdot(\rho\mbV) = 0
\end{equation}
\begin{equation}
  \rho \left(\frac{\partial\mbV}{\partial t}+\mbV\cdot
  \nabla\mbV\right) = \mbJ\times\mbB-\nabla p_{\text{b}}-\nabla
  \cdot\bm P_{\text{h}}
\end{equation}
\begin{equation}
  \frac{n}{\Gamma-1} \left(\frac{\partial T}{\partial t}
  +\mbV\cdot\nabla T\right) = 0
\end{equation}
\begin{equation}
  \frac{\partial\mbB}{\partial t}=-\nabla\times\mbE
\end{equation}
\begin{equation}
  \nabla\times\mbB=\mu_0\mbJ
\end{equation}
\begin{equation}
  \mbE=-\mbV\times\mbB
\end{equation}
where $\rho$, $\mbV$, $\mbJ$, $p_{\text{b}}$, $n$ and $T$ are
the mass density, center of mass velocity, current density,
plasma pressure, number density, and temperature of the main species plasma,
$\mbE$ and $\mbB$ are the electric and magnetic fields,
$\Gamma$ and $\mu_0$ are the specific heat ratio and the vacuum permeability,
respectively. Static equilibrium is considered in order to
exclude effects from the equilibrium rotations. For $n_{\text{h}}\ll n_{\text{b}}$
and $\beta_{\text{h}}\sim \beta_{\text{b}}$, where $n_{\text{b}}$ ($n_{\text{h}}$) is
the main species plasma (energetic particle) number density,
 and $\beta_{\text{b}}$ ($\beta_{\text{h}}$) is the ratio
of main species background plasma (energetic particle) pressure
to magnetic pressure, the pressure tensor $\bm P_{\text{h}}$
in the momentum equation can couple the kinetic effects from EPs,
which are governed by the drift-kinetic equation\cite{Kim08}.
Here $\bm P_{\text{h}} = \bm P_{\text{h0}}+\delta \bm P_{\text{h}}$,
where $\bm P_{\text{h0}}$ is assumed isotropic, and
$\delta \bm P_{\text{h}}$ is defined as
\begin{equation}
  \delta \bm P_{\text{h}} = \begin{pmatrix}
    \delta p_{\bot} & 0 & 0 \\
    0 & \delta p_{\bot} & 0 \\
    0 & 0 & \delta p_{\parallel}
  \end{pmatrix}
\end{equation}
$\delta p_{\bot}$ is pressure due to hot particle motions perpendicular
to the magnetic field, and $\delta p_{\parallel}$ is pressure due
to the hot particle motions parallel to the magnetic field. \par
\section{Simulation setup}\label{sec:setup}
\begin{figure}[ht]
  \centering
  \includegraphics[width=6.7cm]{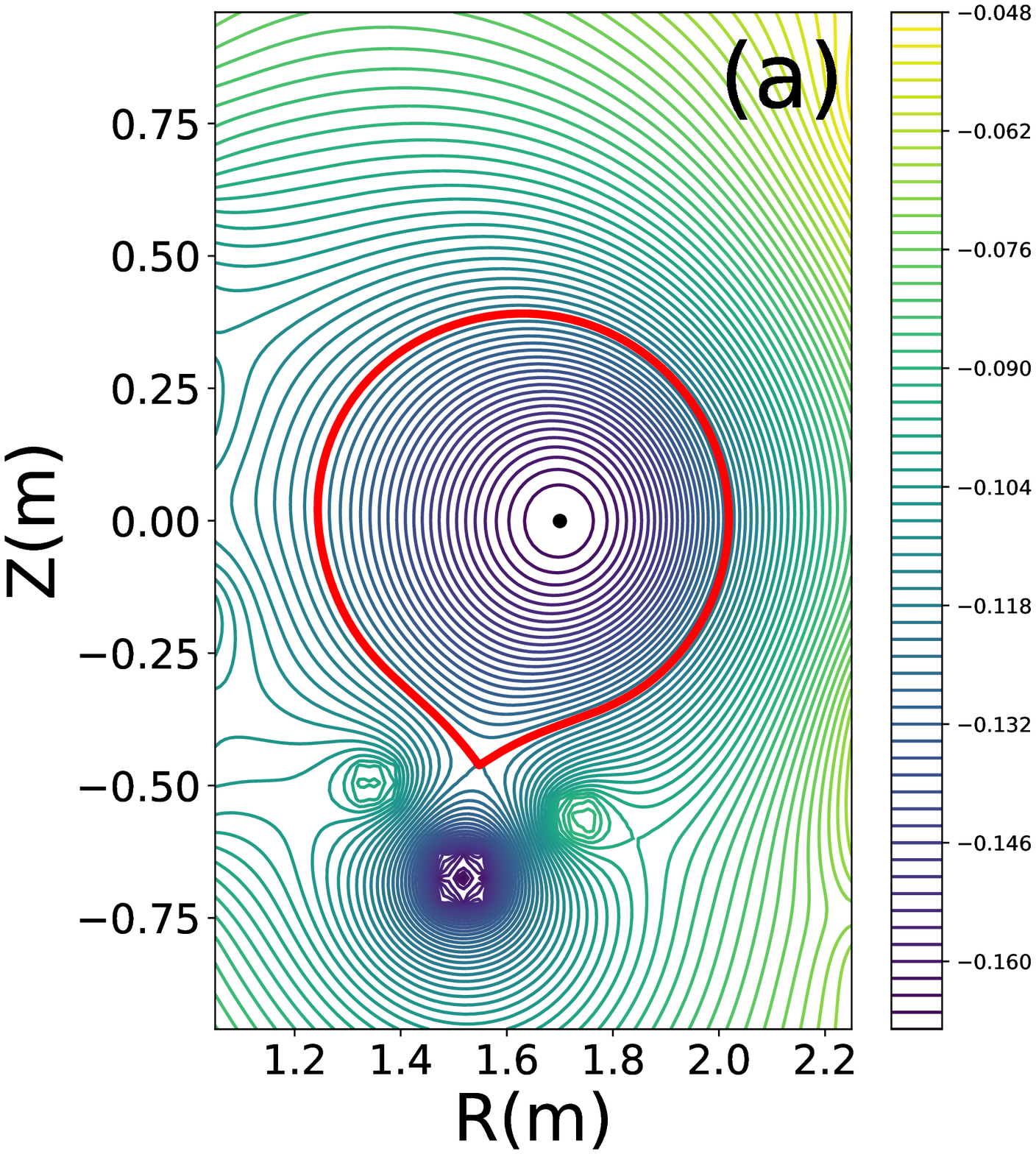}
  \includegraphics[width=7.3cm]{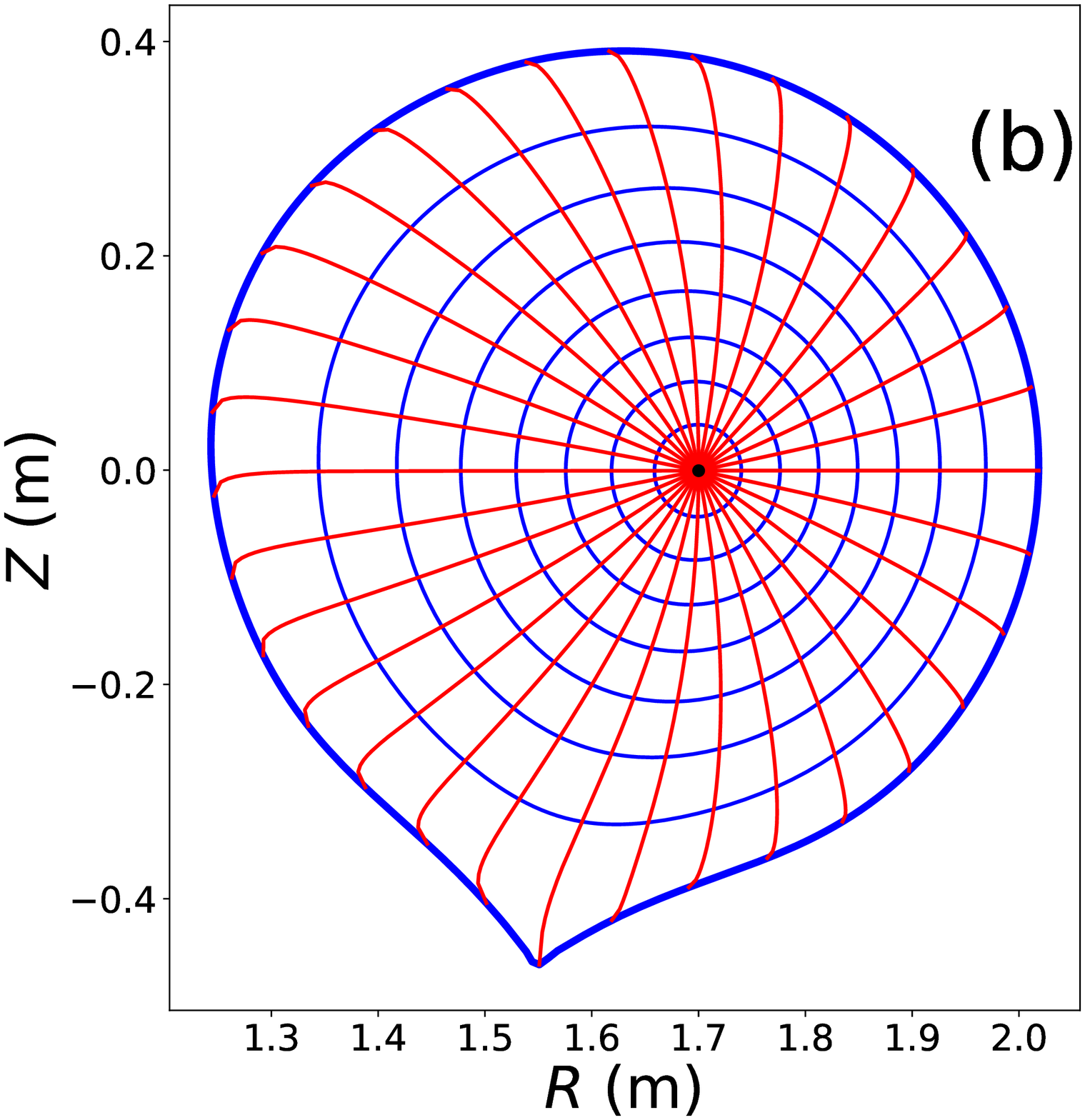}
  \caption{\label{fig:fig1} (a) Contour plot of equilibrium poloidal
    flux with LCFS (the red curve).
    (b) The sketch map of flux coordinates with uniform poloidal
    flux and equal poloidal arc length.}
\end{figure}

\begin{figure}[ht]
  \centering
  \includegraphics[width=8.0cm]{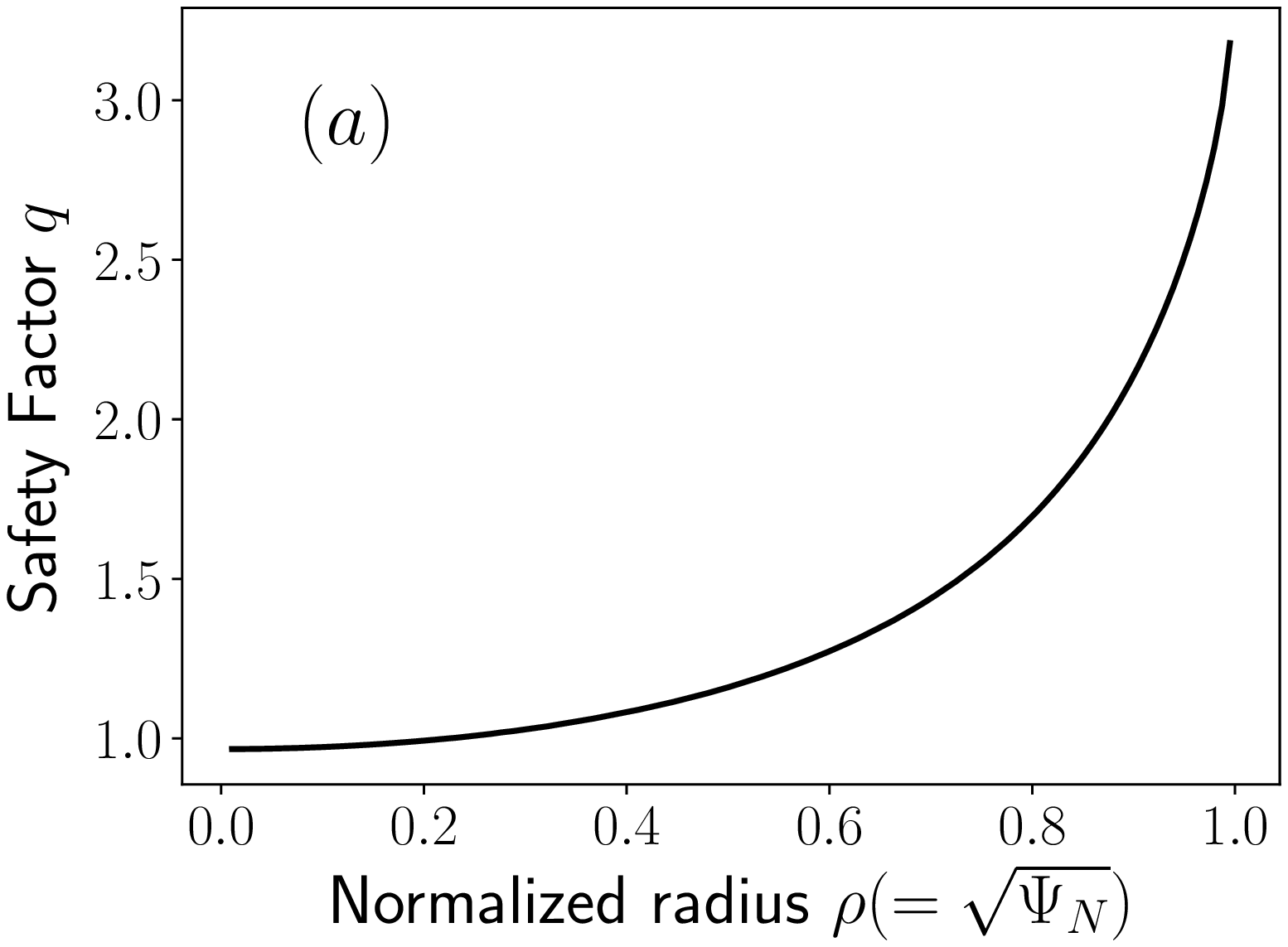}
  \includegraphics[width=8.0cm]{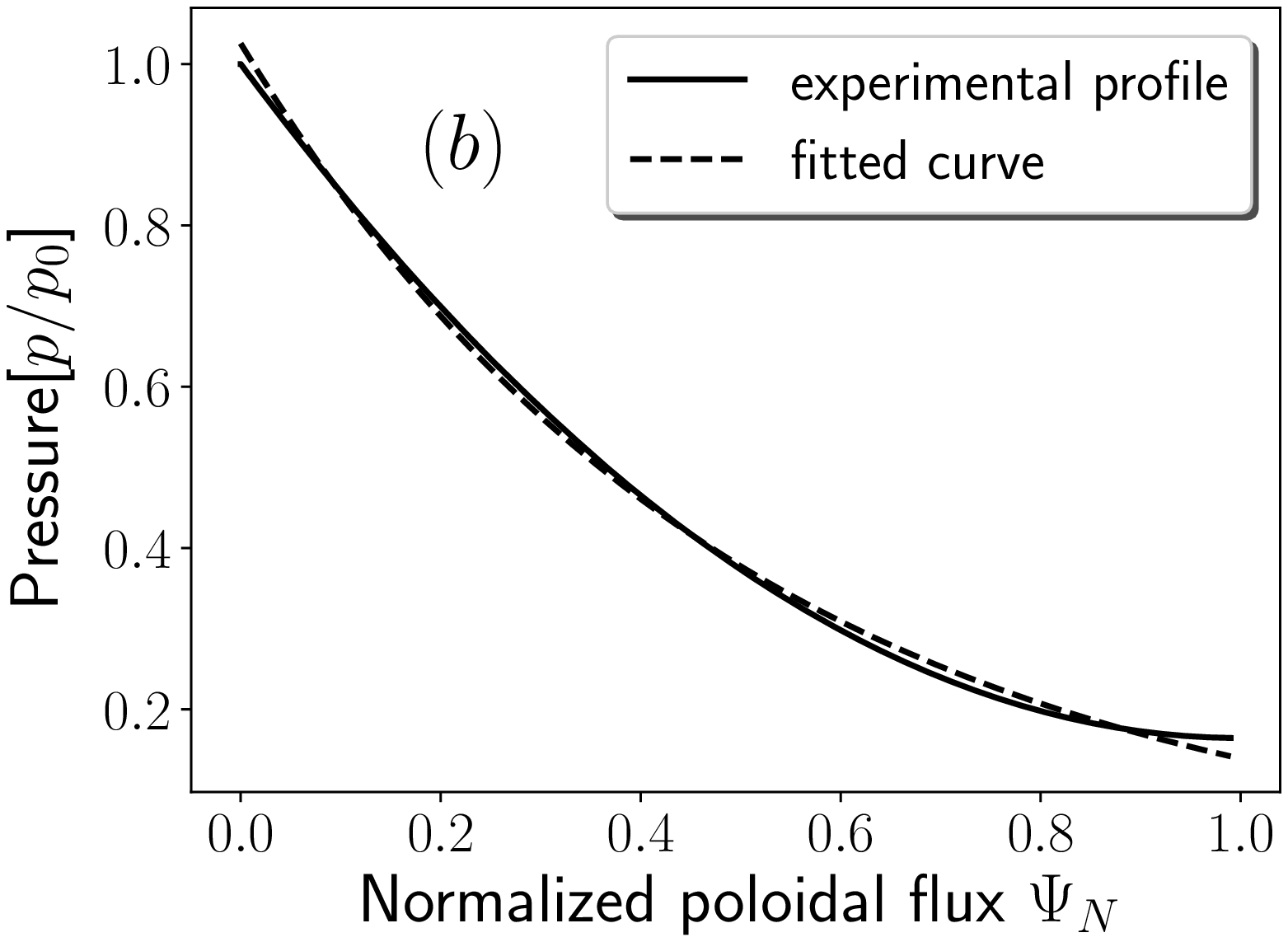}
  \caption{\label{fig:fig2}(a) The safety factor profile and
  (b) pressure profile in HL-2A discharge \#16074 at moment 452 (M452).}
\end{figure}

An \efit equilibrium reconstruction based on
HL-2A discharge \#16074, is used in our simulation\cite{Deng14}.
The equilibrium flux surfaces and the mesh in
magnetic flux coordinates within the last
closed flux surface (LCFS) are shown in figure \ref{fig:fig1}.
Although the LCFS is up-down asymmetric,
the flux surfaces are close to circles in the core region.
The internal kink/fishbone mode locates
in the region $0<\sqrt{\psi/\psi_0}<0.2$, where the safety factor $q$-profile
is close to unity along with weak magnetic shear [figure \ref{fig:fig2}(a)].
Here $\psi$ is the poloidal magnetic flux
and $\psi_0$ is the total poloidal magnetic flux within the LCFS. \par

The EPs from NBI are initialized with the slowing-down
distribution function\cite{Kim08},
\begin{equation}
  f_0 = \frac{P_0\exp \left(\frac{P_{\zeta}}{\psi_{\text{n}}}\right)}{
    \varepsilon^{3/2}+\varepsilon_{\text{c}}^{3/2}}
\end{equation}
where $P_0$ is the normalization constant,
$P_{\zeta}=g\rho_{\parallel}-\psi_{\text{p}}$ is the canonical toroidal momentum,
$g=RB_{\phi}$, $\rho_{\parallel}=mv_{\parallel}/qB$, $\psi_{\text{p}}$ is
the poloidal flux, $\psi_{\text{n}} = h\psi_0$ ,
$\psi_0$ is the total flux and the parameter $h$
is used  to match the spatial
profile of the equilibrium, $\varepsilon$ is the particle energy,
and $\varepsilon_{\text{c}}$ is the critical slowing
down energy\cite{Goldston00}
\begin{equation}\label{eq:critdef}
  \varepsilon_{\text{c}} = \left(\frac{3}{4}\right)^{2/3}\left
  (\frac{\pi m_{\text{i}}}{m_{\text{e}}} \right)^{1/3} T_{\text{e}}
\end{equation}
with $m_{\text{i}}$ being the ion mass,
$m_{\text{e}}$ the electron mass, and $T_{\text{e}}$ the electron
temperature. Beam ions collide dominantly with the
background electrons (ions)
for $\varepsilon>\varepsilon_{\text{c}}$
($\varepsilon<\varepsilon_{\text{c}}$).\par
The EPs are loaded into the physical space following
the profile $p=p_0\exp(-h\psi/\psi_0)$,
where $p_0$ is pressure at magnetic axis,
EP pressure gradient becomes higher as $h$ decreases
[figure \ref{fig:fig2}(b)].
For linear simulations, we set resistivity $\eta = 0$,
and a total $10^6$ simulation particles are
prescribed in the poloidal plane with $64\times 64$ finite elements.
Other main parameters are input from equilibrium \cite{Zhang14},
with the major radius $R=1.65\,\mm$, the minor
radius $a=0.40\,\mm$, the toroidal magnetic field
$B_0=1.37\,\mT$, and the number density is set to be constant
in the radial direction  with $n=2.44\times 10^{19}\,\mm^{-3}$.
The case with $q_0=0.9$, $\beta_f=\beta_h/\beta_0=0.1$, $h=0.25$
and beam energy $\varepsilon_b=10\mkeV$
is specified as the standard reference case for comparisons. \par
\section{Simulation results}
\begin{figure}[ht]
  \centering
  \includegraphics[width=8.0cm]{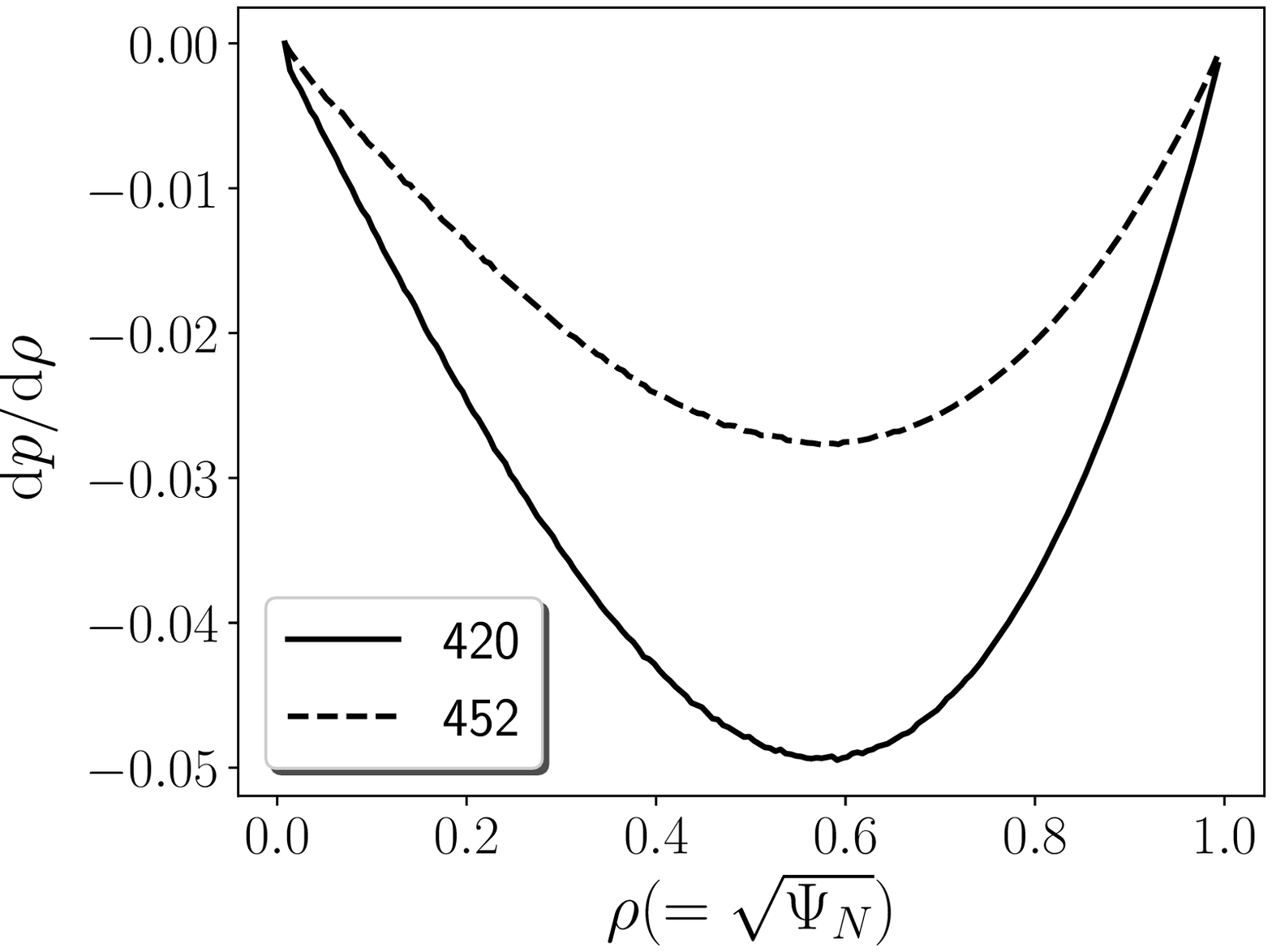}
  \caption{\label{fig:fig3} The pressure gradient profiles
    of equilibriums M420 and M452 from discharge \#016074, respectively.}
\end{figure}

LLMs are observed on the experiments with NBI heating,
and they can be controlled by ECRH and SMBI, subject
to the influences from magnetic shear,  pressure gradient and
NBI beam properties\cite{Deng14}. In order to investigate these effects
in our simulations, we scan $q_0$ to study effects of magnetic
shear on FM for LLMs, and the EP beta fraction $\beta_f$,
beam energy $\varepsilon_b$ and EP pressure gradient to study
the effects of EPs on FM. We choose two representative time moments of the
discharge \#016074, one is at $420\,\mm\ms$ (M420), and the another
is at $452\,\mm\ms$ (M452). The M420 case is at early stage
of the mode growth, and the M452 case is near the saturation of the mode,
as shown in the figure 1 of reference \cite{Deng14}.
The pressure gradient profiles of the M420
and M452 cases are shown in figure \ref{fig:fig3}.
The pressure gradient of the M420 case is almost two times of that
of the M452 case. We use the two pressure profiles to study
the effects of background  plasma pressure gradient on FM.\par

\subsection{$q_0$ effects in absence of EPs}\label{subsec:qeff}
\begin{figure}[ht]
  \centering
  \includegraphics[width=8.0cm]{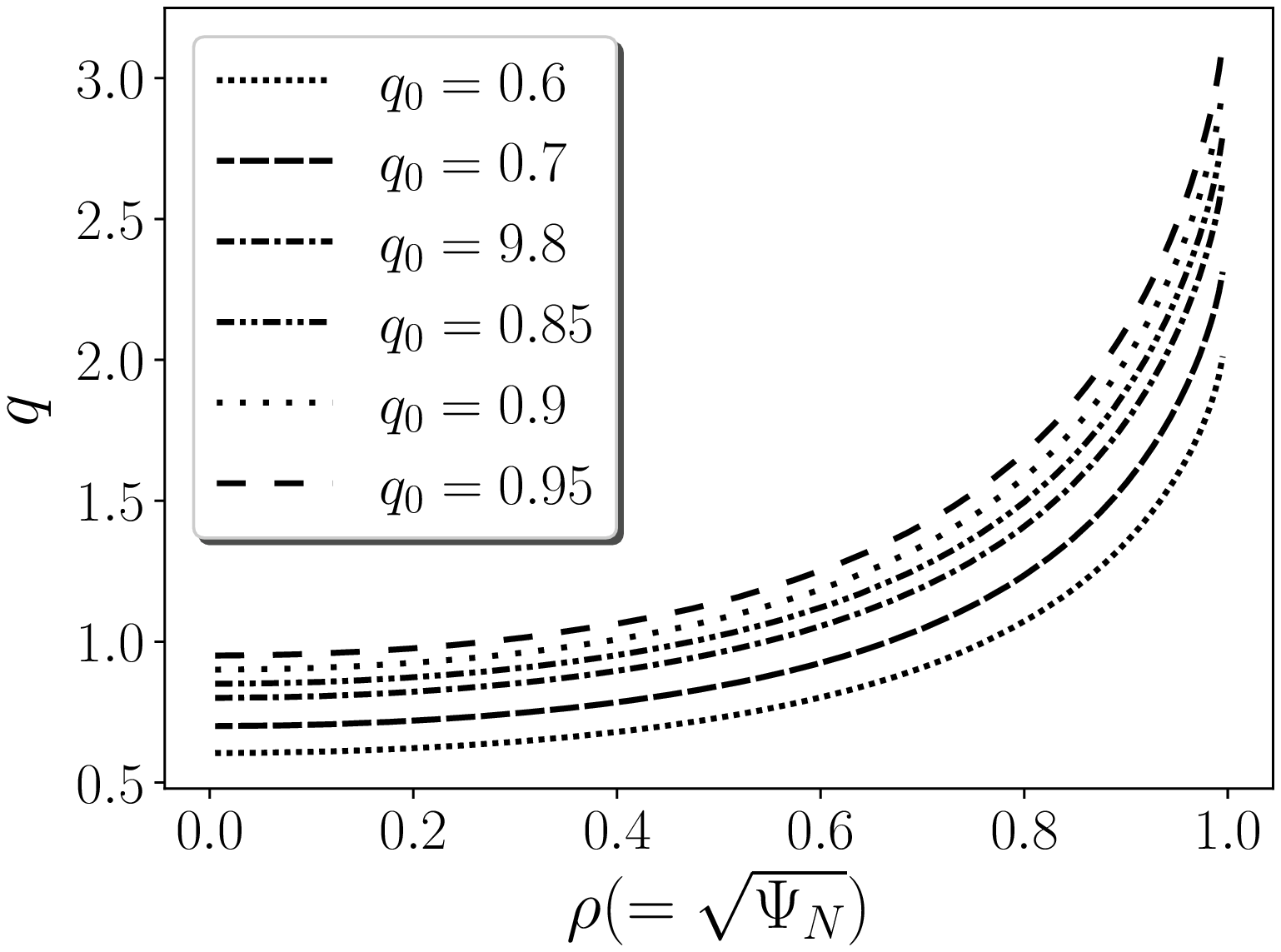}
  \caption{\label{fig:fig4} $q$-profiles with different $q_0$.}
\end{figure}

First we scan $q_0$ to study its effects on the modes without
EP. As shown in figure \ref{fig:fig4}, varying $q_0$ along shifts the
$q$-profile up or down entirely with minimal change in its shape.

\begin{figure}[ht]
  \centering
  \includegraphics[width=8.0cm]{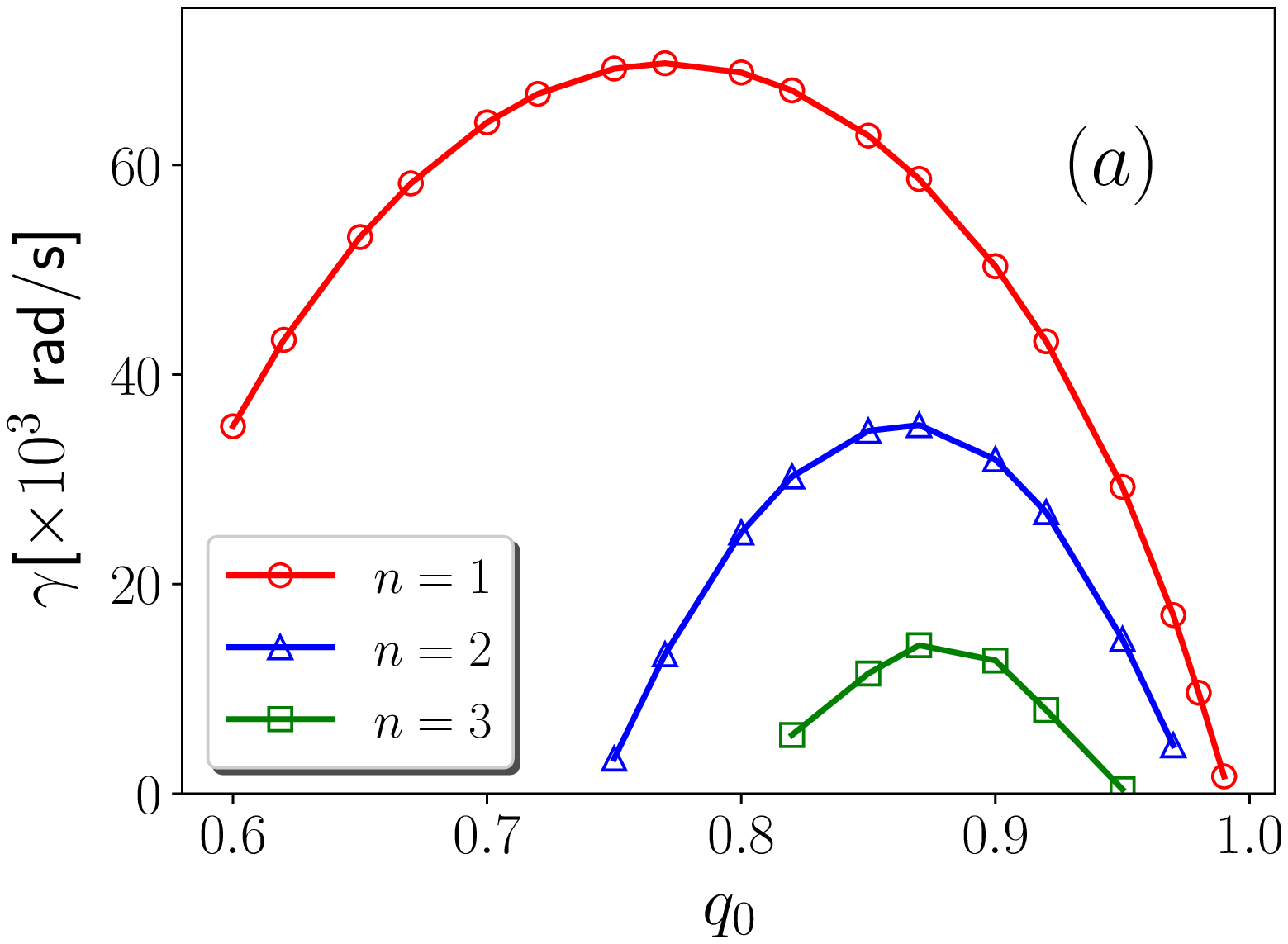}
  \includegraphics[width=8.0cm]{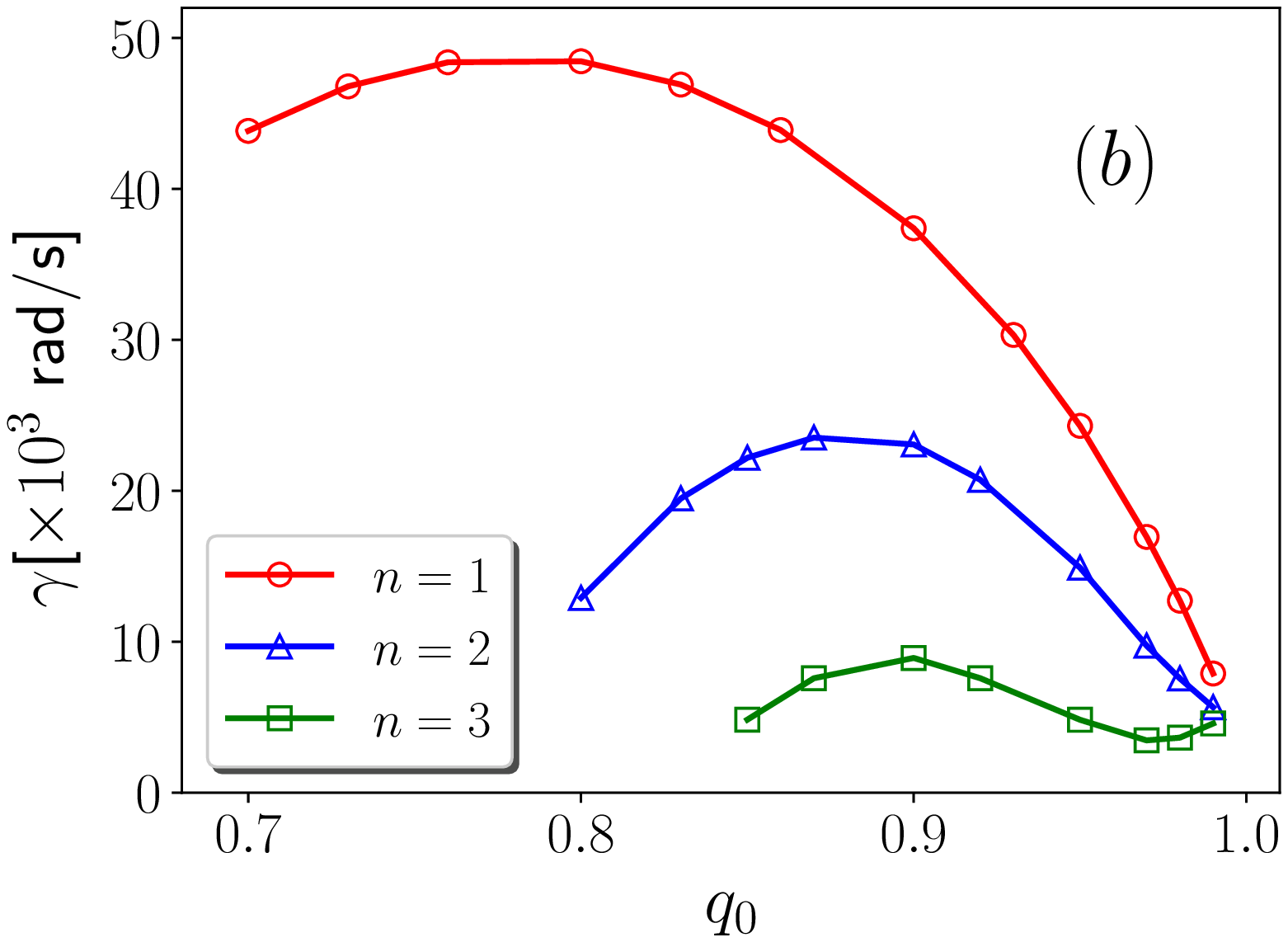}
  \caption{\label{fig:fig5} The growth rate
    as a function of $q_0$ for $m/n=1/1, 2/2, 3/3$ modes
    in (a) the M420 and (b) the M452 equilibriums, respectively.}
\end{figure}

\begin{figure}[ht]
  \centering
  \begin{overpic}[scale=0.21]{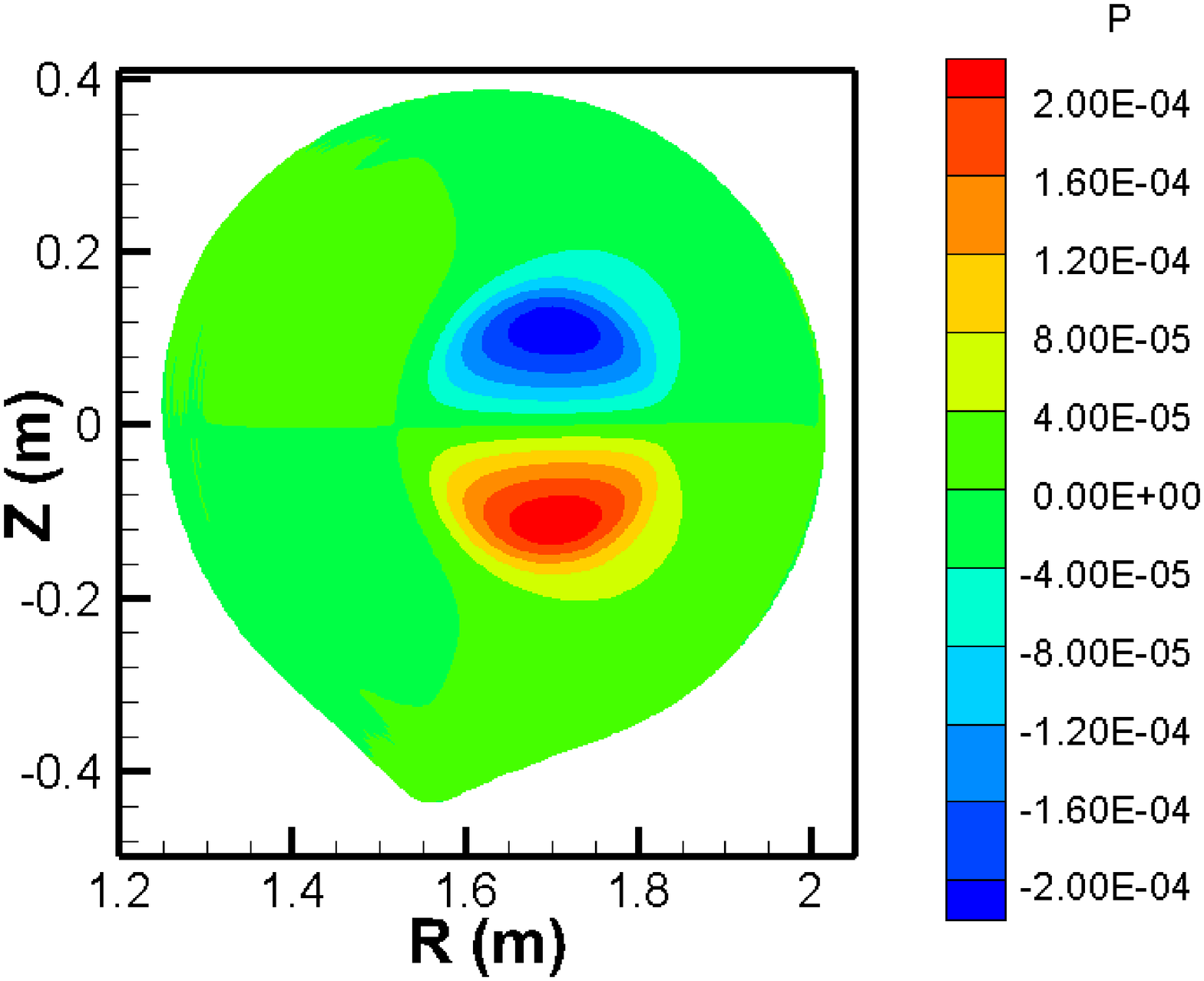}
    \put(13, 68){$(a1)$}
  \end{overpic}
  \begin{overpic}[scale=0.21]{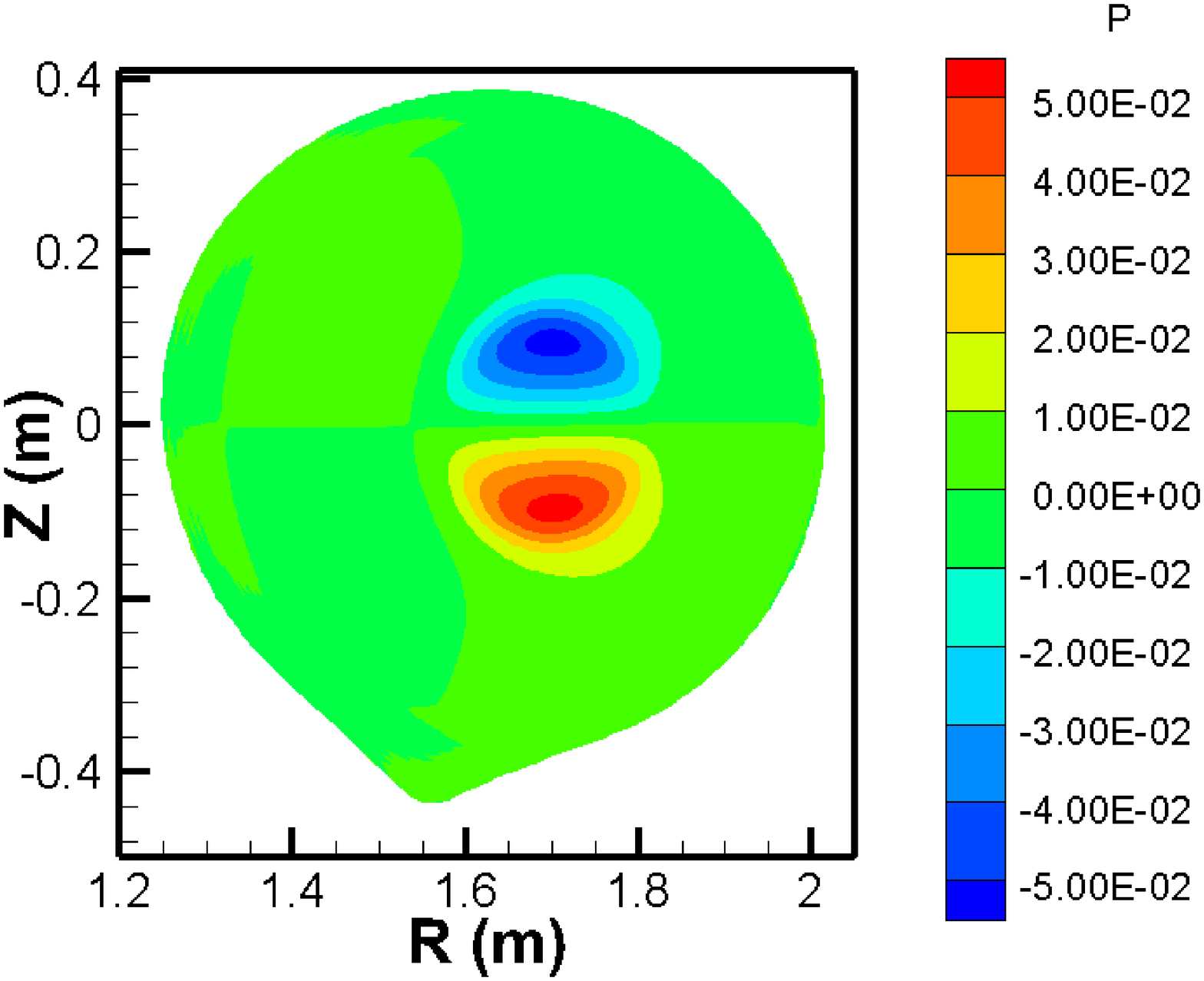}
    \put(13, 68){$(b1)$}
  \end{overpic}
  \begin{overpic}[scale=0.21]{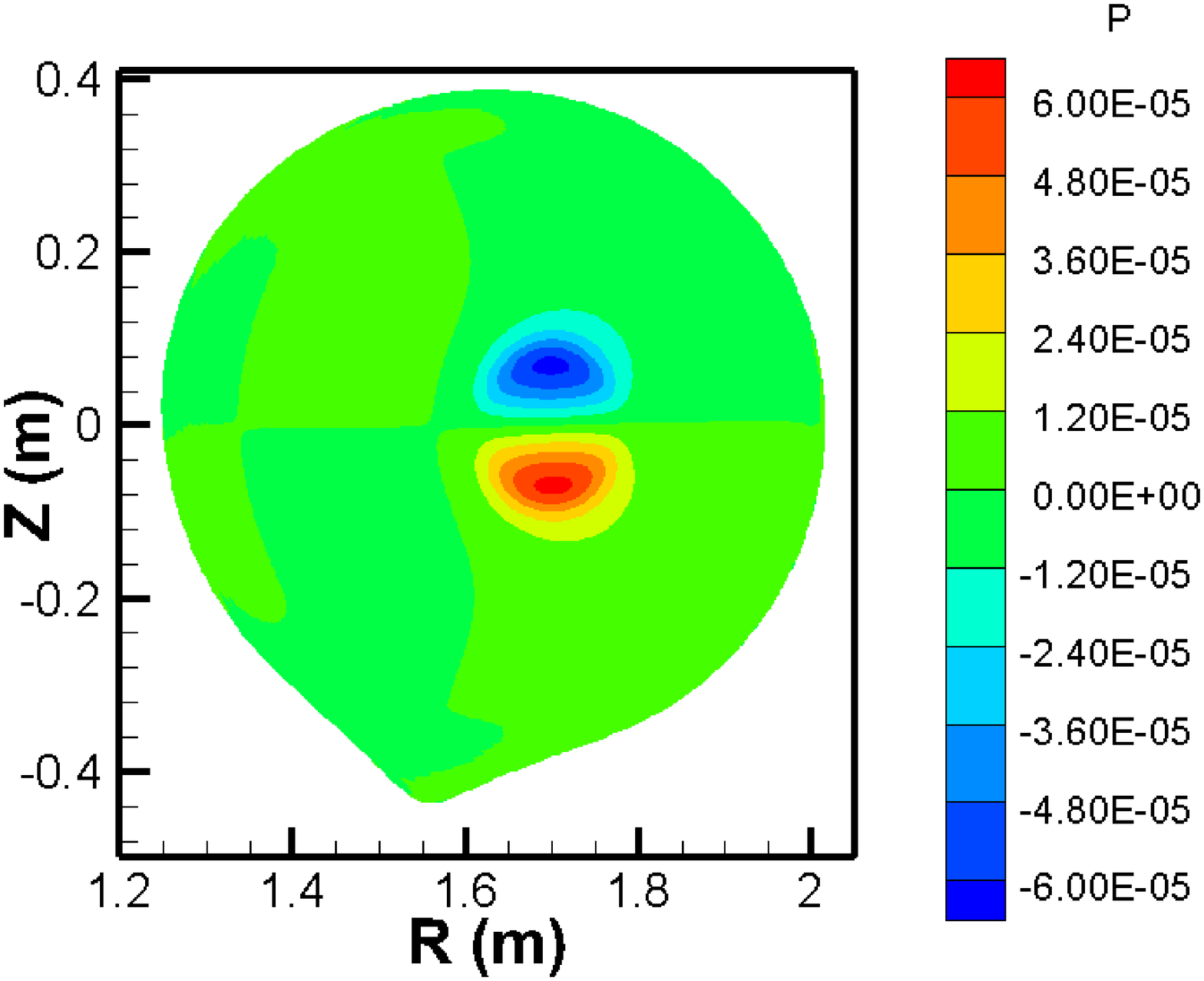}
    \put(13, 68){$(c1)$}
  \end{overpic}
  \begin{overpic}[scale=0.21]{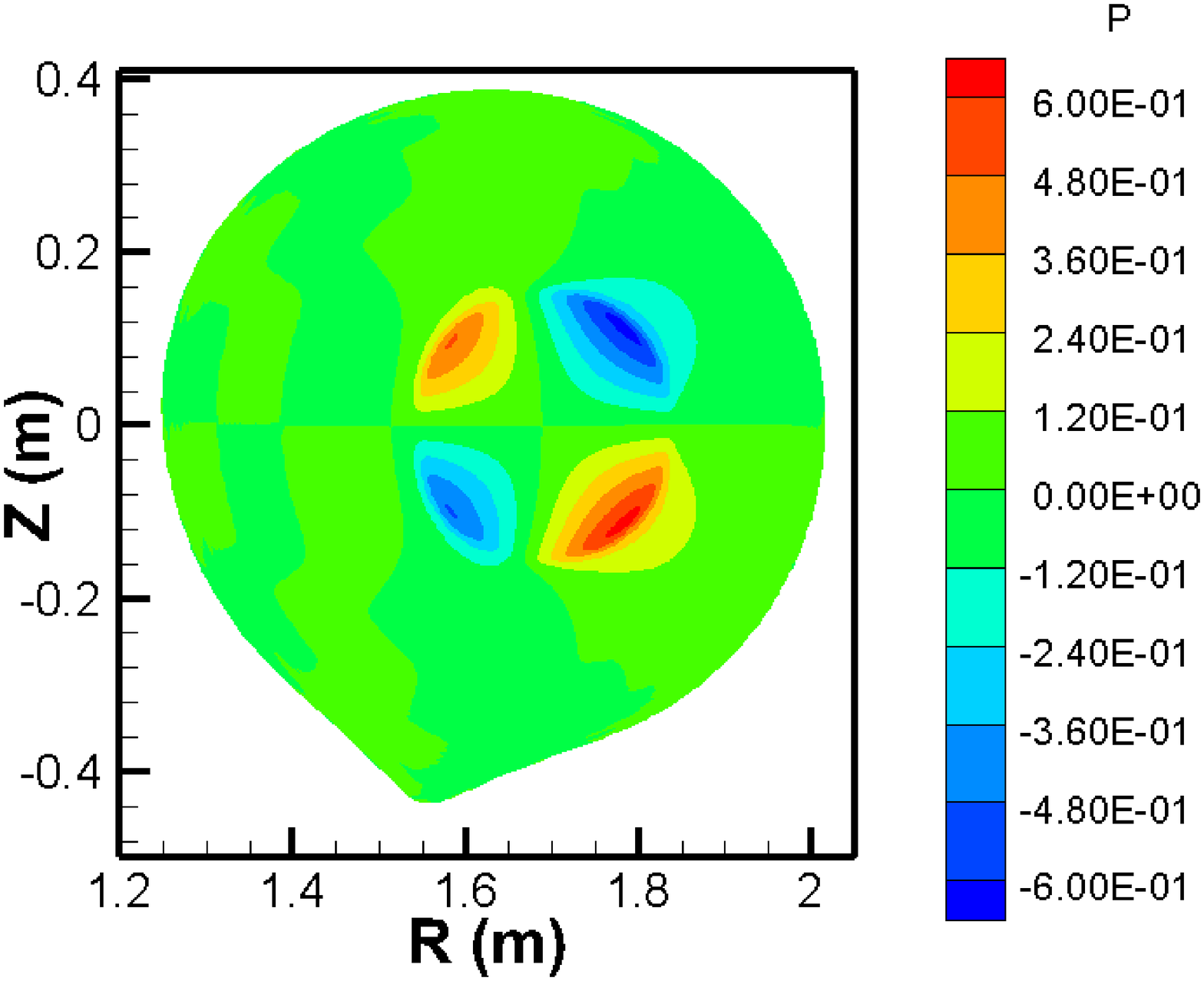}
    \put(13, 68){$(a2)$}
  \end{overpic}
  \begin{overpic}[scale=0.21]{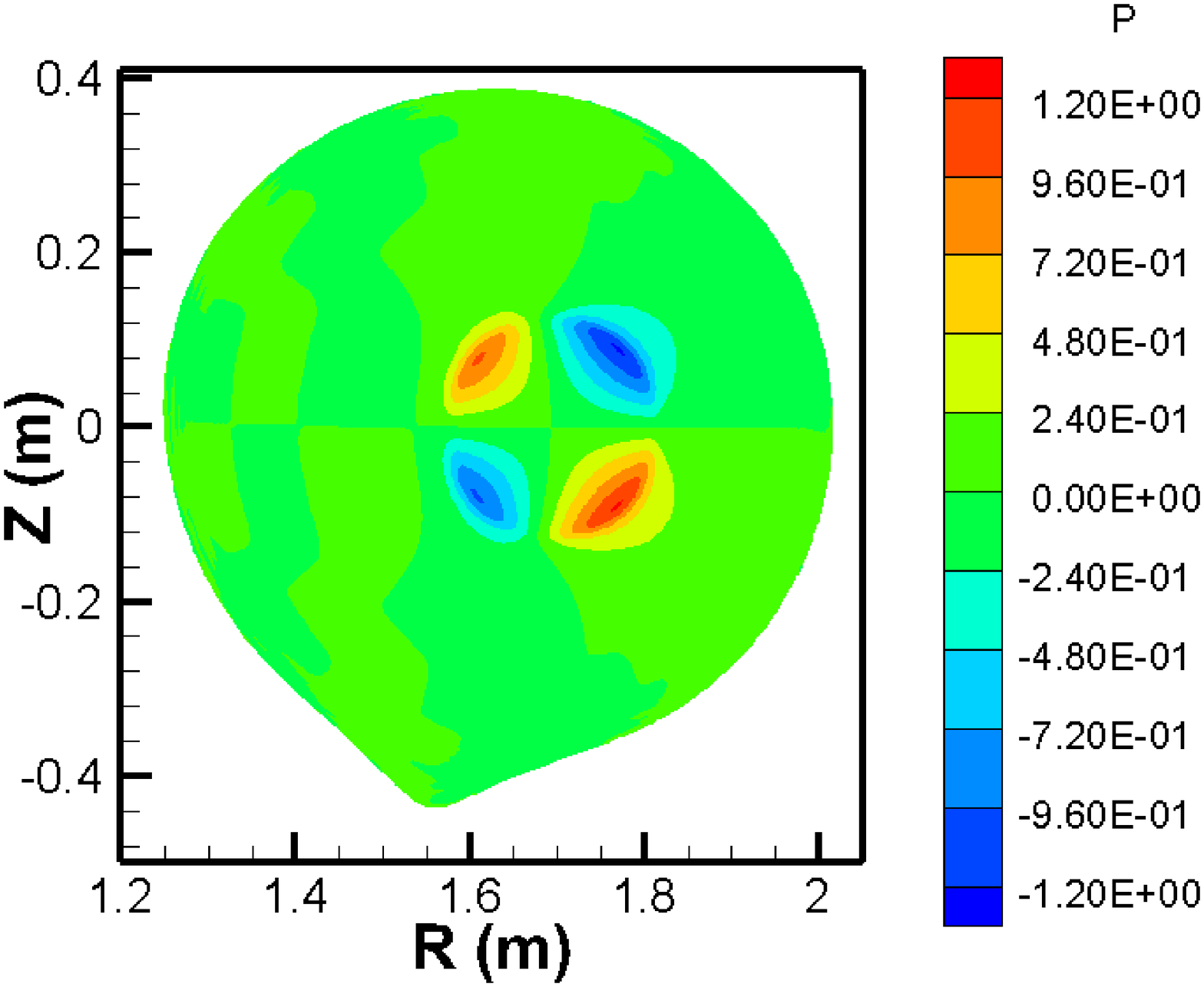}
    \put(13, 68){$(b2)$}
  \end{overpic}
  \begin{overpic}[scale=0.21]{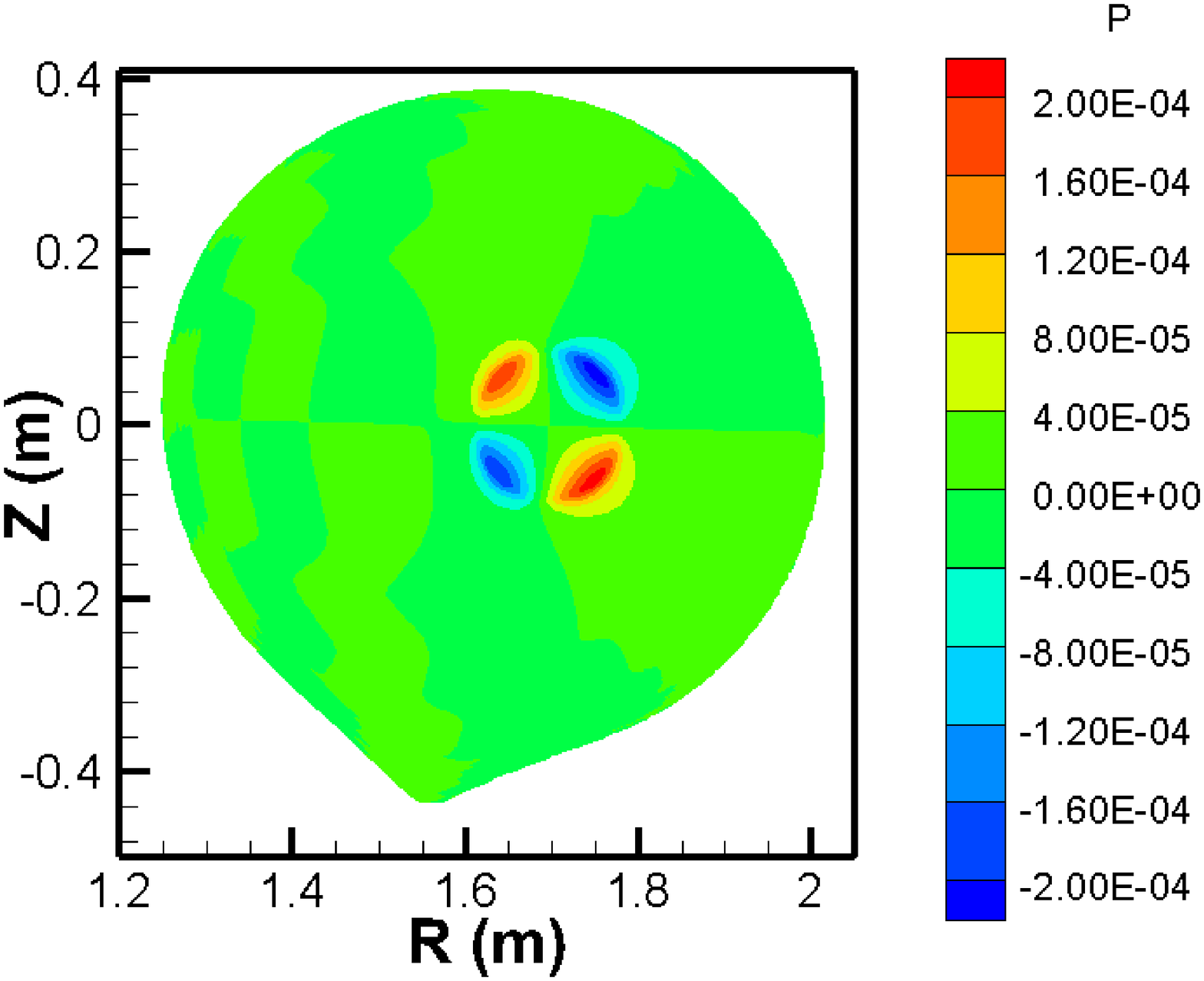}
    \put(13, 68){$(c2)$}
  \end{overpic}
  \begin{overpic}[scale=0.21]{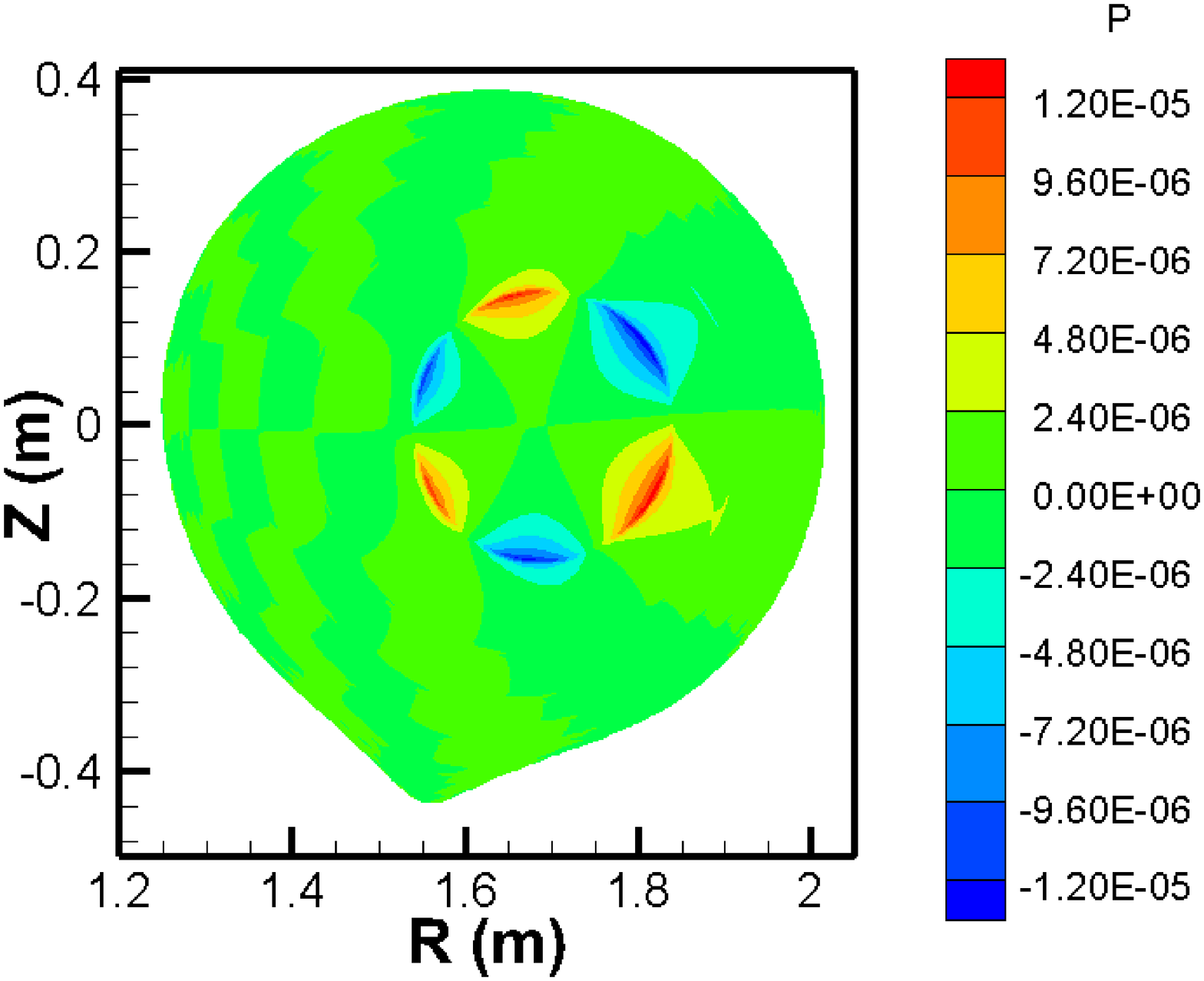}
    \put(13, 68){$(a3)$}
  \end{overpic}
  \begin{overpic}[scale=0.21]{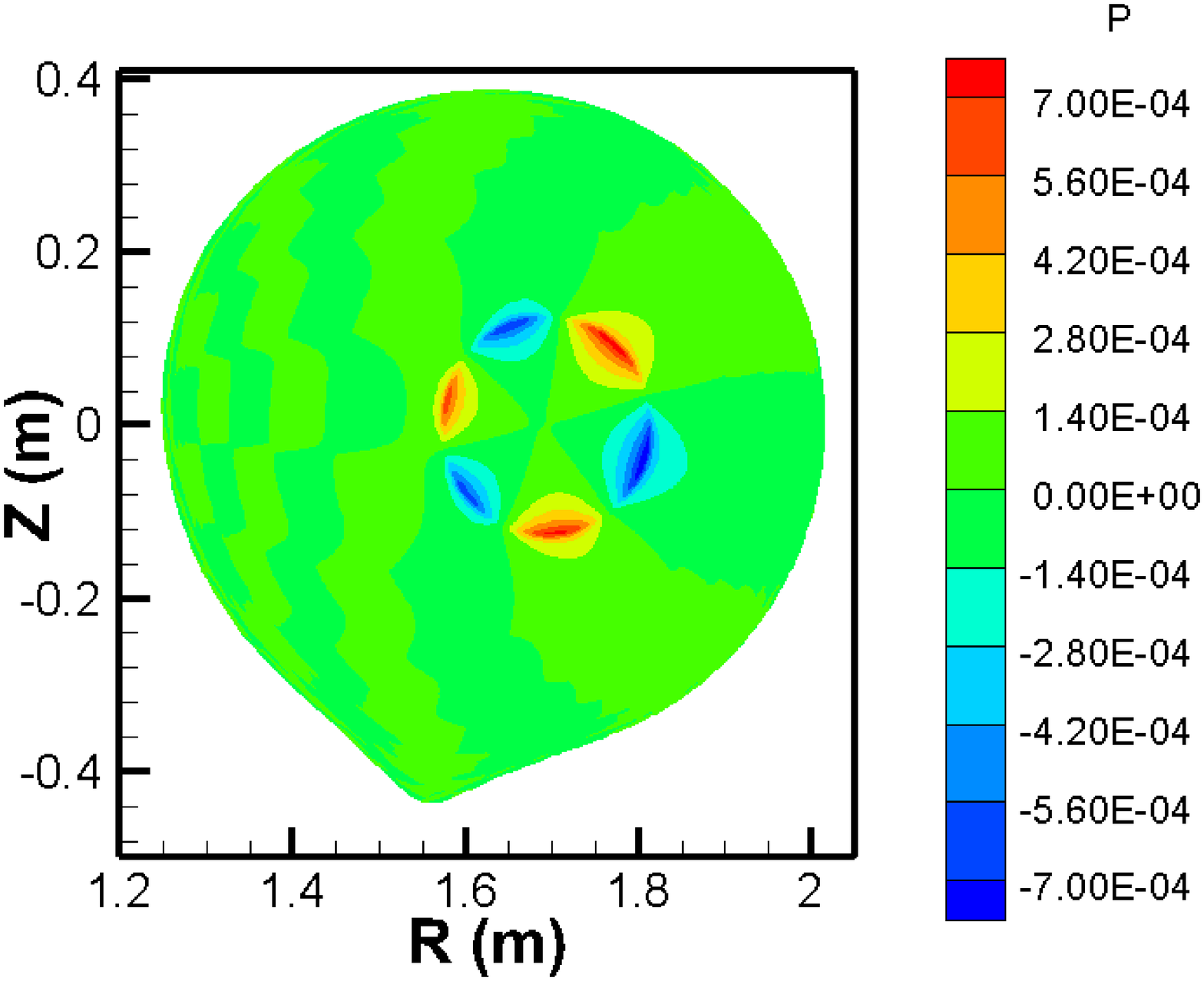}
    \put(13, 68){$(b3)$}
  \end{overpic}
  \begin{overpic}[scale=0.21]{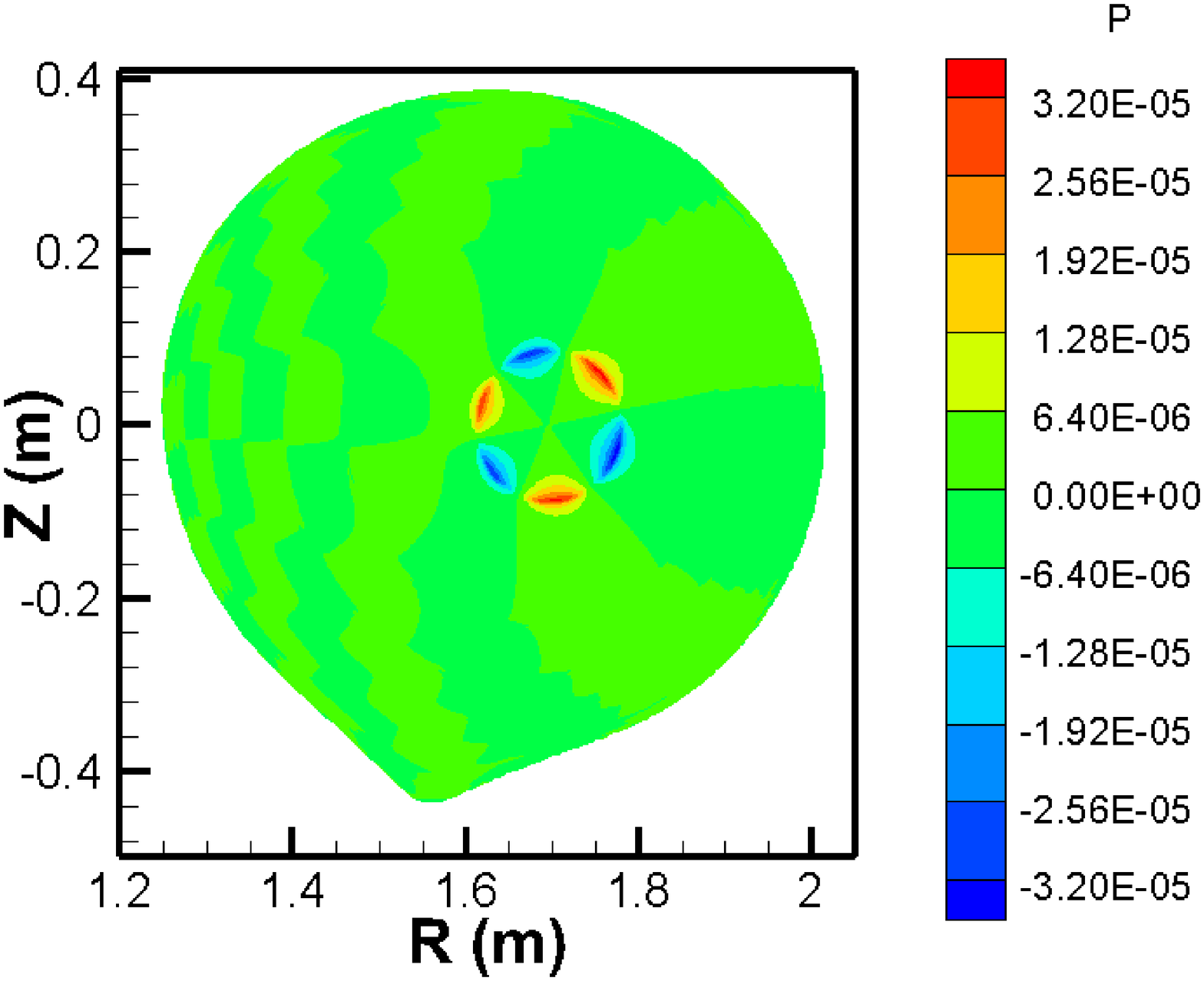}
    \put(13, 68){$(c3)$}
  \end{overpic}
  \caption{\label{fig:fig6} Pressure perturbation contours of
    the $1/1$ modes [(a1), (b1) and (c1)],
    the $2/2$ modes [(a2), (b2) and (c2)] and
    the $3/3$ modes [(a3), (b3) and (c3)] with
    $q_0=0.85$ [(a1), (a2) and (a3)],
    $q_0=0.9$ [(b1), (b2) and (b3)] and
    $q_0=0.95$ [(c1), (c2) and (c3)]
    based on the M452 equilibrium.}
\end{figure}

For the two equilibriums from the M420 and M452 case,
linear NIMROD calculations show that
the $m/n$ = $1/1$, $2/2$ and $3/3$ kink modes are unstable
when $q_0<1$ in absence of energetic particles,
where $m$ is the poloidal mode number.
The linear growth rate of kink modes decreases with $n$,
namely, $\gamma_{1/1}>\gamma_{2/2}>\gamma_{3/3}$.
As $q_0$ increases, the growth rate increases
first and then decreases to zero as $q_0$ approaches
unity. The $q_{0max}$, the value of $q_0$ at which the
growth rate reaches the maximum,
increases with $n$ (figure \ref{fig:fig5}).
The contour plots of the plasma pressure perturbation
for the M452 case (figure \ref{fig:fig6}) show that the mode structure
shrinks in size as $q_0$ approaches to unity.
The contour plots for the M420 case are similar, and thus are not repeated
here. For the $1/1$ kink mode, the mode structure is global within
the $q<1$ region. For $2/2$ and $3/3$ kink modes, the mode structures
become more localized around $q=1$ surface, which is consistent with the
theory prediction \cite{Rosenbluth75}.\par
In absence of EPs, these MHD modes are purely growing instabilities
without real frequency or FM phenomenon. This is in agreement
with the observation that LLMs only occur in presence of NBI heating
or other auxiliary heating. \par

\subsection{$q_0$ effects in presence of EPs}\label{subsec:qep}
\begin{figure}[ht]
  \centering
  \includegraphics[width=8.0cm]{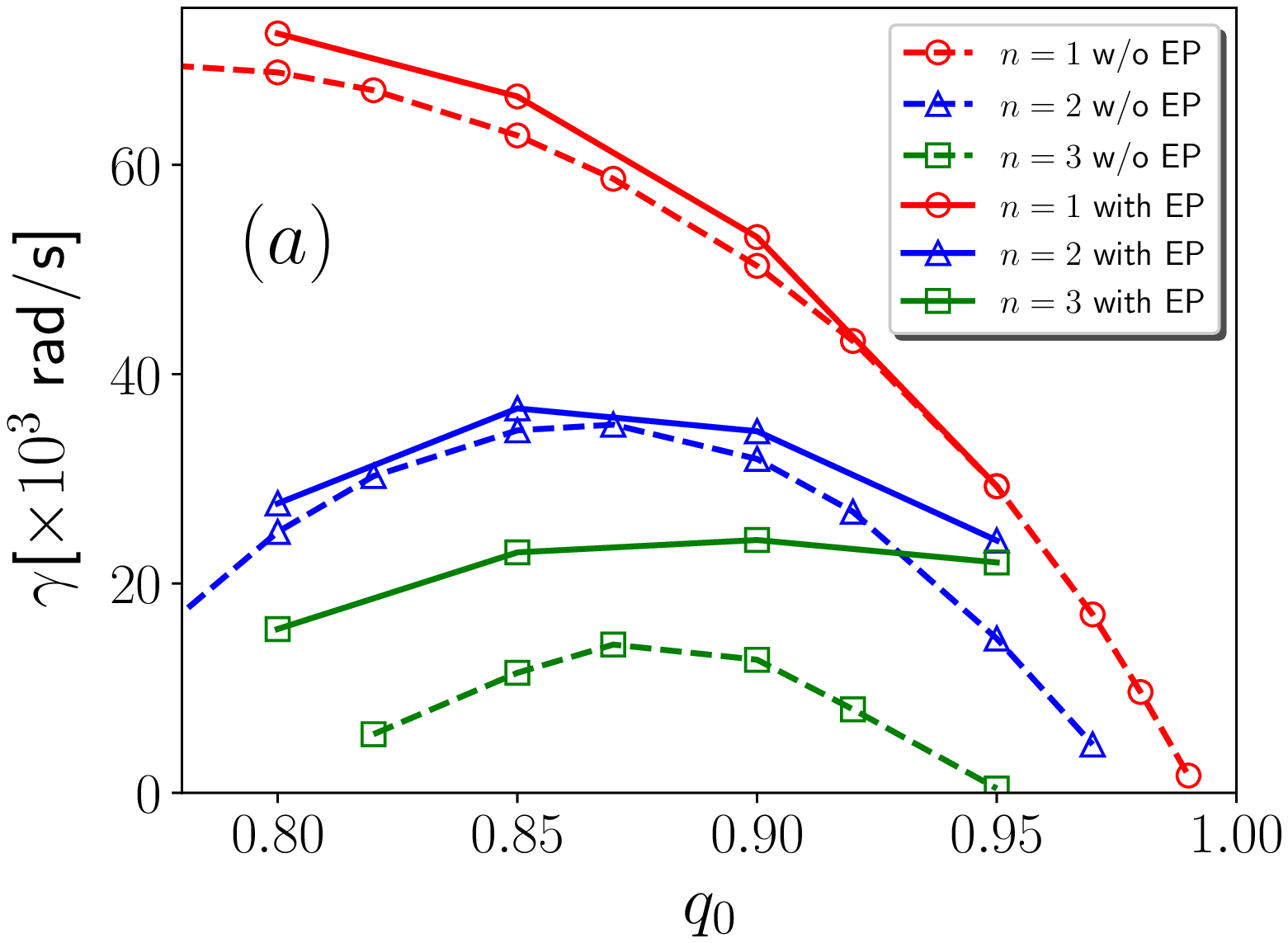}
  \includegraphics[width=8.0cm]{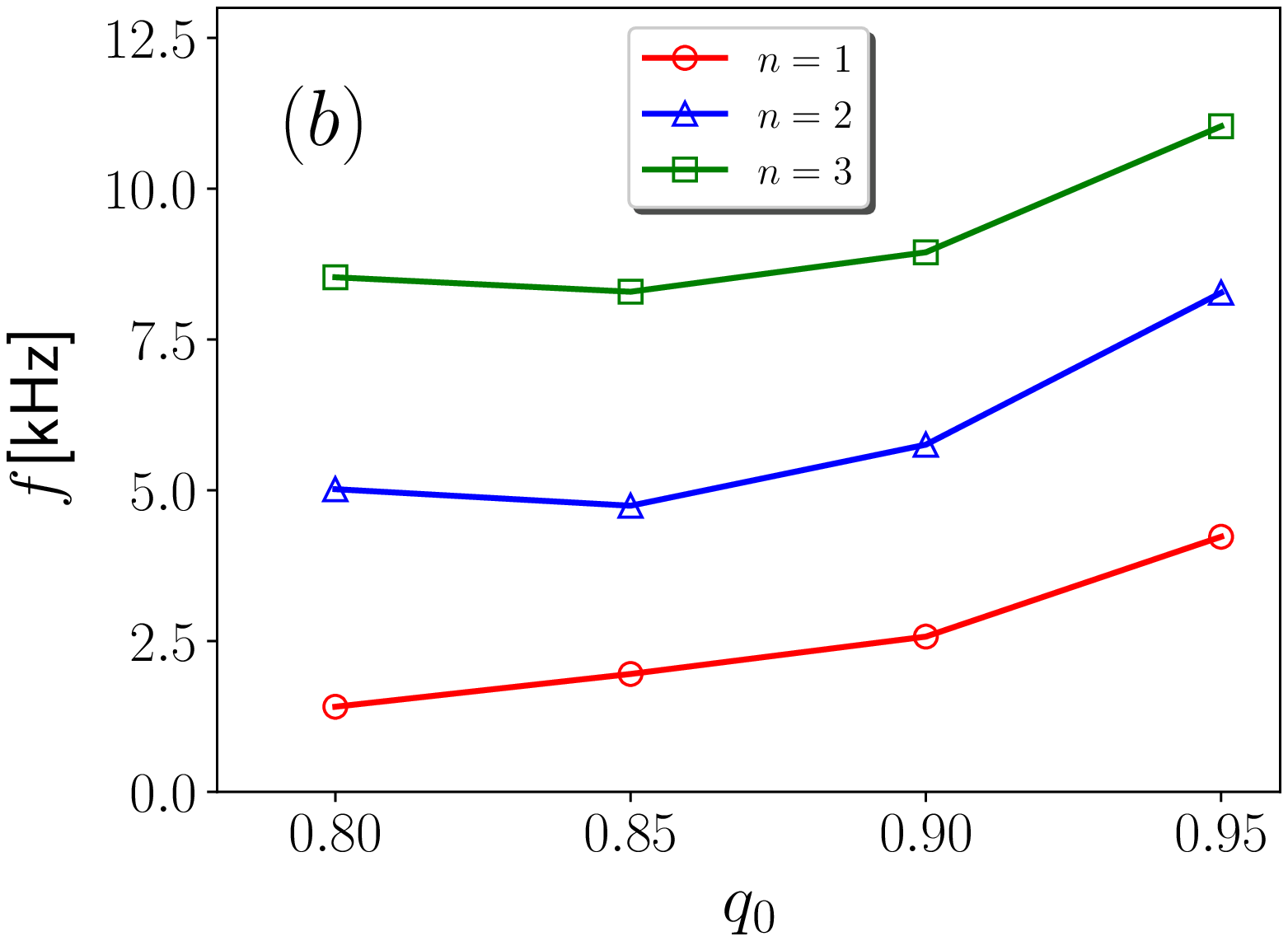}
  \includegraphics[width=8.0cm]{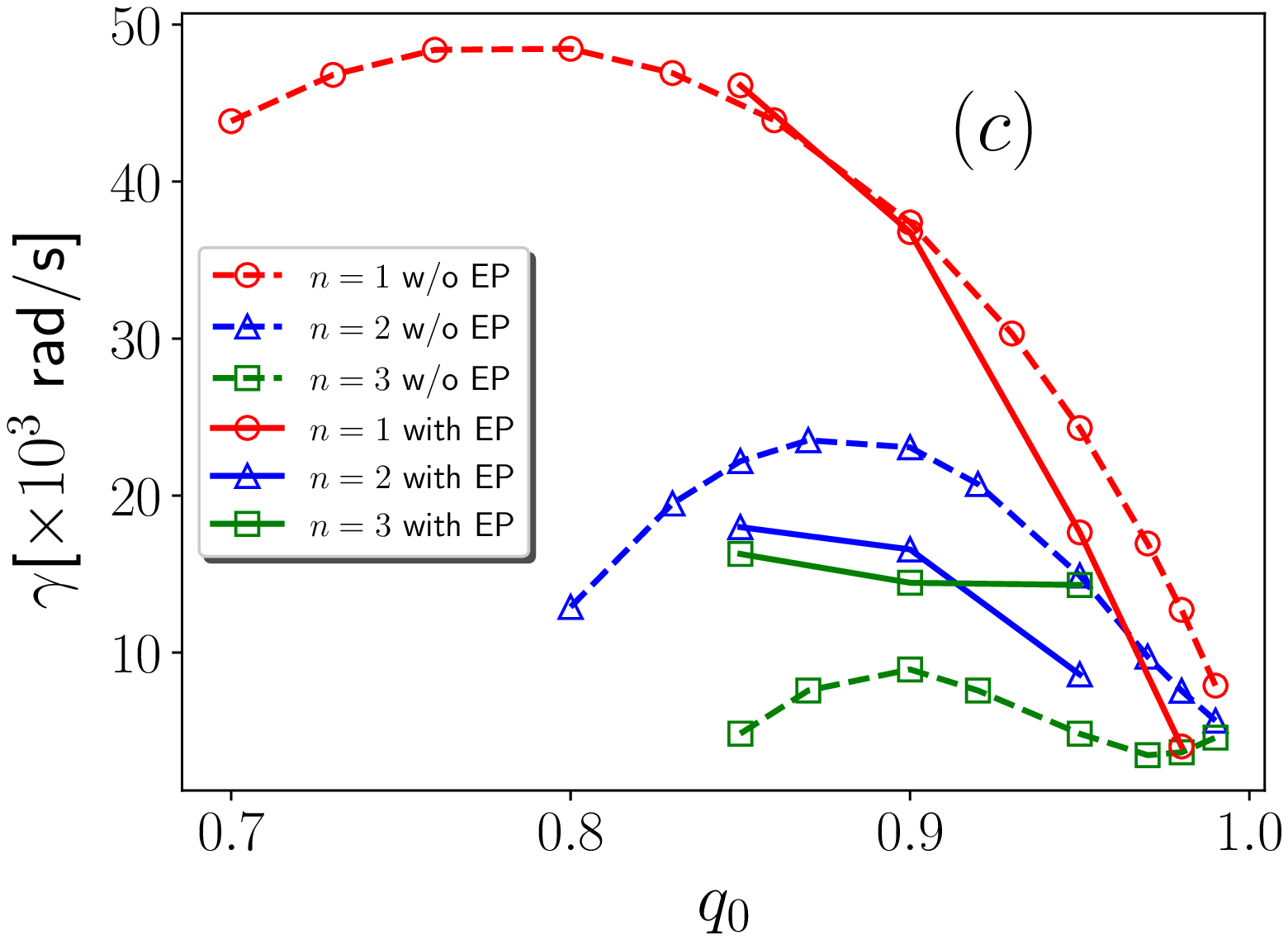}
  \includegraphics[width=8.0cm]{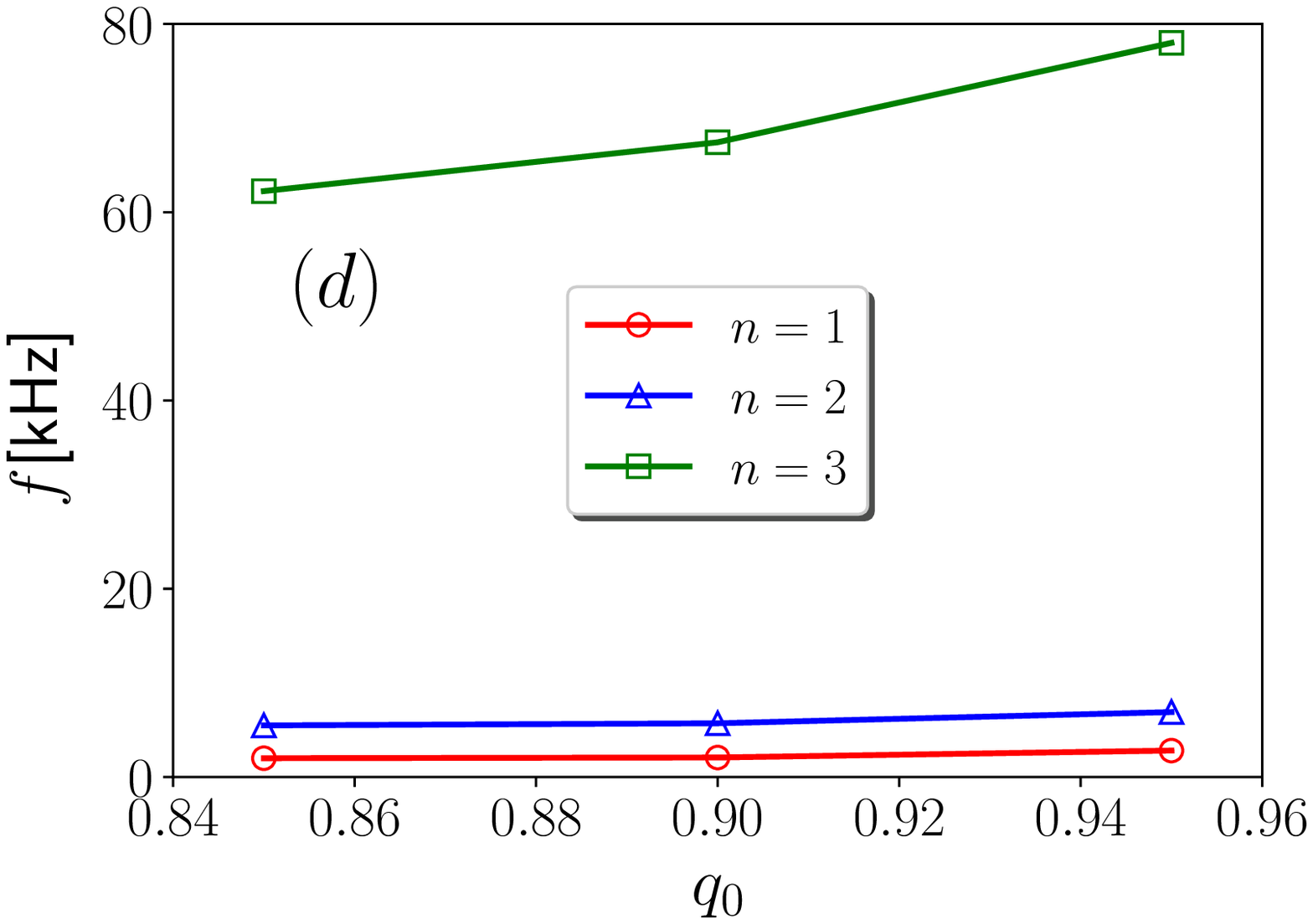}
  \caption{\label{fig:fig7}  Growth rates [(a) and (c)] and
    mode frequencies [(b) and (d)] as functions of $q_0$
    for the $1/1$, $2/2$ and $3/3$ modes, where $\beta_f=0.1$.
    (a) and (b) are based on the M420 equilibrium,
    (c) and (d) are based on the M452 equilibrium.}
\end{figure}

Now we study the $q_0$ effects on the modes in
presence of EPs, and set $\beta_f=0.1$.
For the M420 case in presence of EPs, the overall
growth rates become higher [figure \ref{fig:fig7}(a)].
The mode frequency increases slightly as $q_0$ increases,
and more importantly, the FM exists for $n=1$, $n=2$ and
$n=3$ [figure \ref{fig:fig7}(b)],
which is consistent with the experiment observation in HL-2A \cite{Zhang14}.
For the M452 case, as shown in figure \ref{fig:fig7}(c),
the $1/1$ and $2/2$ kink modes are more stable in presence of EPs,
whereas only the $3/3$ mode is driven more unstable by EPs.
The FM remains with $n=1$ and $n=2$. However for the $3/3$ mode,
the mode frequency difference from the $2/2$ mode is
significantly larger than the frequency interval between
$2/2$ and $1/1$ modes [figure \ref{fig:fig7}(d)].
Comparing the results of M420 and M452, we find that stronger
background plasma pressure gradient maintains the FM
for $n=1, 2, 3$ (as in the M420 case),
which becomes weakened with weaker background plasma pressure gradient
(as in the M452 case). This is consistent with experimental findings that LLMs
are characteristic of pressure-driven modes\cite{Deng14}.
Comparing the results from section \ref{subsec:qeff}, we have
found FM in presence of EPs, and FM is weakened or broken with reduced
background plasma pressure gradient. \par

Comparing figure \ref{fig:fig7} in this paper and
figure 1 in reference \cite{Deng14} carefully, we find that
the frequency range does not match, although FM exists in both figures.
The main reason is that we have not considered plasma rotation. According
to $f_{\text{exp}}=f_{\text{EP}}+nf_{\text{rot}}$, where $f_{\text{exp}}$
is the frequency measured in the experiments, $f_{\text{EP}}$ is
the frequency caused by EPs, and $f_{\text{rot}}$ is the
frequency of plasma rotation. Adding the frequencies
from the simulations ($f_{1/1}^{\text{sim}}\simeq 2.5\,\mkHz$,
$f_{2/2}^{\text{sim}}\simeq 6\,\mkHz$ and
$f_{3/3}^{\text{sim}}\simeq 9\,\mkHz$) and
frequency of plasma rotation ($f_{\text{rot}}\simeq 7\,\mkHz$ as measured in
the experiments\cite{Deng14}), we get $f_{1/1}\simeq 10\,\mkHz$,
$f_{2/2}\simeq 20\,\mkHz$ and $f_{3/3}\simeq 30\,\mkHz$, which are
close to the frequencies measured in the experiments. \par
\begin{figure}[ht]
  \centering
  \begin{overpic}[scale=0.15]{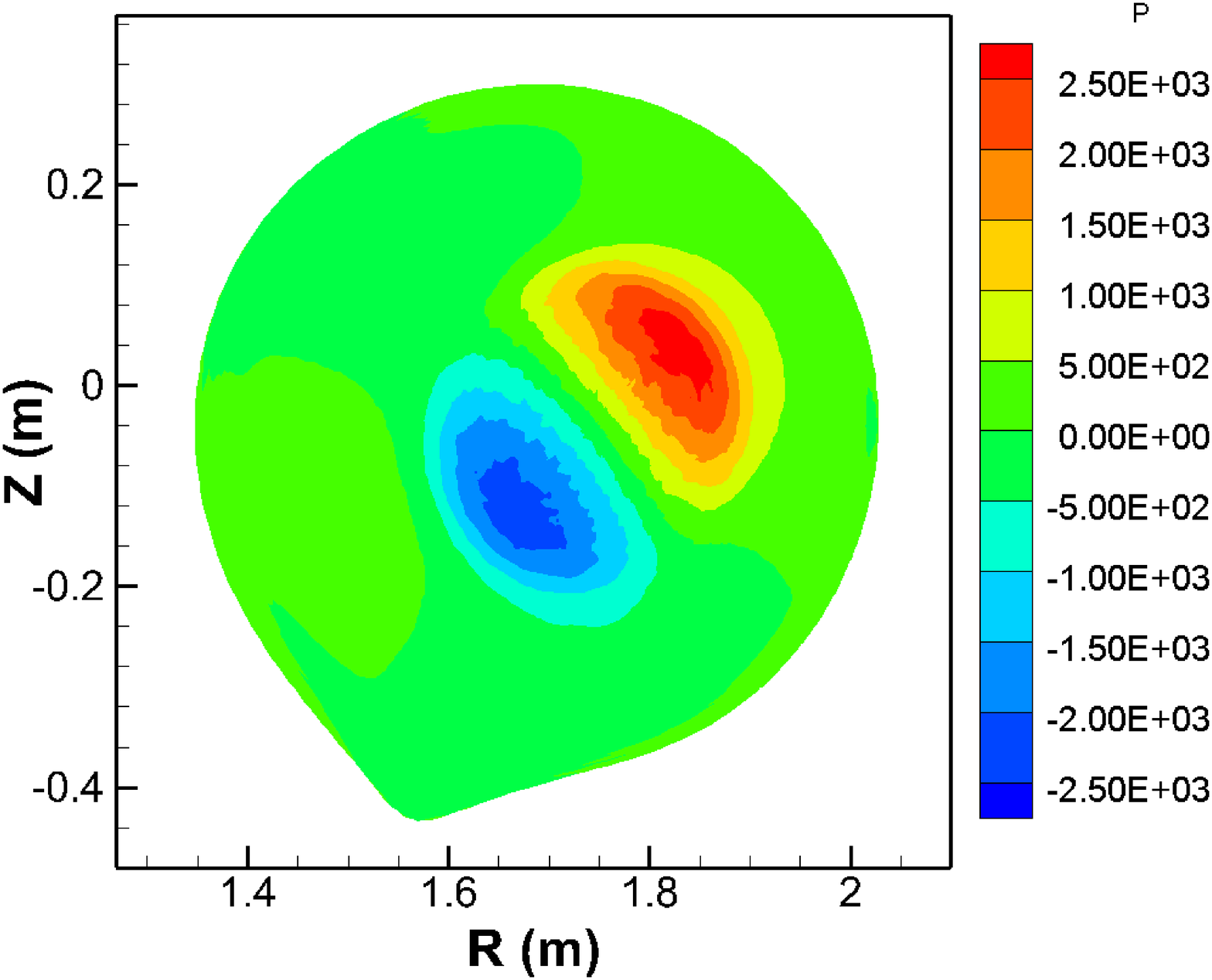}
    \put(13, 68){$(a1)$}
  \end{overpic}
  \begin{overpic}[scale=0.15]{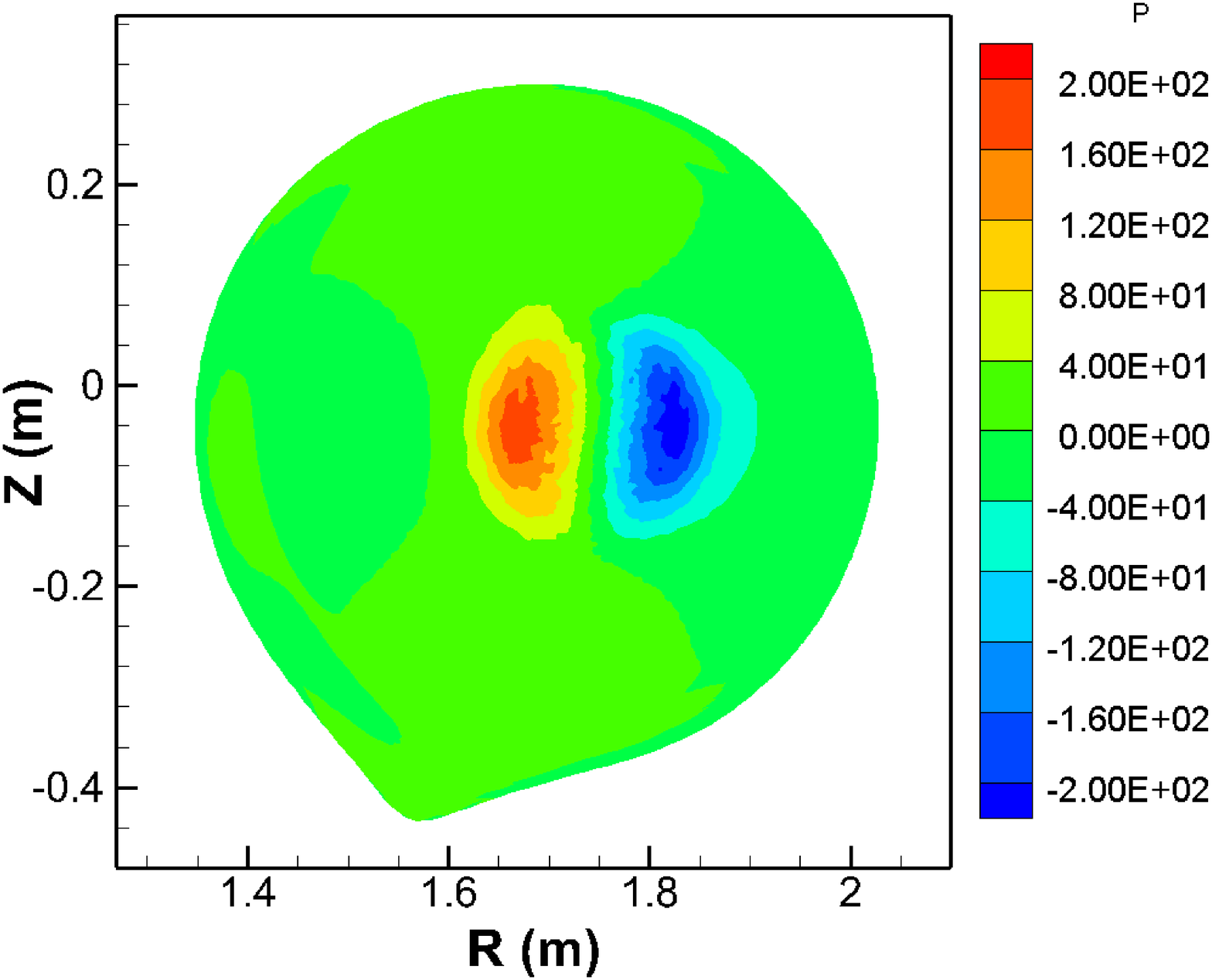}
    \put(13, 68){$(b1)$}
  \end{overpic}
  \begin{overpic}[scale=0.15]{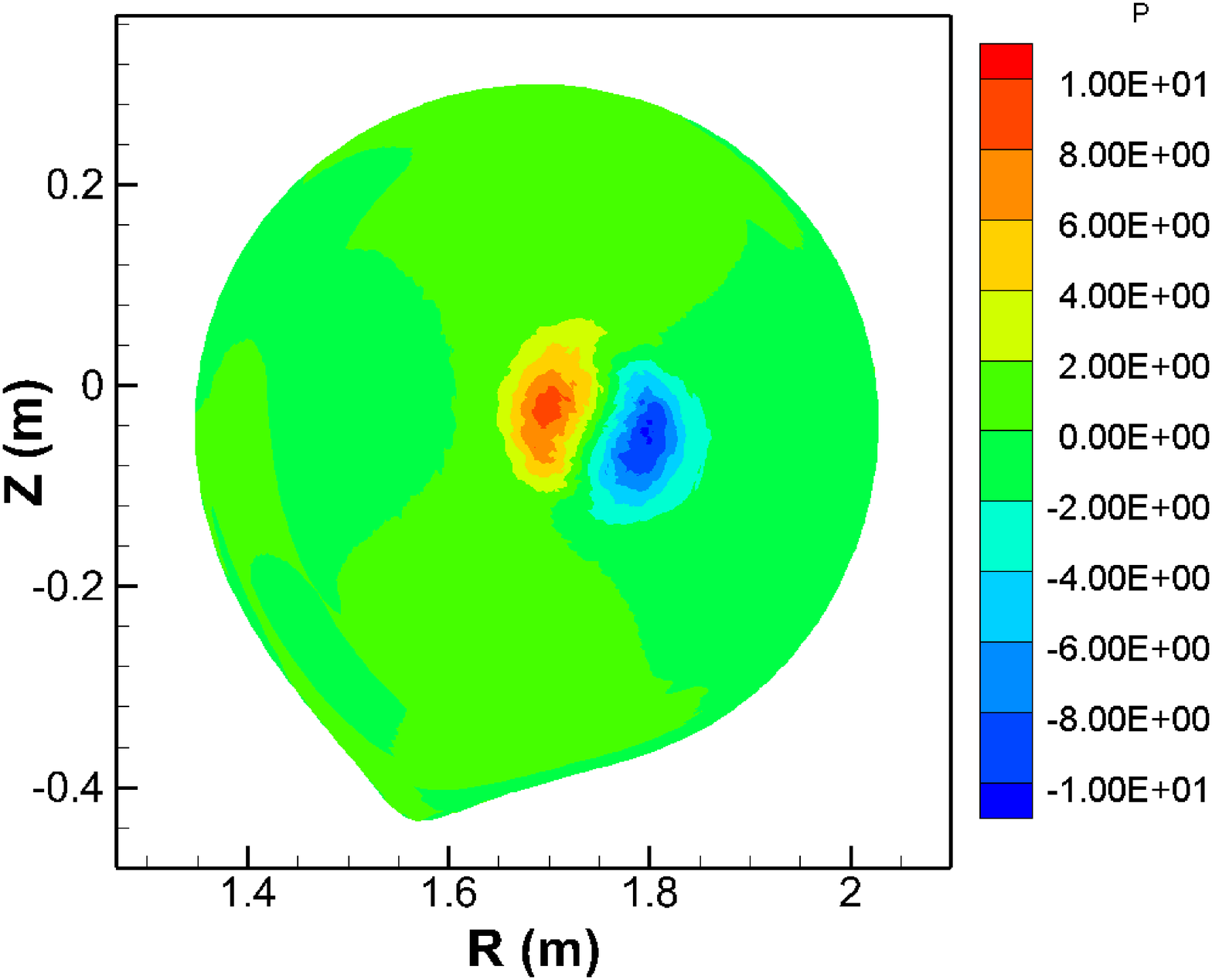}
    \put(13, 68){$(c1)$}
  \end{overpic}
  \begin{overpic}[scale=0.15]{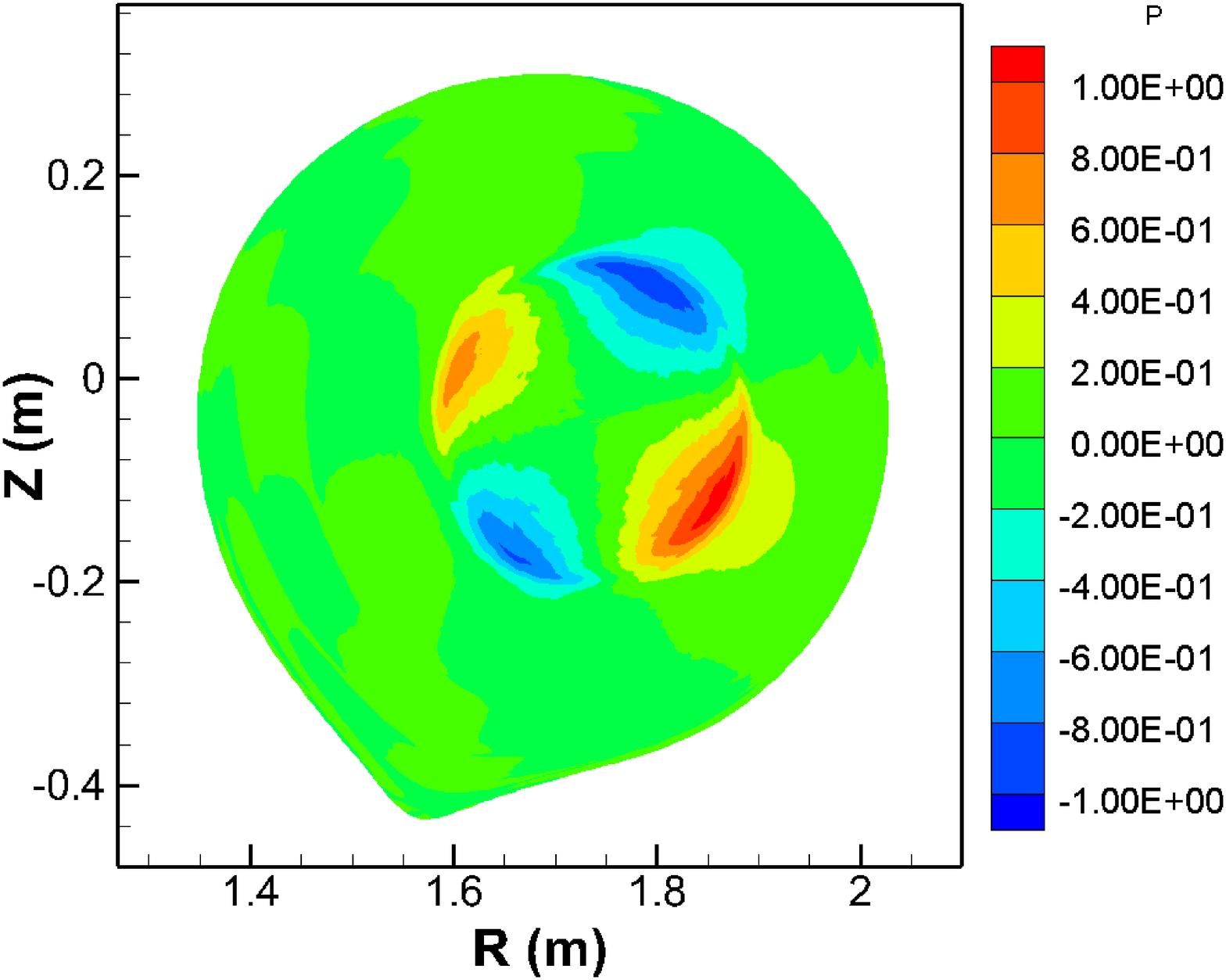}
    \put(13, 68){$(a2)$}
  \end{overpic}
  \begin{overpic}[scale=0.15]{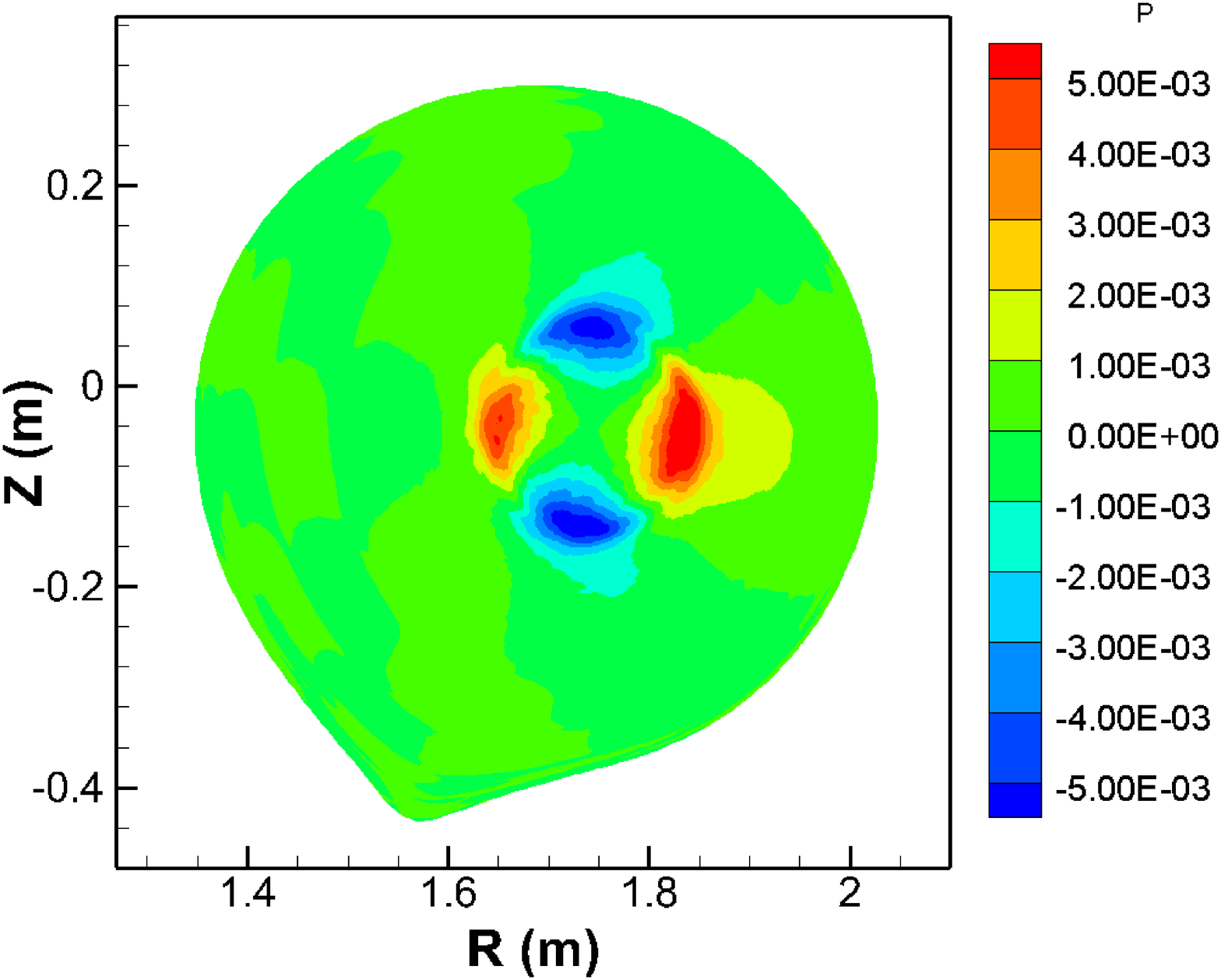}
    \put(13, 68){$(b2)$}
  \end{overpic}
  \begin{overpic}[scale=0.15]{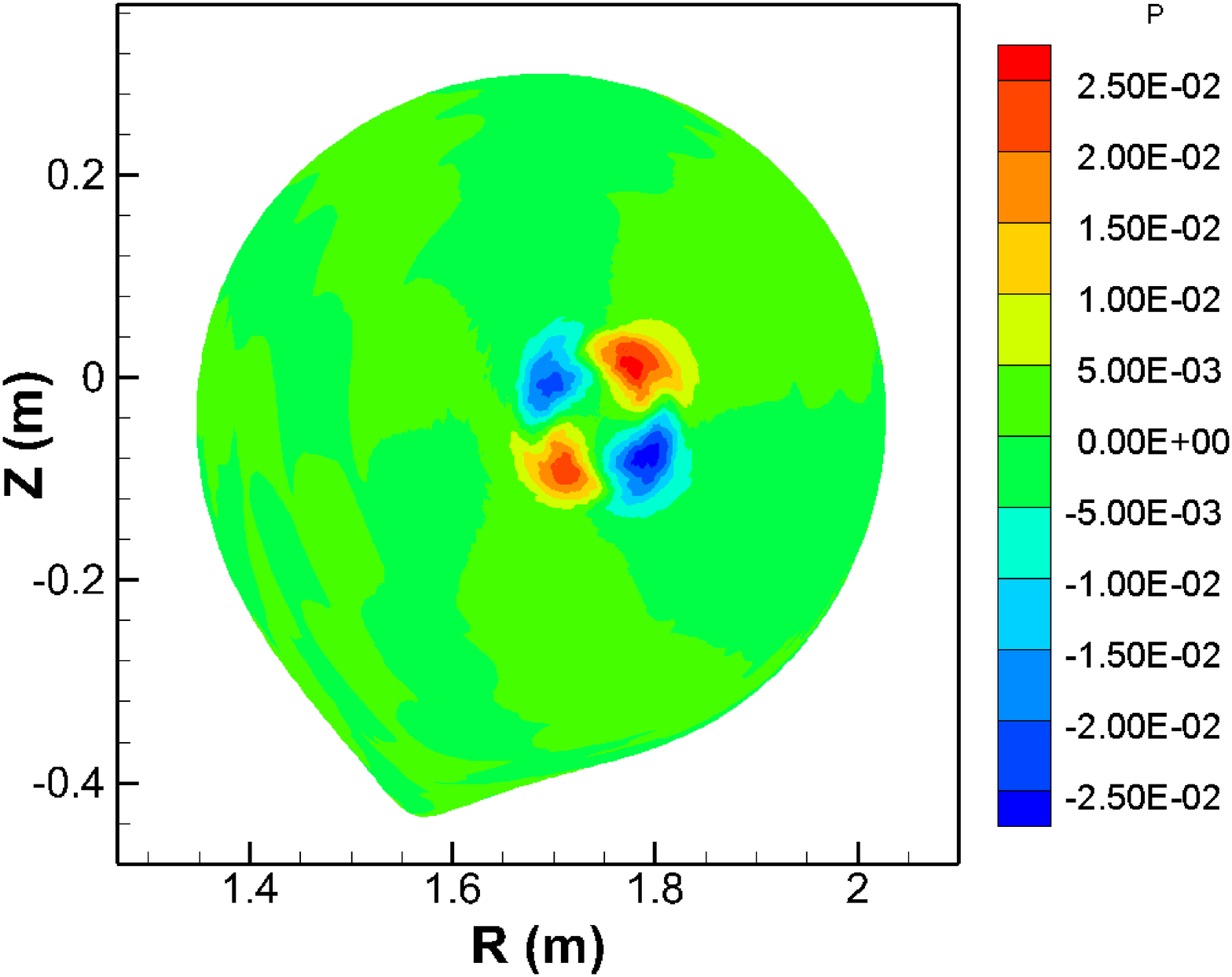}
    \put(13, 68){$(c2)$}
  \end{overpic}
  \begin{overpic}[scale=0.15]{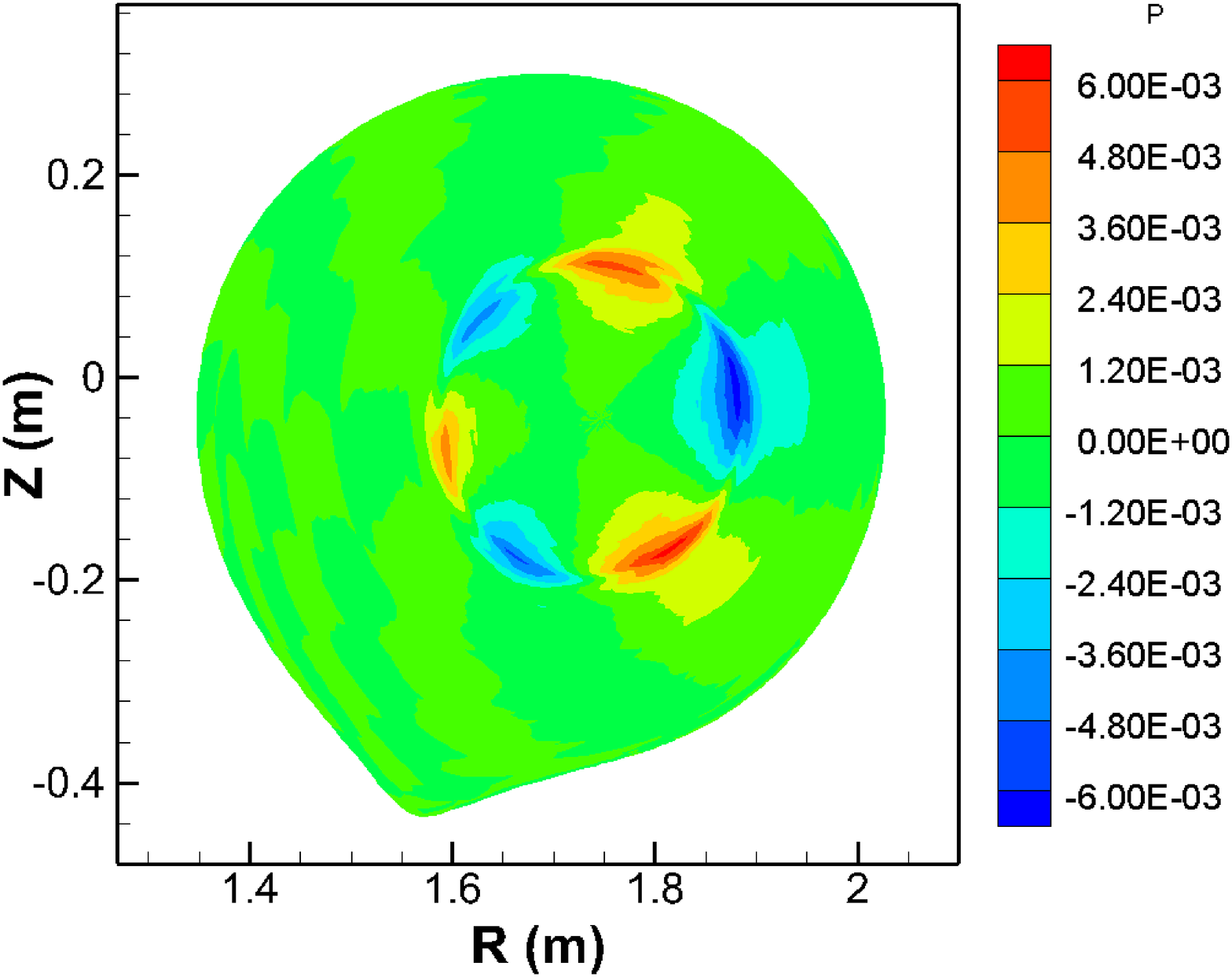}
    \put(13, 68){$(a3)$}
  \end{overpic}
  \begin{overpic}[scale=0.15]{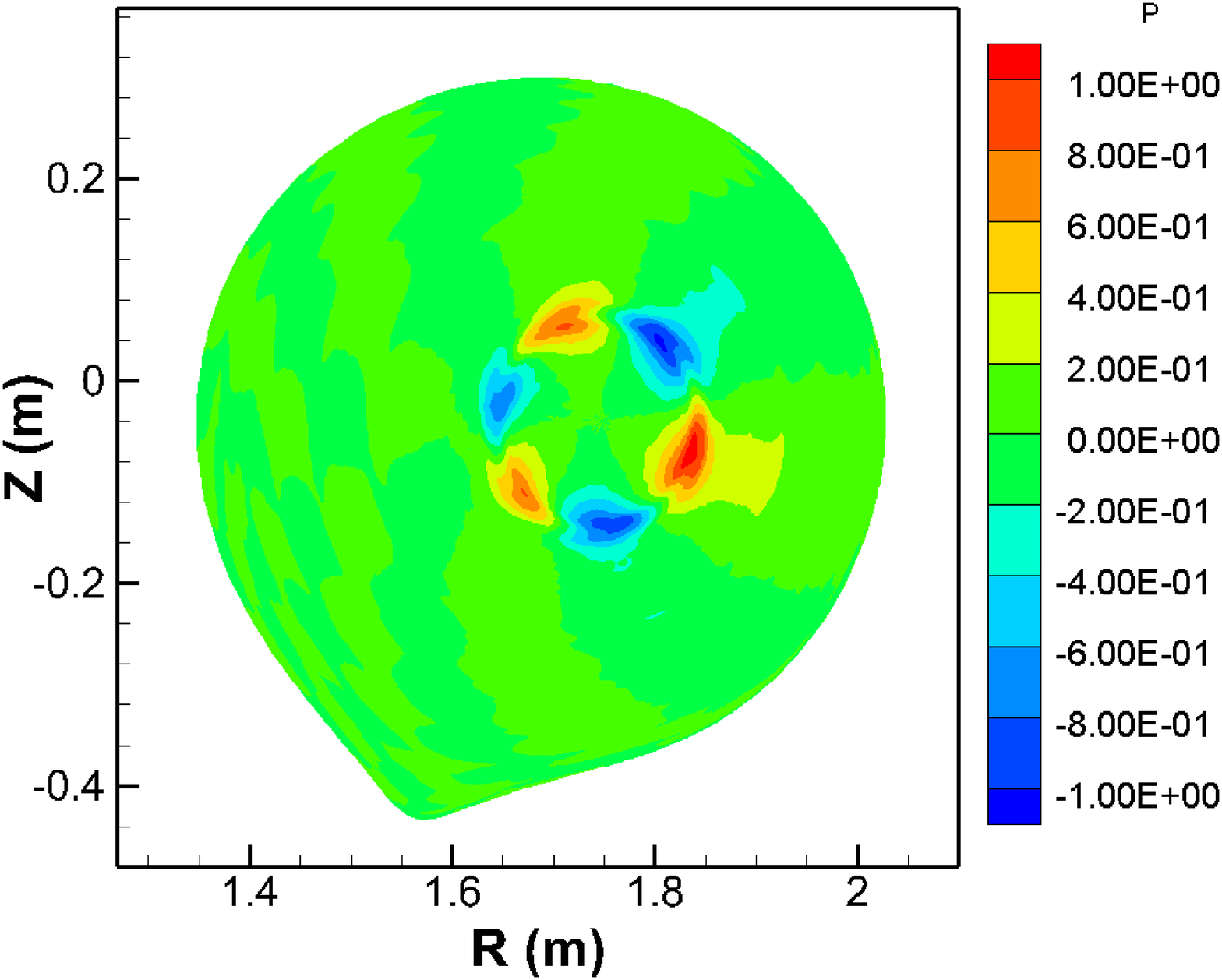}
    \put(13, 68){$(b3)$}
  \end{overpic}
  \begin{overpic}[scale=0.15]{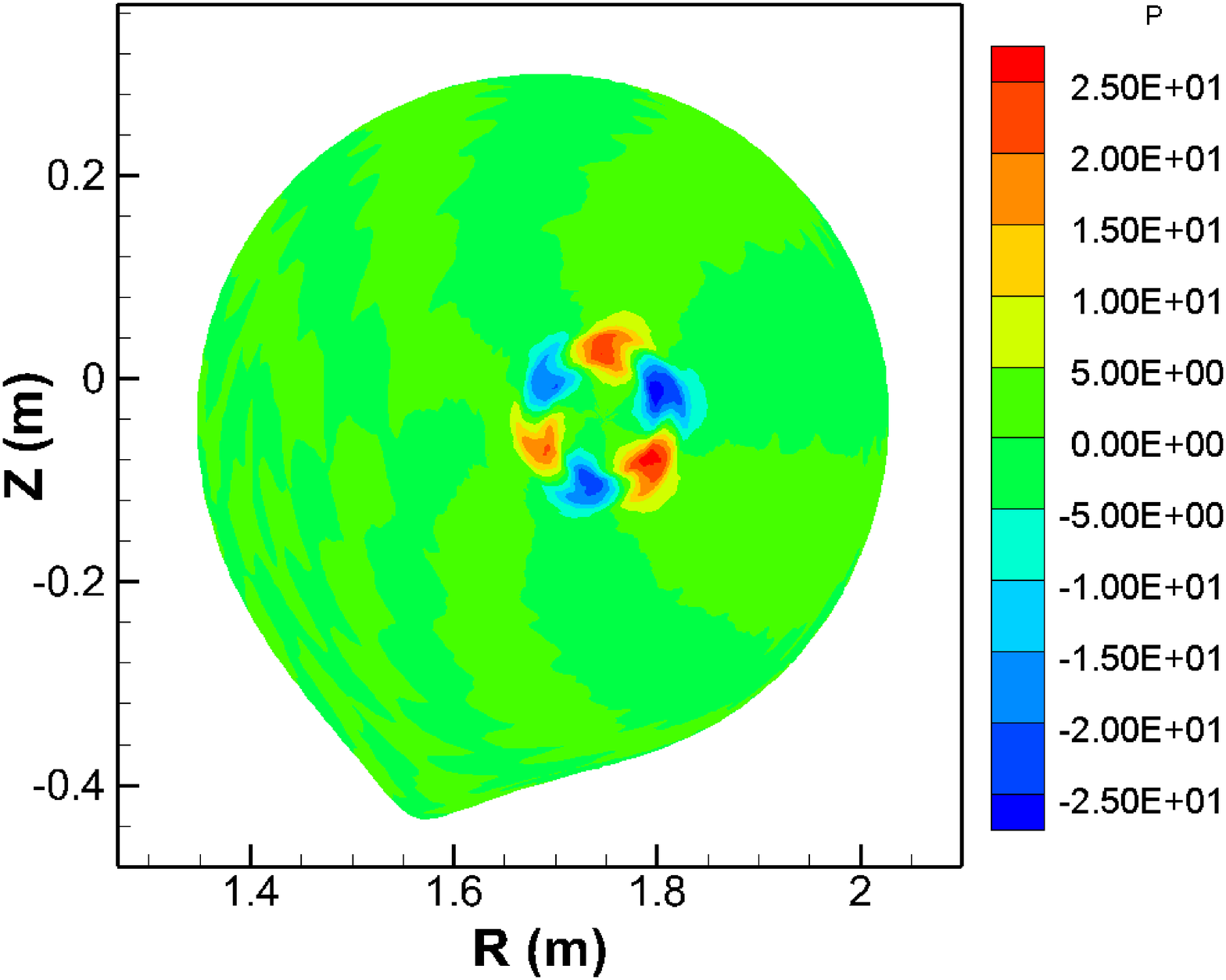}
    \put(13, 68){$(c3)$}
  \end{overpic}
  \caption{\label{fig:fig8} Pressure perturbation contours of
    the $1/1$ modes [(a1), (b1) and (c1)],
    the $2/2$ modes [(a2), (b2) and (c2)] and
    the $3/3$ modes [(a3), (b3) and (c3)] with
    $q_0=0.8$ [(a1), (a2) and (a3)],
    $q_0=0.9$ [(b1), (b2) and (b3)] and
    $q_0=0.95$ [(c1), (c2) and (c3)],
    where $\beta_f=0.1$ based on the M420 equilibrium.}
\end{figure}
\begin{figure}[ht]
  \centering
  \begin{overpic}[scale=0.15]{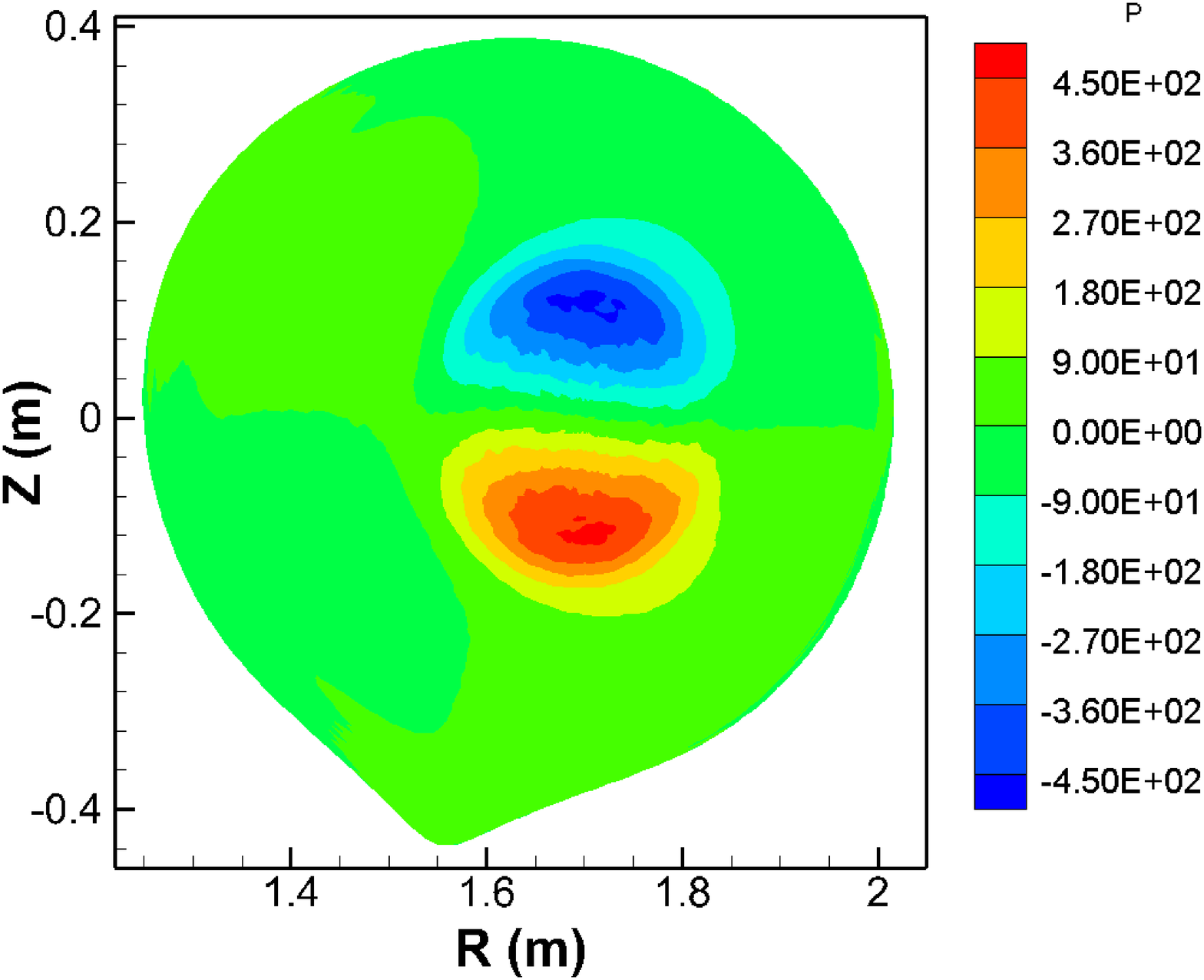}
    \put(13, 68){$(a1)$}
  \end{overpic}
  \begin{overpic}[scale=0.15]{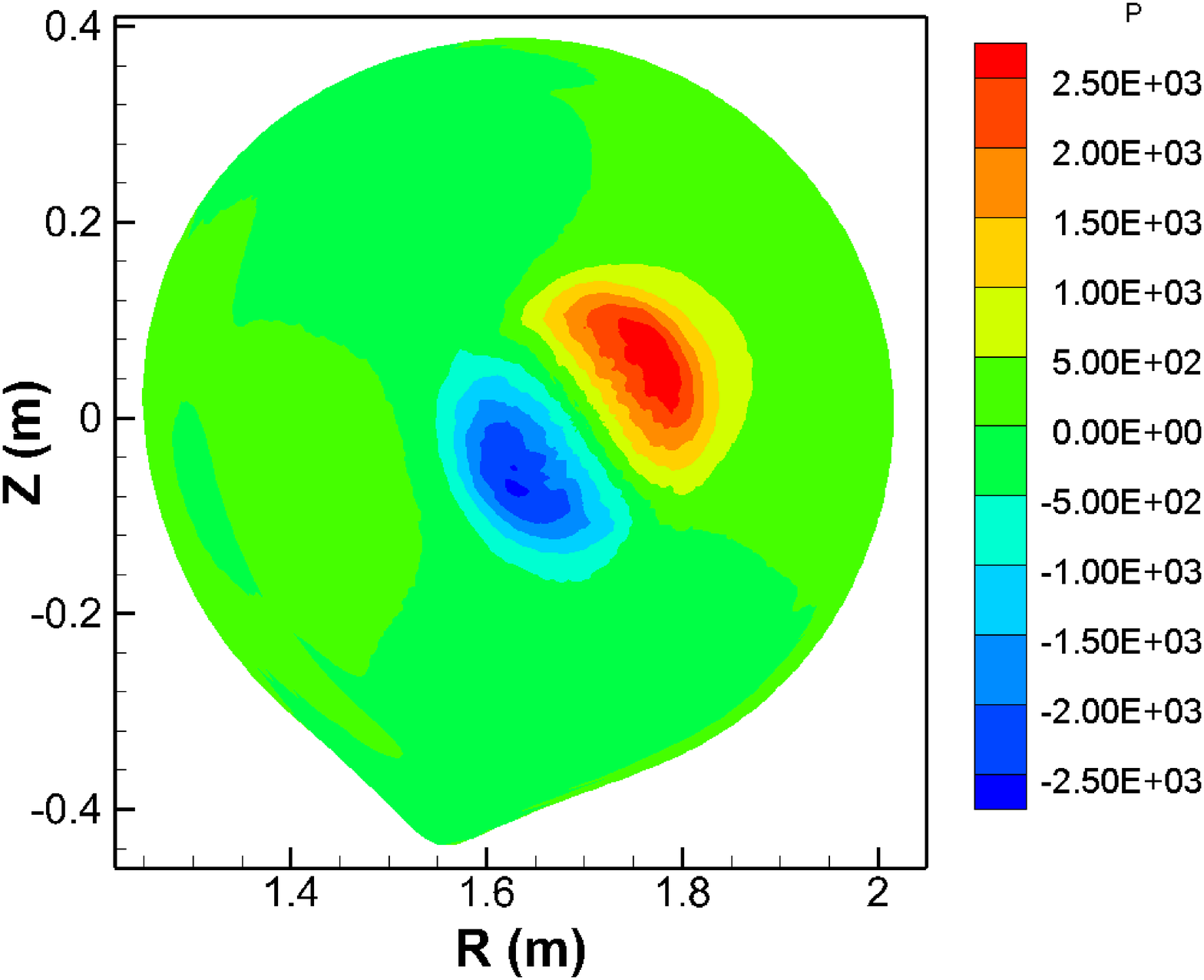}
    \put(13, 68){$(b1)$}
  \end{overpic}
  \begin{overpic}[scale=0.15]{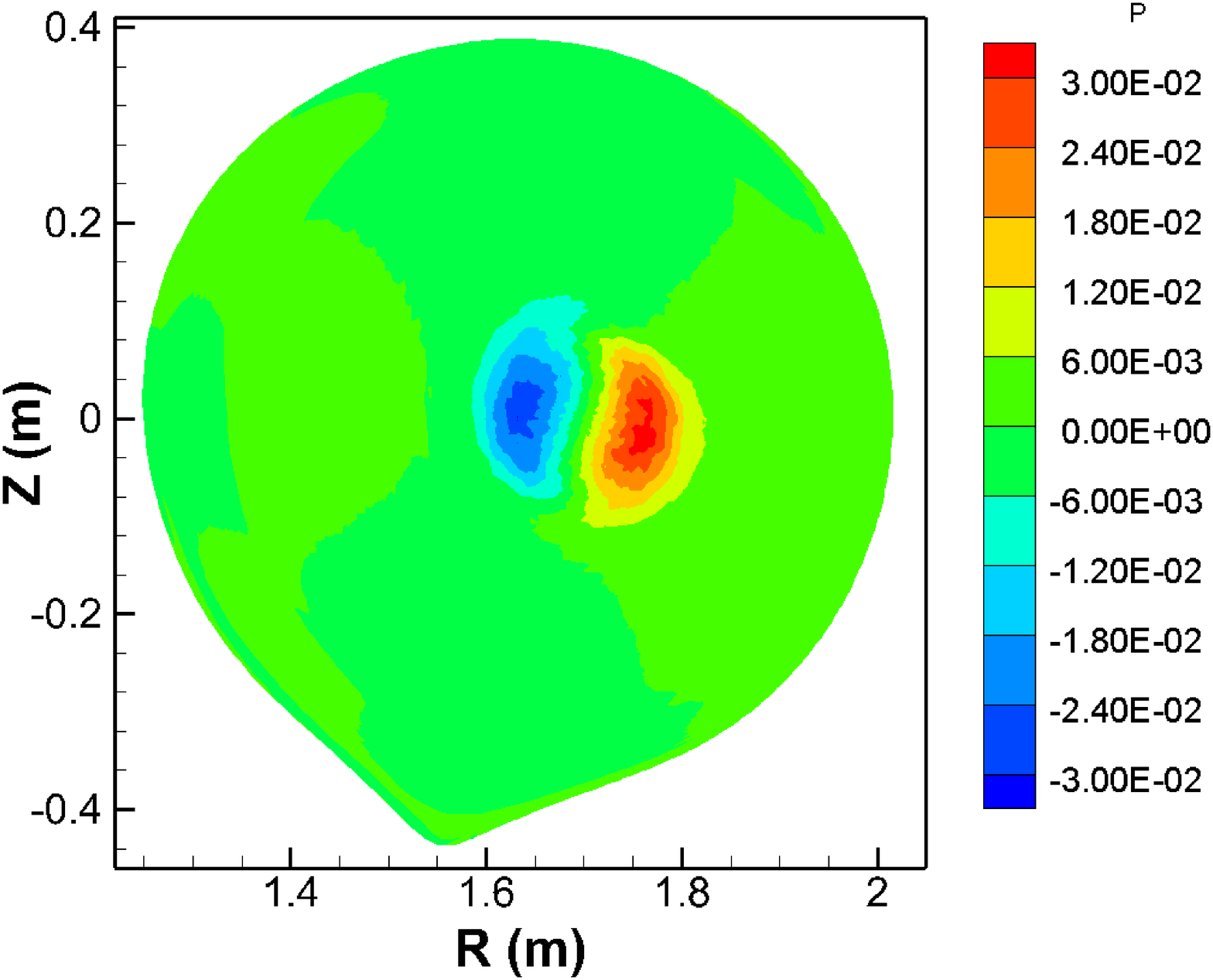}
    \put(13, 68){$(c1)$}
  \end{overpic}
  \begin{overpic}[scale=0.15]{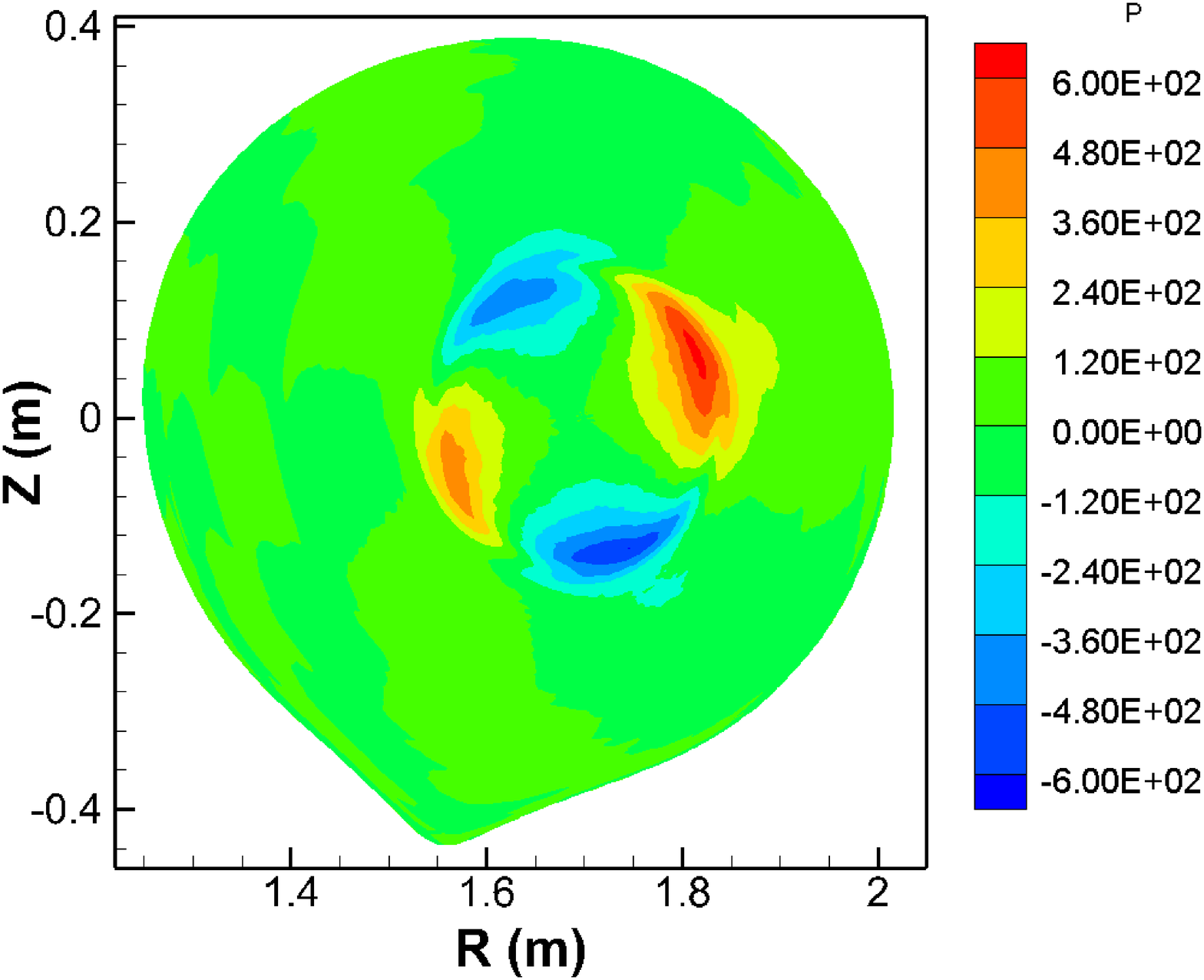}
    \put(13, 68){$(a2)$}
  \end{overpic}
  \begin{overpic}[scale=0.15]{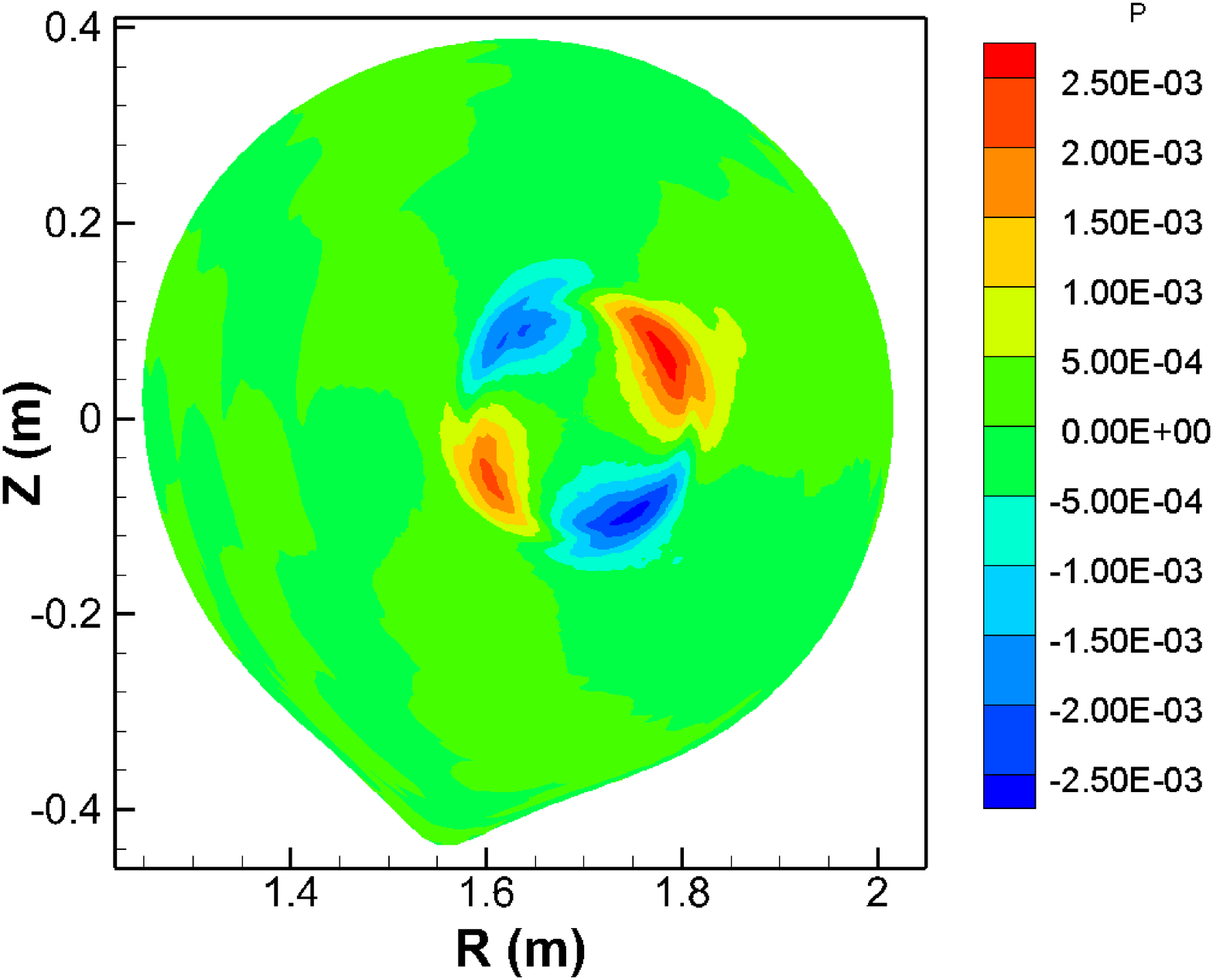}
    \put(13, 68){$(b2)$}
  \end{overpic}
  \begin{overpic}[scale=0.15]{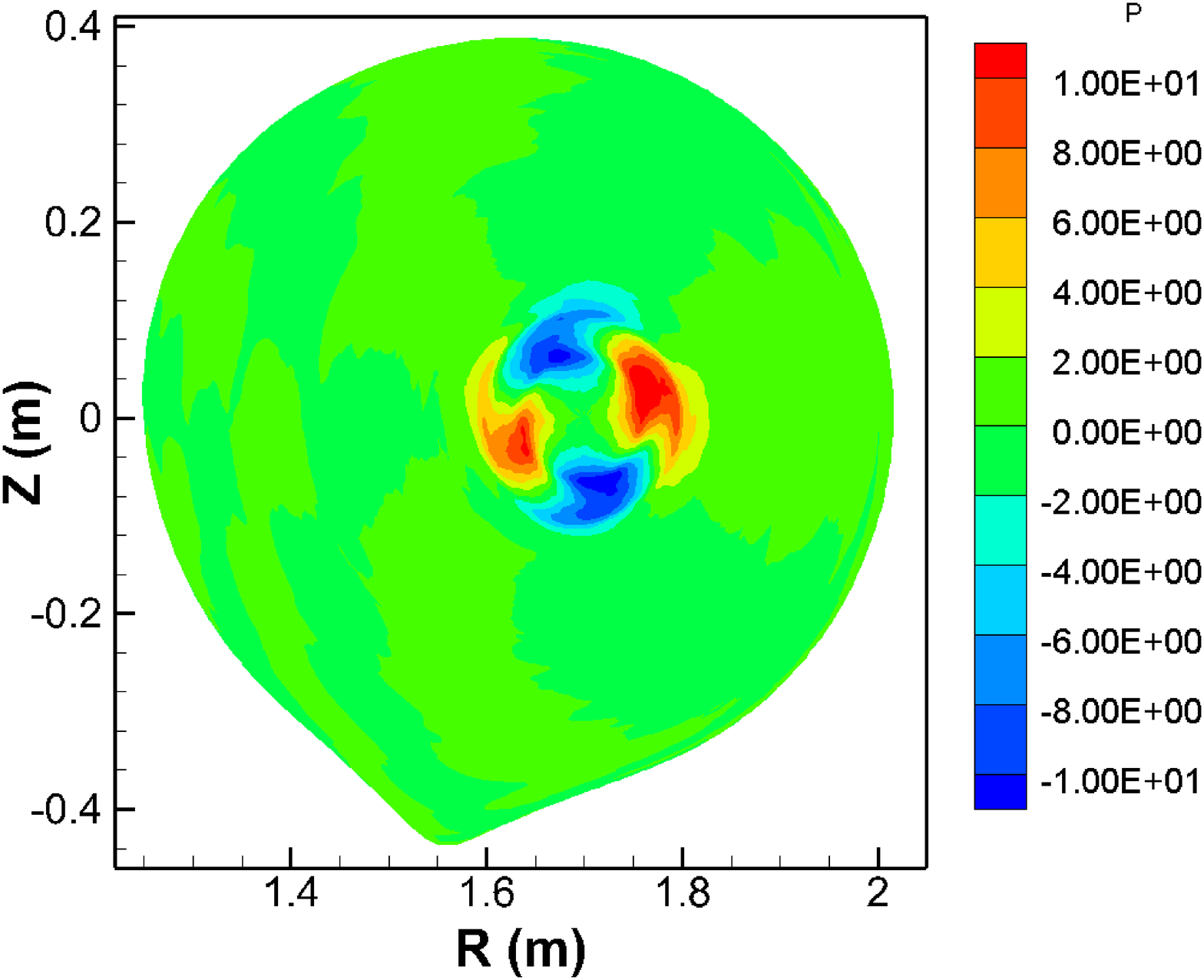}
    \put(13, 68){$(c2)$}
  \end{overpic}
  \begin{overpic}[scale=0.15]{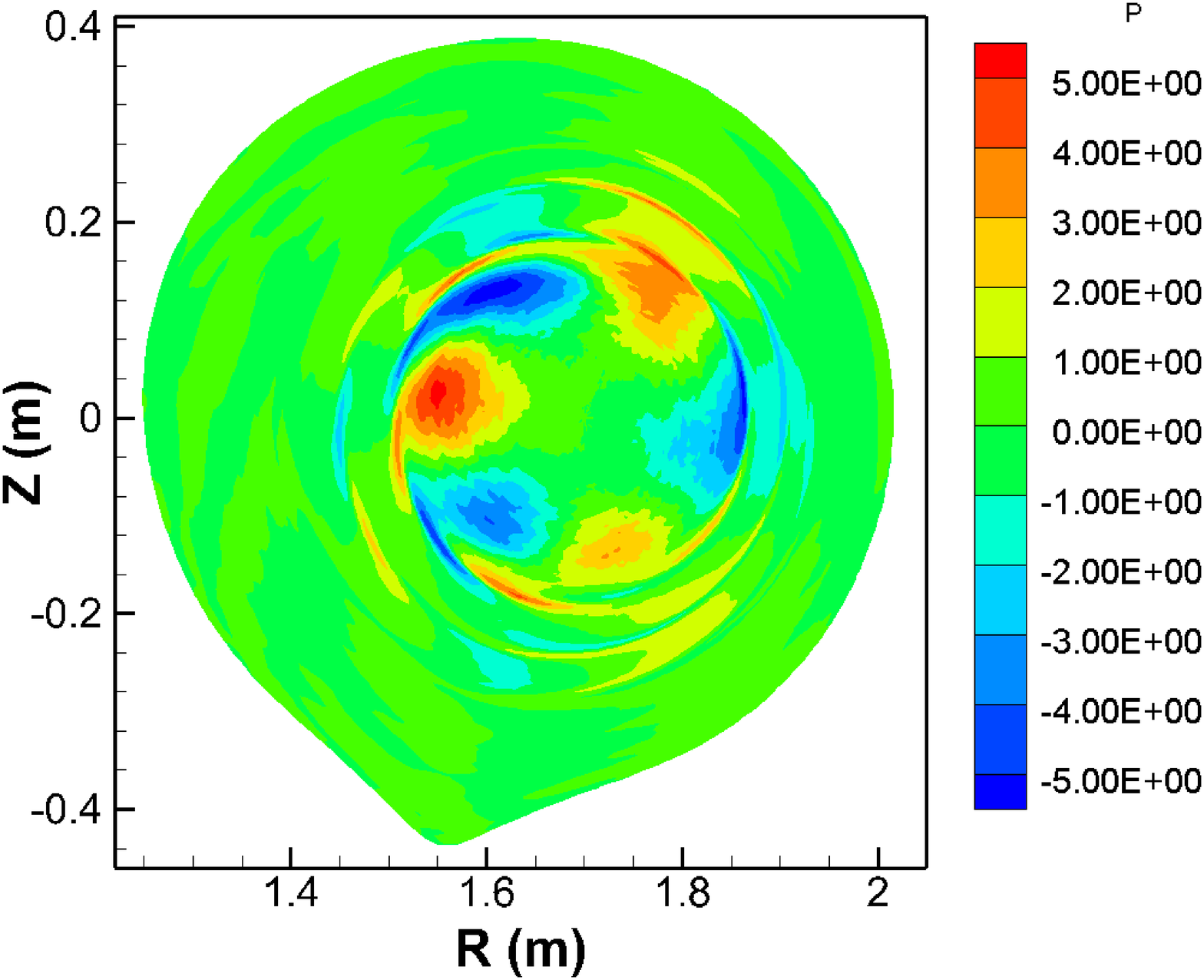}
    \put(13, 68){$(a3)$}
  \end{overpic}
  \begin{overpic}[scale=0.15]{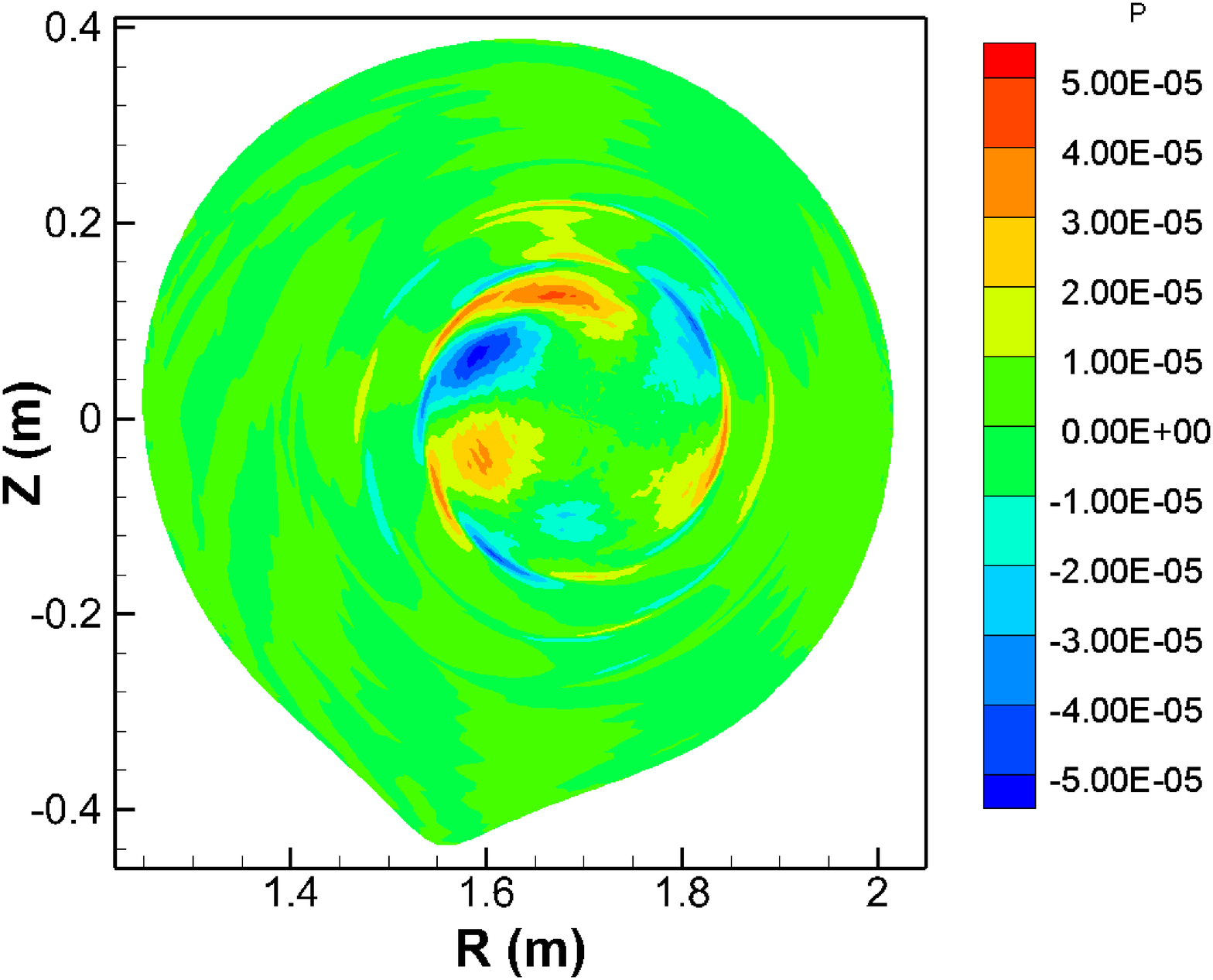}
    \put(13, 68){$(b3)$}
  \end{overpic}
  \begin{overpic}[scale=0.15]{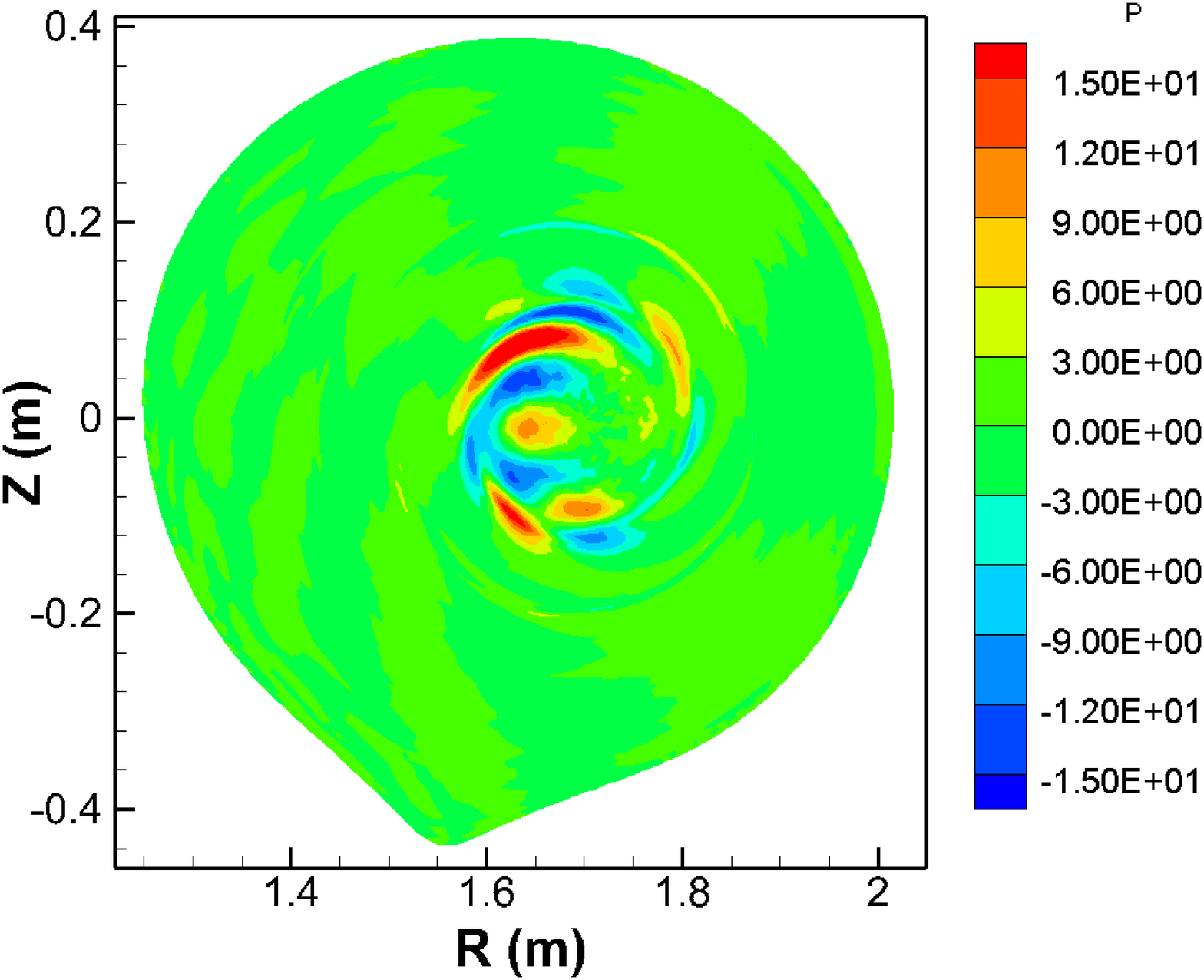}
    \put(13, 68){$(c3)$}
  \end{overpic}
  \caption{\label{fig:fig9} Pressure perturbation contours of
    the $1/1$ modes [(a1), (b1) and (c1)],
    the $2/2$ modes [(a2), (b2) and (c2)] and
    the $3/3$ modes [(a3), (b3) and (c3)] with
    $q_0=0.85$ [(a1), (a2) and (a3)],
    $q_0=0.9$ [(b1), (b2) and (b3)] and
    $q_0=0.95$ [(c1), (c2) and (c3)],
    where $\beta_f=0.1$ based on the M452 equilibrium.}
\end{figure}

For the M420 case, the mode structures shown in figure \ref{fig:fig8}
are similar to the cases in absence of EPs
as shown in figure \ref{fig:fig6}, except that they are now
twisted by EPs,  which become more significant as $n$ increases
and $q_0$ approaches to unity. For the M452 case,
the poloidal mode structures twisting by EPs are more apparent in comparison
to the M420 case. For the $n=3$ mode, the structure of mode coupling
appears [figure \ref{fig:fig9} (a3), (b3) and (c3)],
which is indicative of the toroidal Alfv\'en eigenmode (TAE). \par
\subsection{Effects of EP $\beta$ fraction  $\beta_f$}\label{subsec:bfrac}
\begin{figure}[ht]
  \centering
  \includegraphics[width=8.0cm]{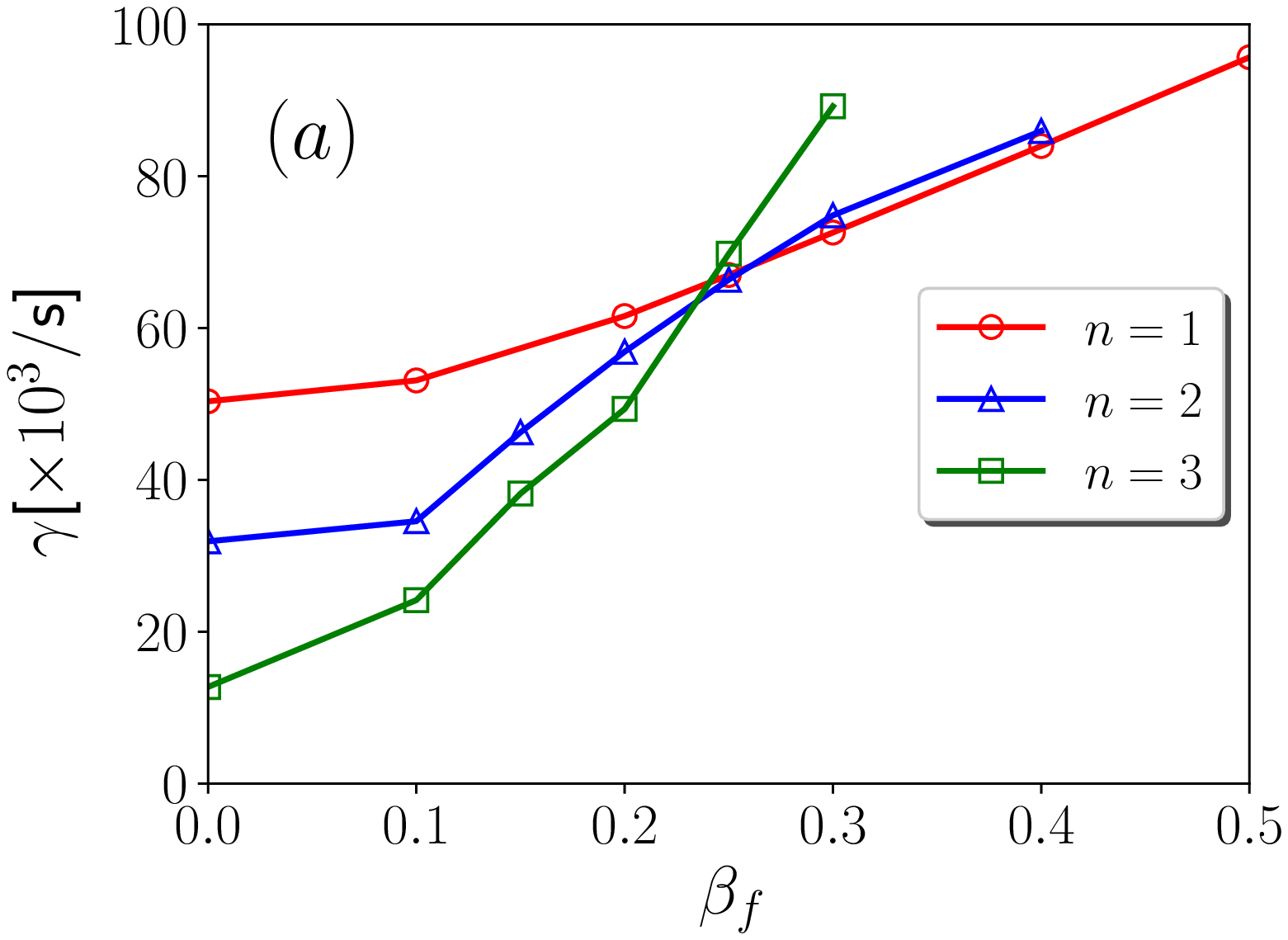}
  \includegraphics[width=8.0cm]{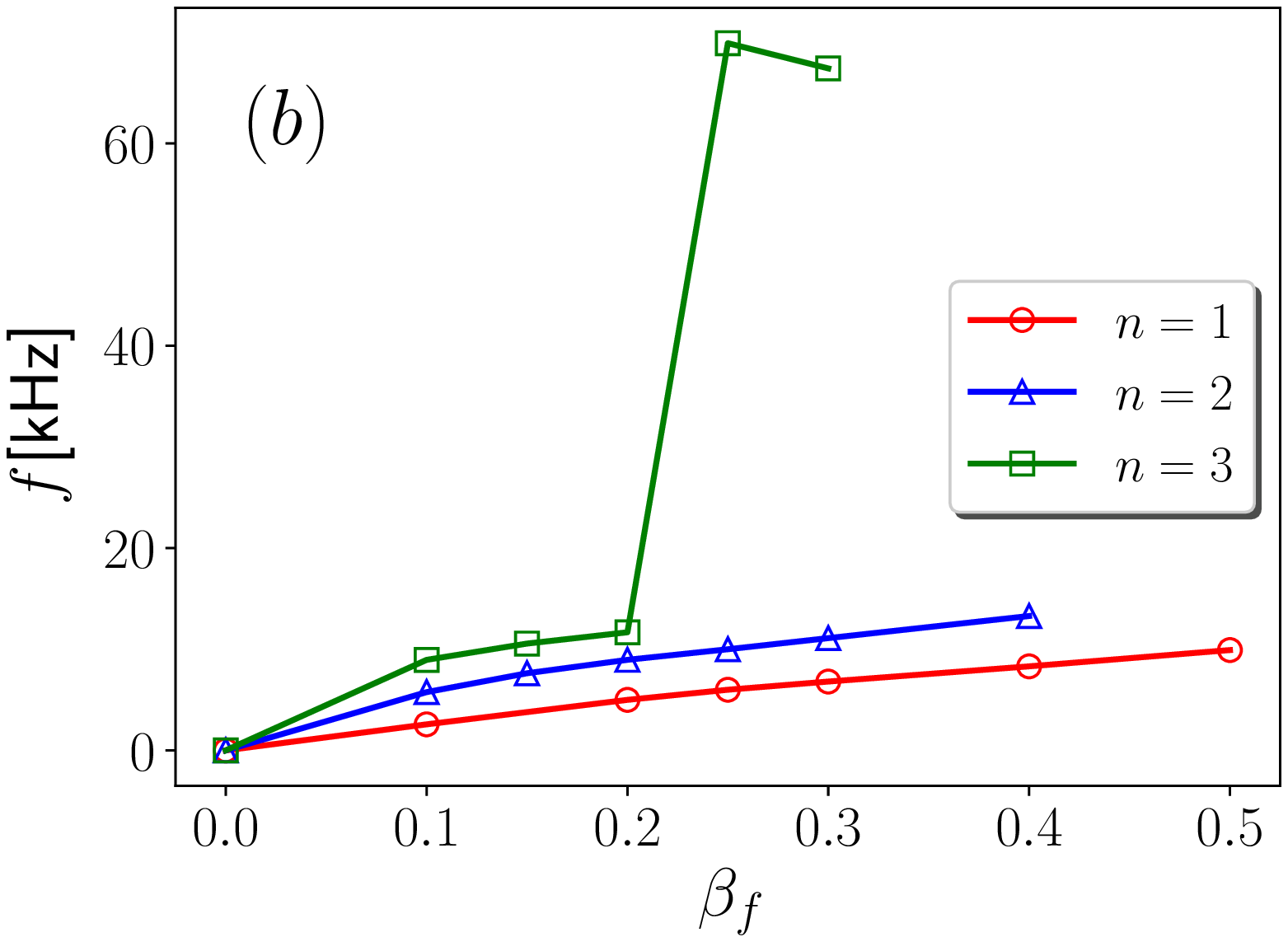}
  \includegraphics[width=8.0cm]{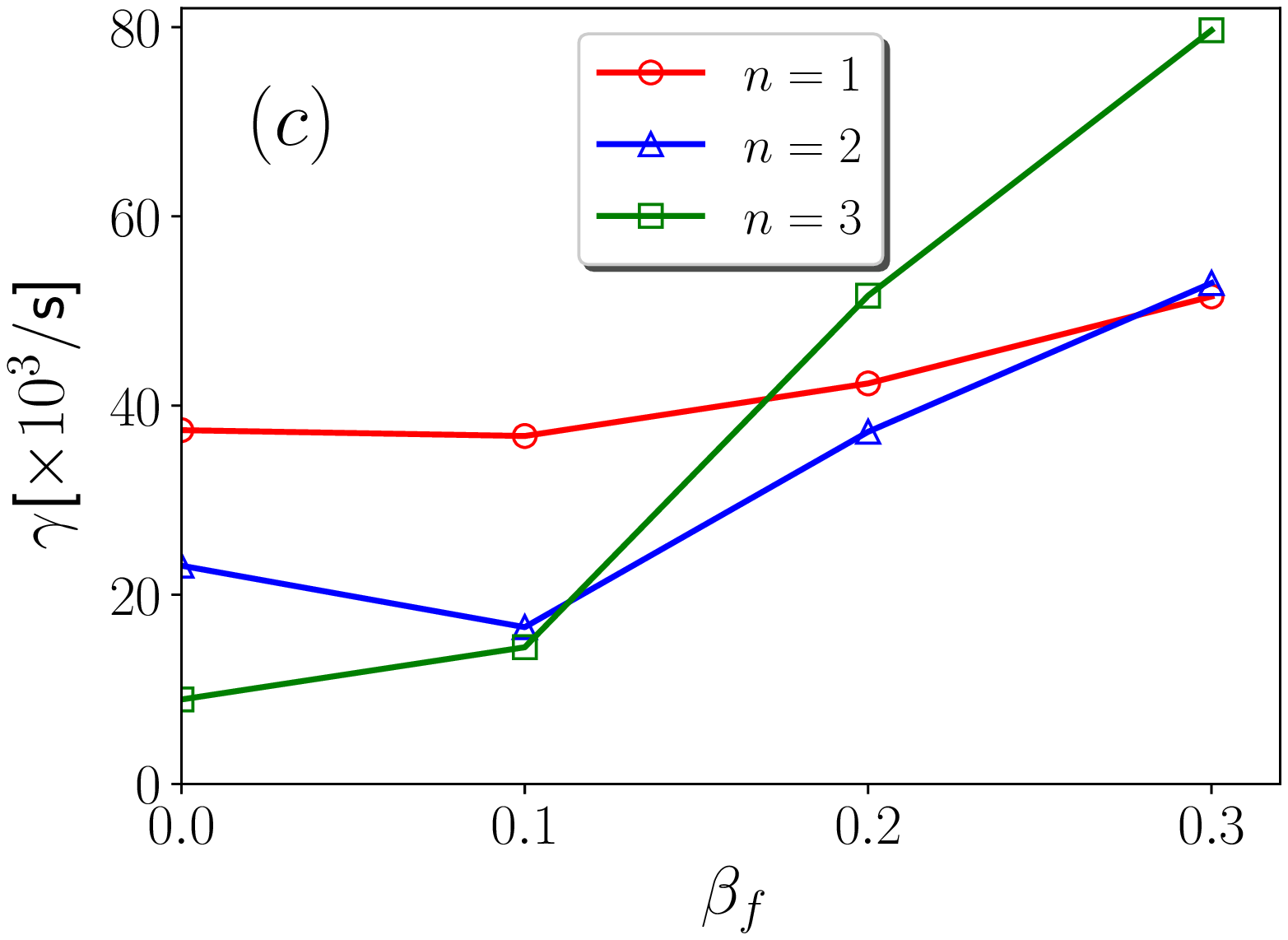}
  \includegraphics[width=8.0cm]{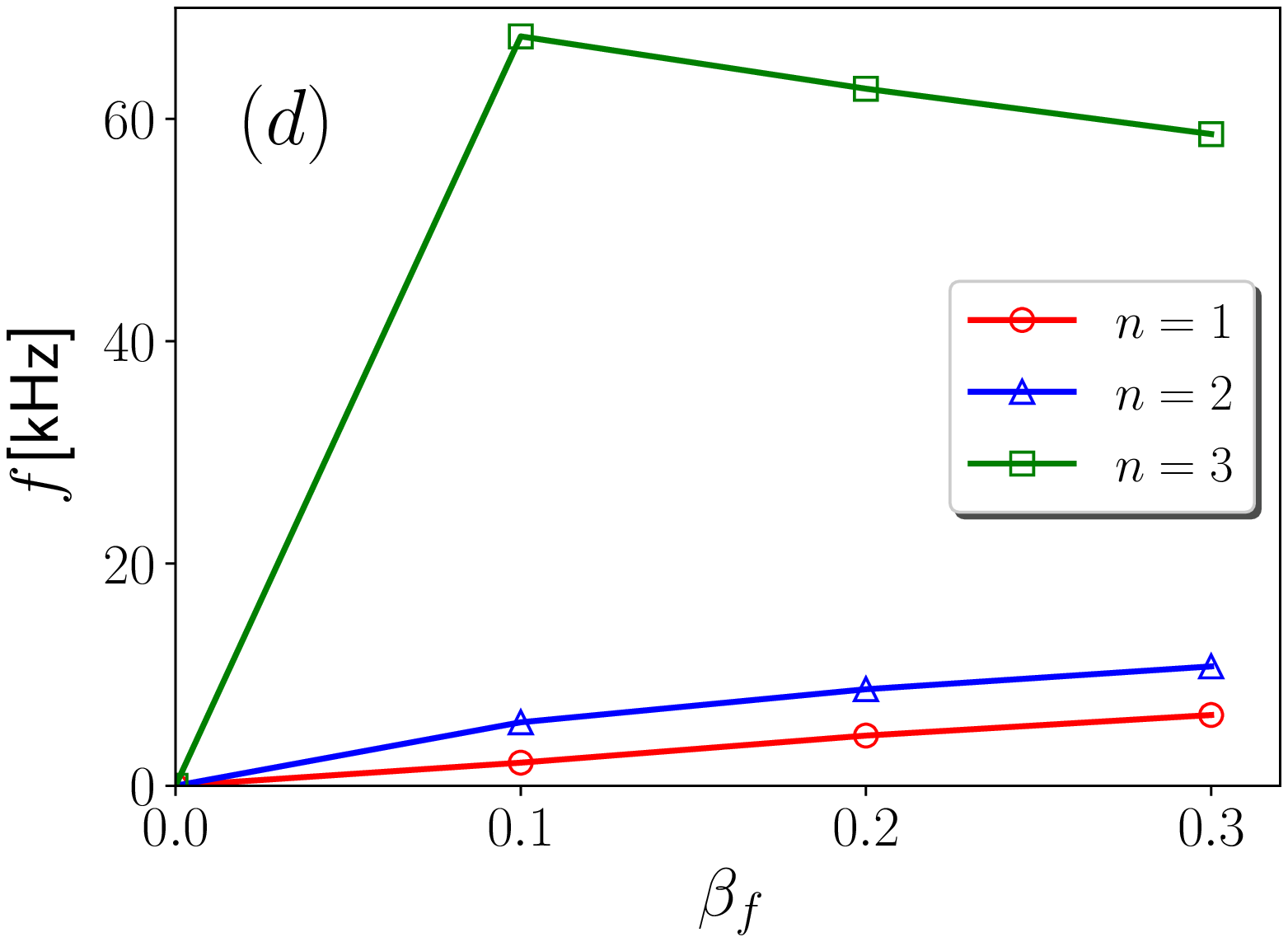}
  \caption{\label{fig:fig10} Growth rates [(a) and (c)] and
    mode frequencies [(b) and (d)] as functions of $\beta_f$
    for the $1/1$, $2/2$ and $3/3$ modes, where $q_0=0.9$.
    (a) and (b) are based on the M420 equilibrium,
    (c) and (d) are based on the M452 equilibrium.}
\end{figure}

We set $q_0=0.9$, and scan $\beta_f$ to study EP pressure effects
on the modes. For the M420 case, as the EP $\beta_f$ increases,
the growth rate of  higher-$n$ mode increases more rapidly.
As the EP pressure increases, the growth rates of higher-$n$ mode approach to
that of $1/1$ mode, and for $\beta_f > 0.25$, the growth rate of
$3/3$ mode becomes the largest [figure \ref{fig:fig10}(a)].
This indicates that higher-$n$ modes are more vulnerable to
EP $\beta$ effects. When the EP pressure is relatively
low ($\beta_f<0.2$), the mode frequency increases almost linearly with
$\beta_f$, and is roughly proportional to $n$.
As $\beta_f$ increases to $0.25$, the frequency of $3/3$ mode
jumps from $\sim 10\mkHz$ to $\sim 70\mkHz$, and then decreases as
$\beta_f$ increases further [figure \ref{fig:fig10} (b)].\par

\begin{figure}[ht]
  \centering
  \begin{overpic}[scale=0.15]{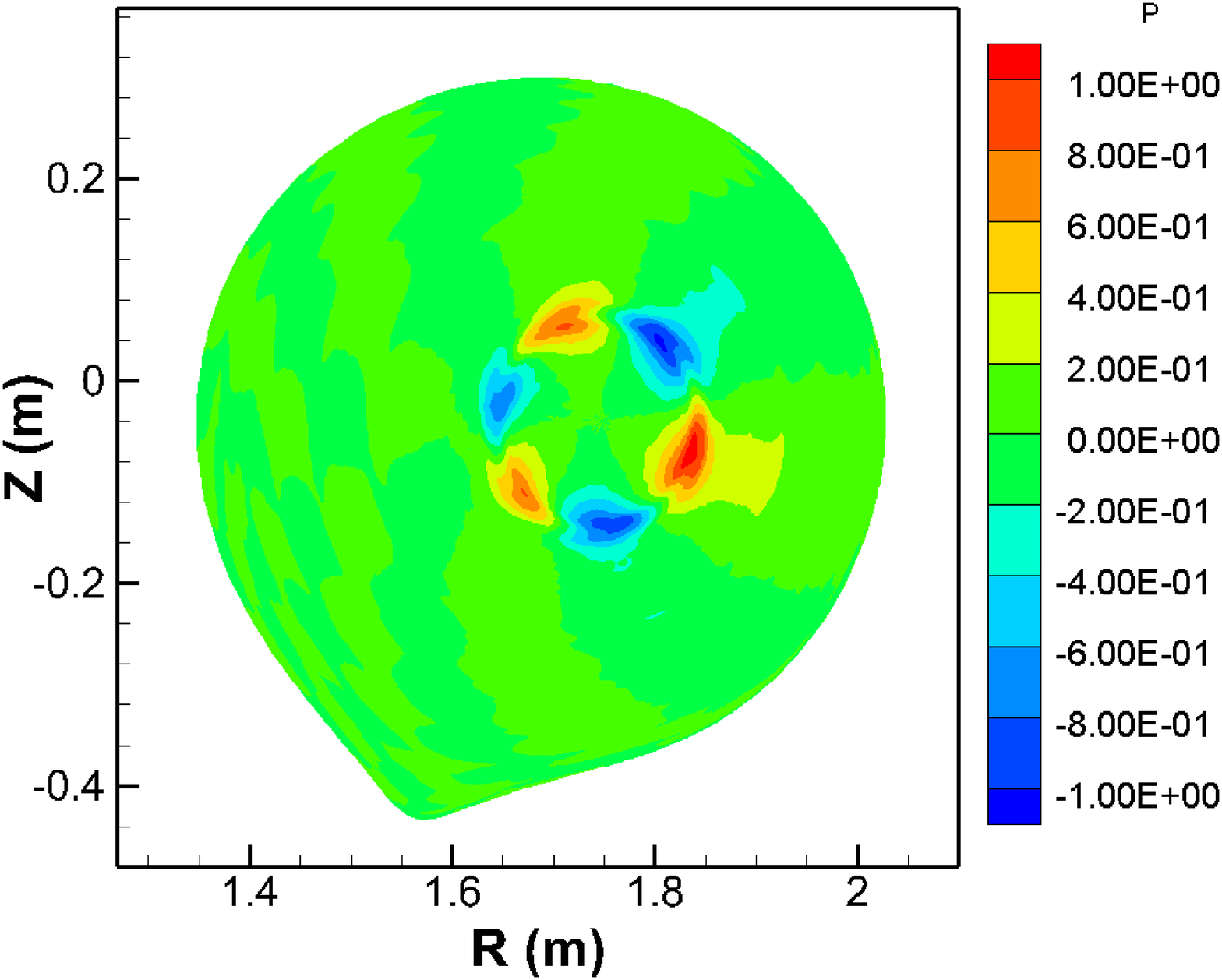}
    \put(13, 68){$(a1)$}
  \end{overpic}
  \begin{overpic}[scale=0.15]{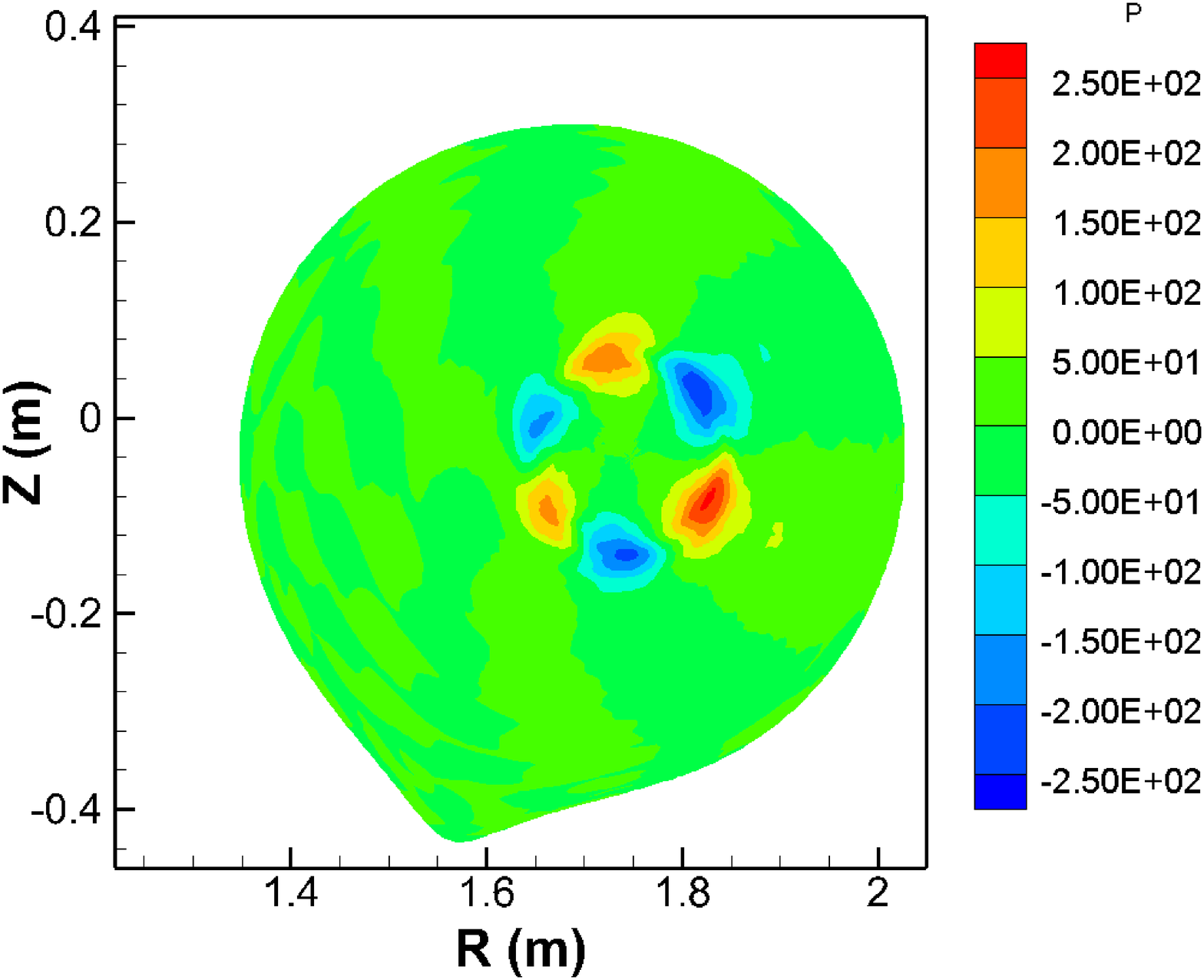}
    \put(13, 68){$(b1)$}
  \end{overpic}
  \begin{overpic}[scale=0.15]{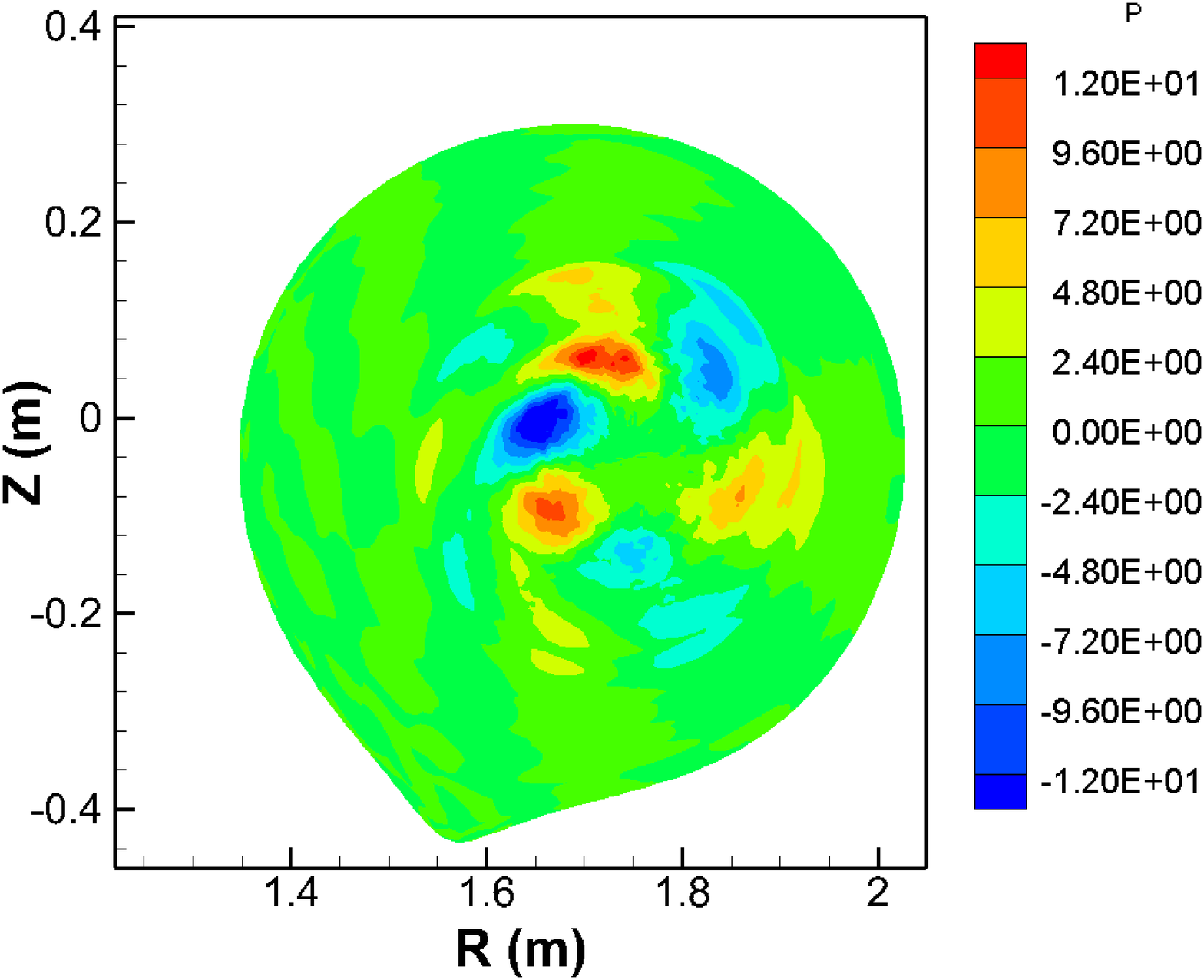}
    \put(13, 68){$(c1)$}
  \end{overpic}
  \begin{overpic}[scale=0.15]{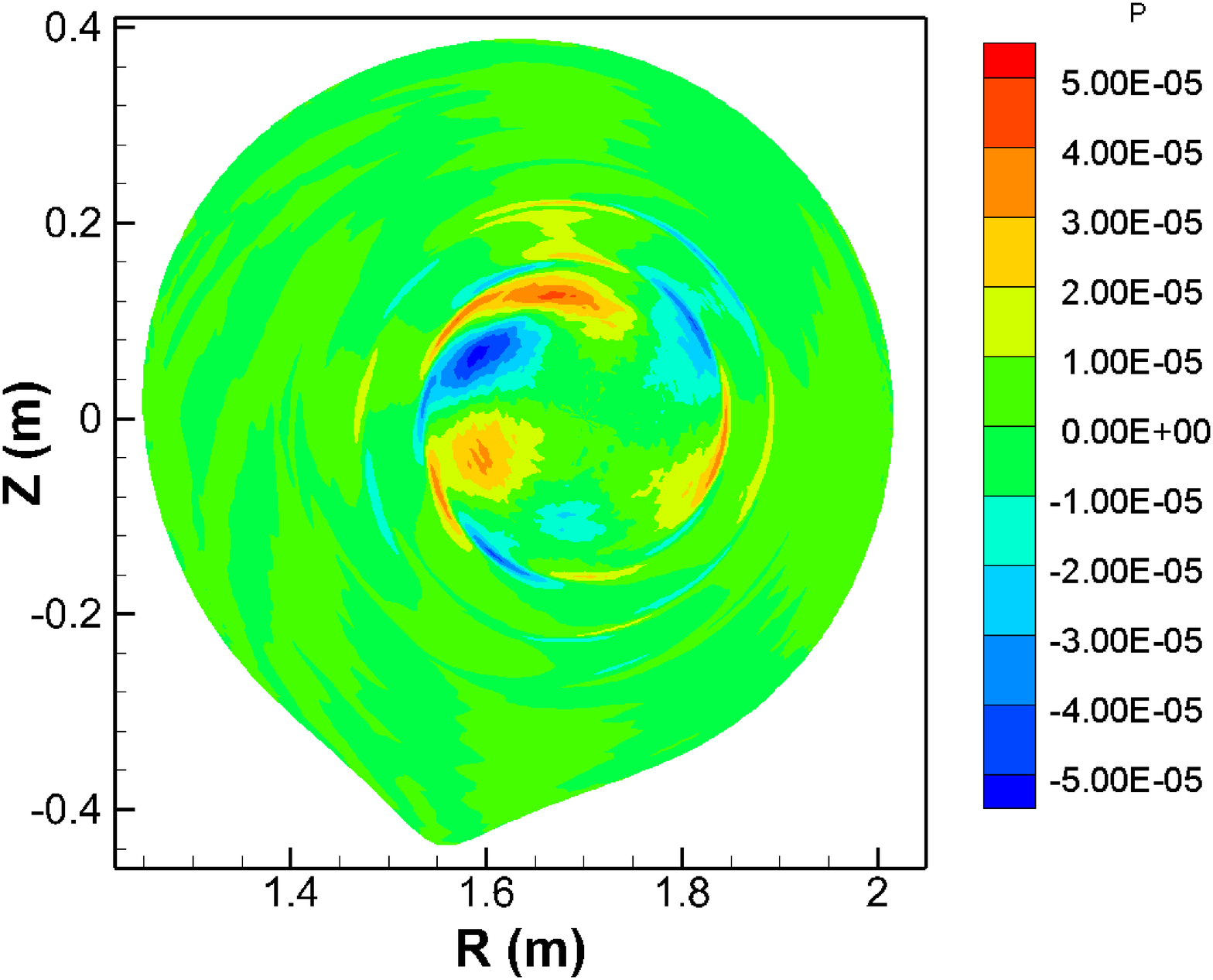}
    \put(13, 68){$(a2)$}
  \end{overpic}
  \begin{overpic}[scale=0.15]{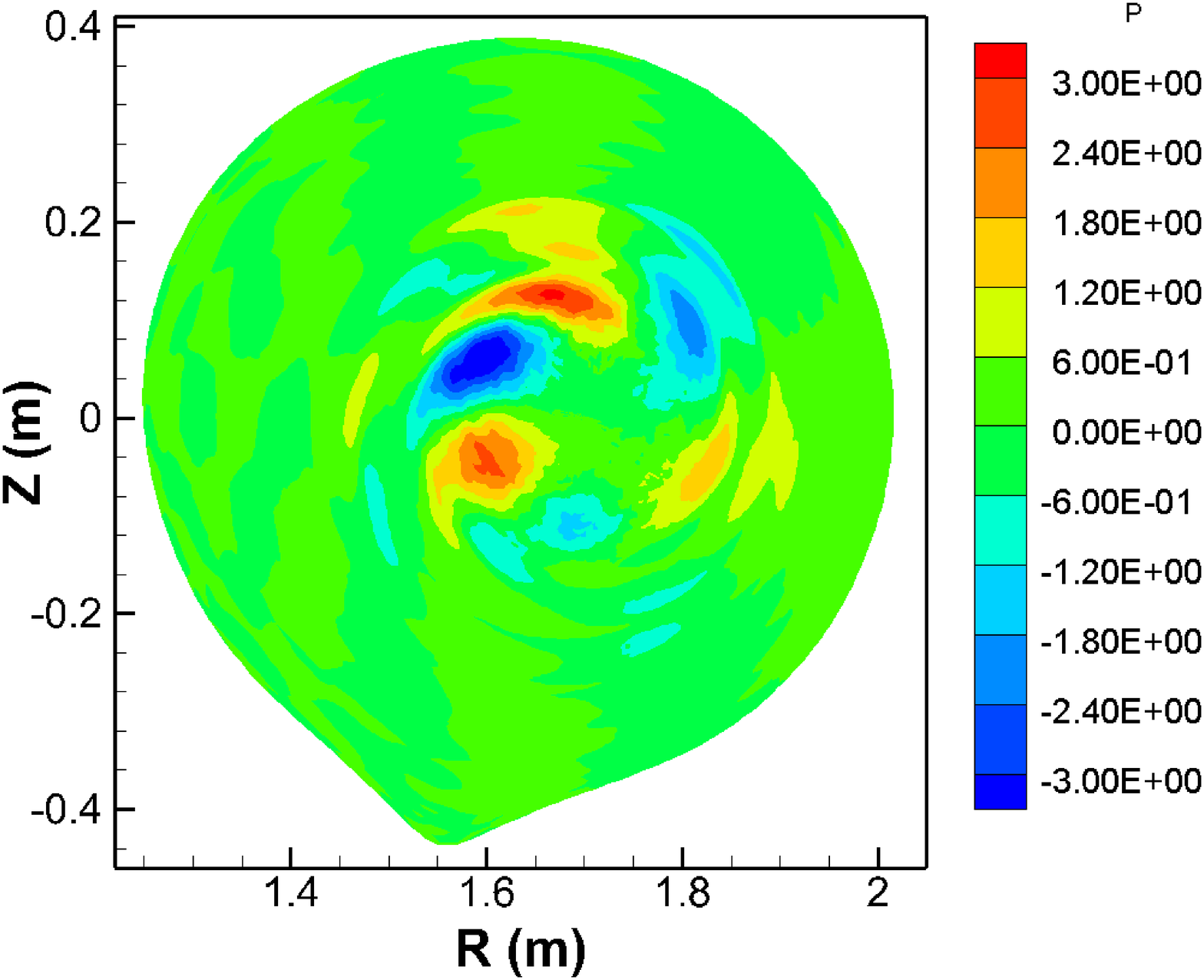}
    \put(13, 68){$(b2)$}
  \end{overpic}
  \begin{overpic}[scale=0.15]{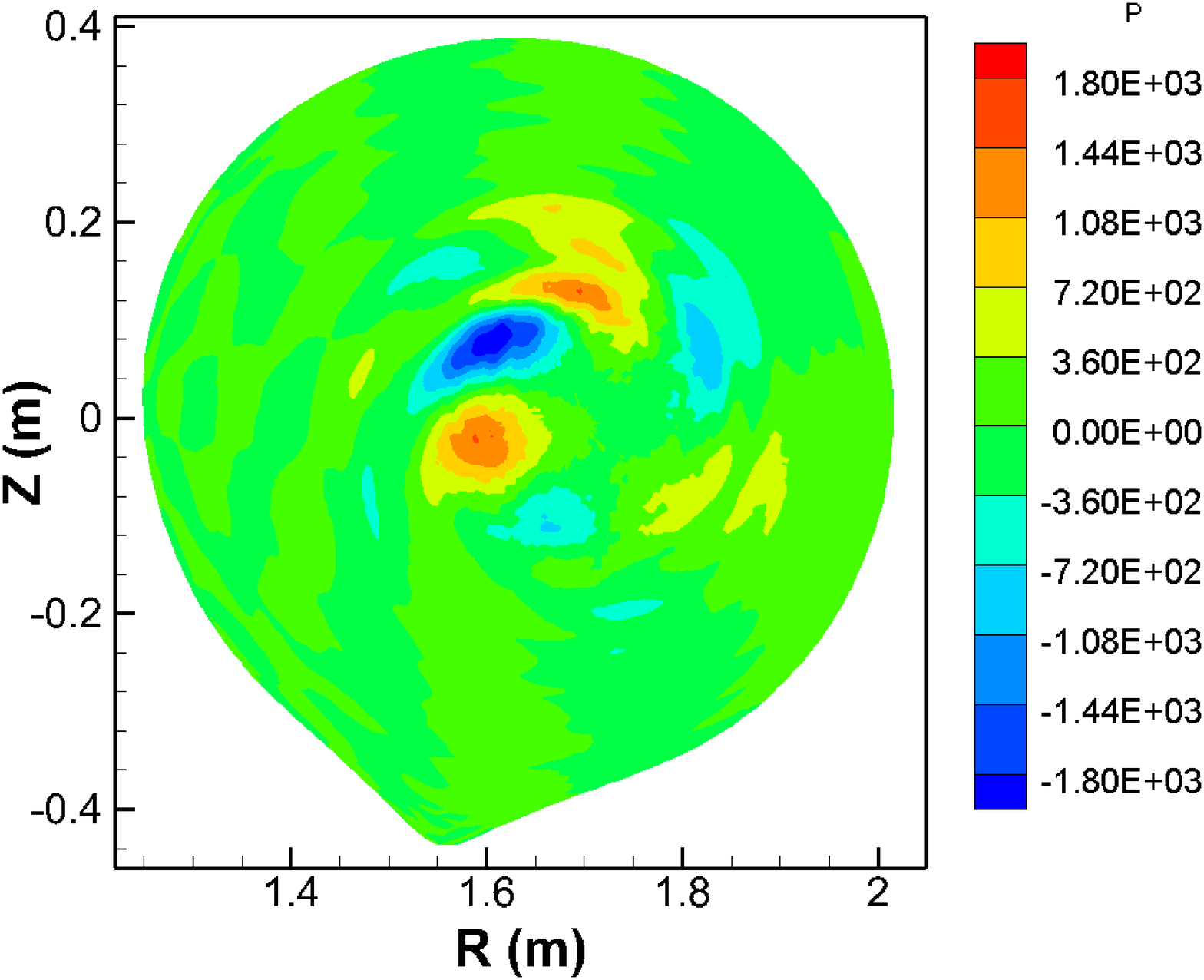}
    \put(13, 68){$(c2)$}
  \end{overpic}
  \caption{\label{fig:fig11} Pressure perturbation contours of
    the $3/3$ modes based on the M420 equilibrium [(a1), (b1) and (c1)],
    and the M452 equilibrium [(a2), (b2) and (c2)],
    where $\beta_f=0.1$ [(a1) and (a2)],
    $\beta_f=0.2$ [(b1) and (b2)], and
    $\beta_f=0.3$ [(c1) and (c2)]. }
\end{figure}

For the M452 case, the $1/1$ modes and  $2/2$ modes are
suppressed slightly by EPs when the EP pressure is relatively
low ($\beta_f<0.1$); however, there is no suppressing effect on the $3/3$ mode.
As the EP pressure increases further ($\beta_f>0.2$),
the growth rate of $3/3$ mode becomes largest [figure \ref{fig:fig10}(c)],
which is similar to the M420 case. For the $1/1$ and $2/2$ modes,
the mode frequency increases almost linearly
with $\beta_f$, and is roughly proportional to $n$,
which is also similar to the M420 case. The frequency of the $3/3$ mode
jumps to a higher branch ($\sim 60\,\mkHz$) when $\beta_f$ increases
from $0$ to $0.1$, and the frequency jumping occurs with
$\beta_f$ smaller than that of the M420 case [figure \ref{fig:fig10}(d)].
For the M420 case, mode structure is twisted by EPs when $\beta_f<0.2$,
and there exists poloidal mode coupling for the $n=3$ blue mode
when $\beta_f=0.3$ [figure \ref{fig:fig11} (a1), (b1) and (c1)].
For the M452 case, the mode structure is twisted by
EPs, and there exists poloidal mode coupling for the $n=3$ mode
when $\beta_f\ge 0.1$ [figure \ref{fig:fig11} (a2), (b2) and (c2)]. \par
Comparing the results of the M420 and M452 cases, we can see that,
with stronger background plasma pressure gradient (M420), the suppressing
effects of EPs on kink modes disappear, and the driving effects of EPs
on higher frequency modes become weaker. Further more,
the FM persists for $n=1,2,3$ up to $\beta_f<0.2$ for
the stronger background plasma pressure gradient (M420 case), while
it is broken for $n=3$ with $\beta_f>0.1$ for the weaker
background plasma pressure gradient (M452 case).
This is consistent with results in Section \ref{subsec:qep}  that the
FM of LLMs becomes more apparent with stronger pressure gradient. \par
\subsection{Effects of beam energy $\varepsilon_b$}\label{subsec:beam}
\begin{figure}[ht]
  \centering
  \includegraphics[width=8.0cm]{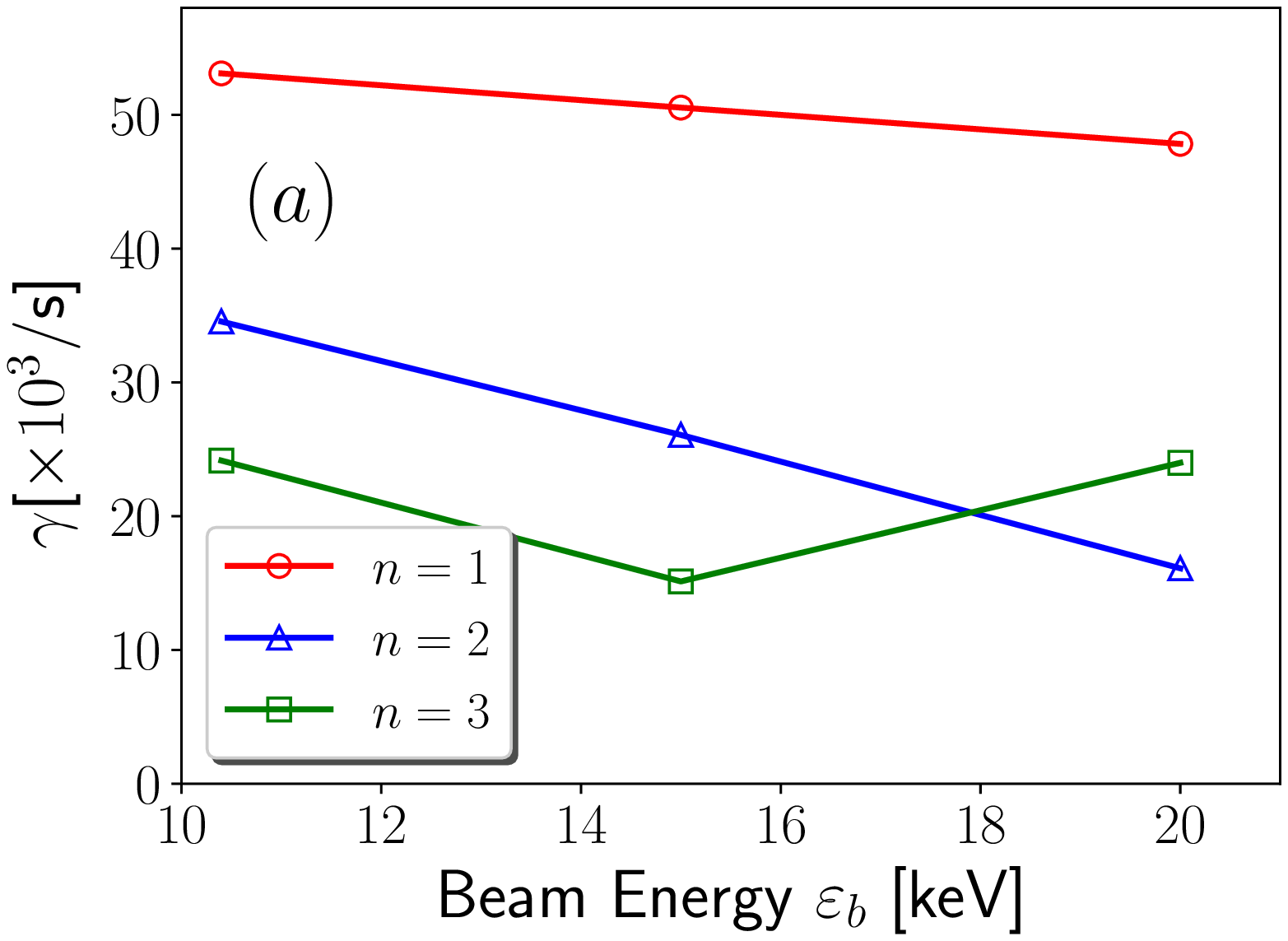}
  \includegraphics[width=8.0cm]{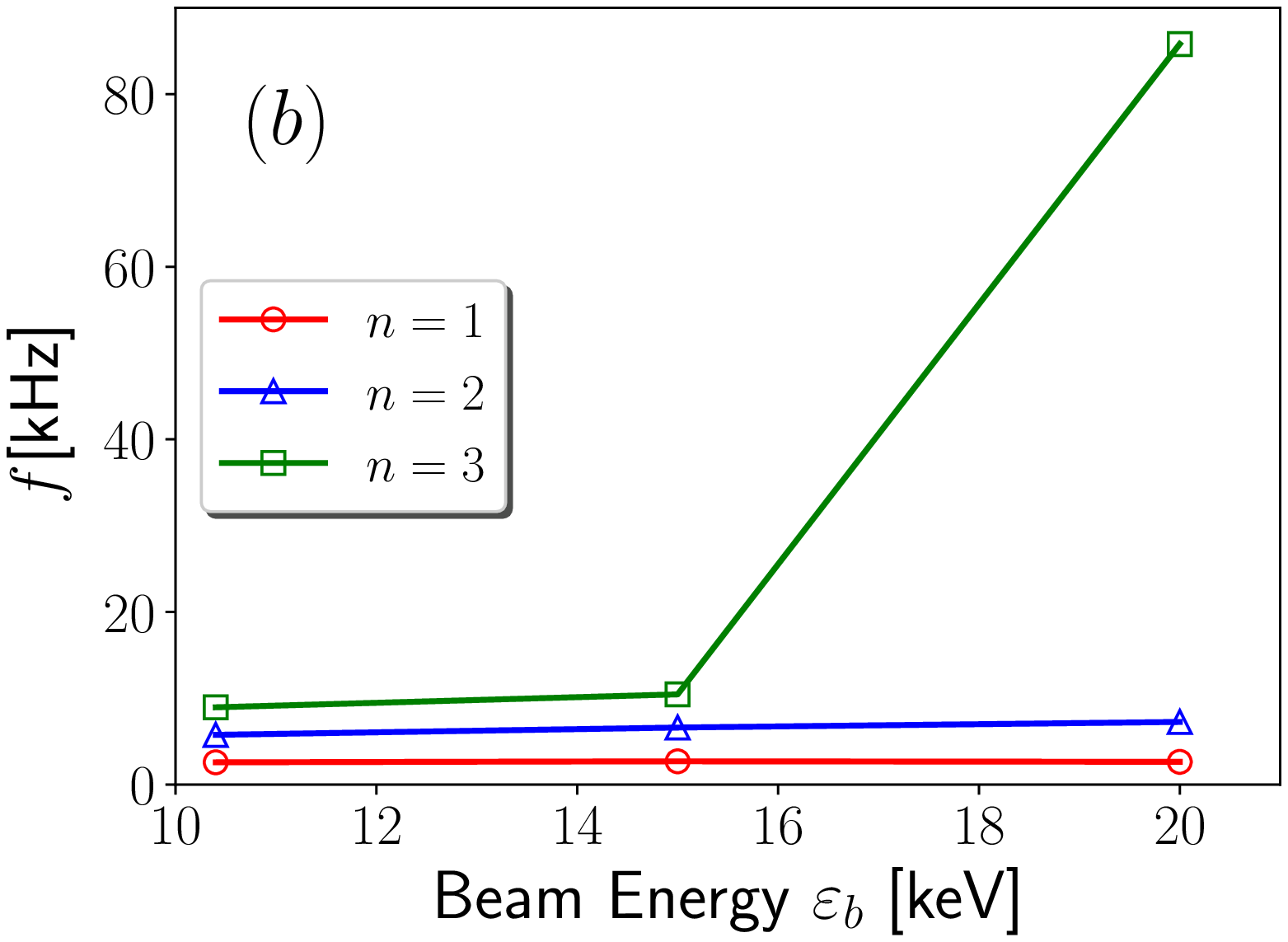}
  \includegraphics[width=8.0cm]{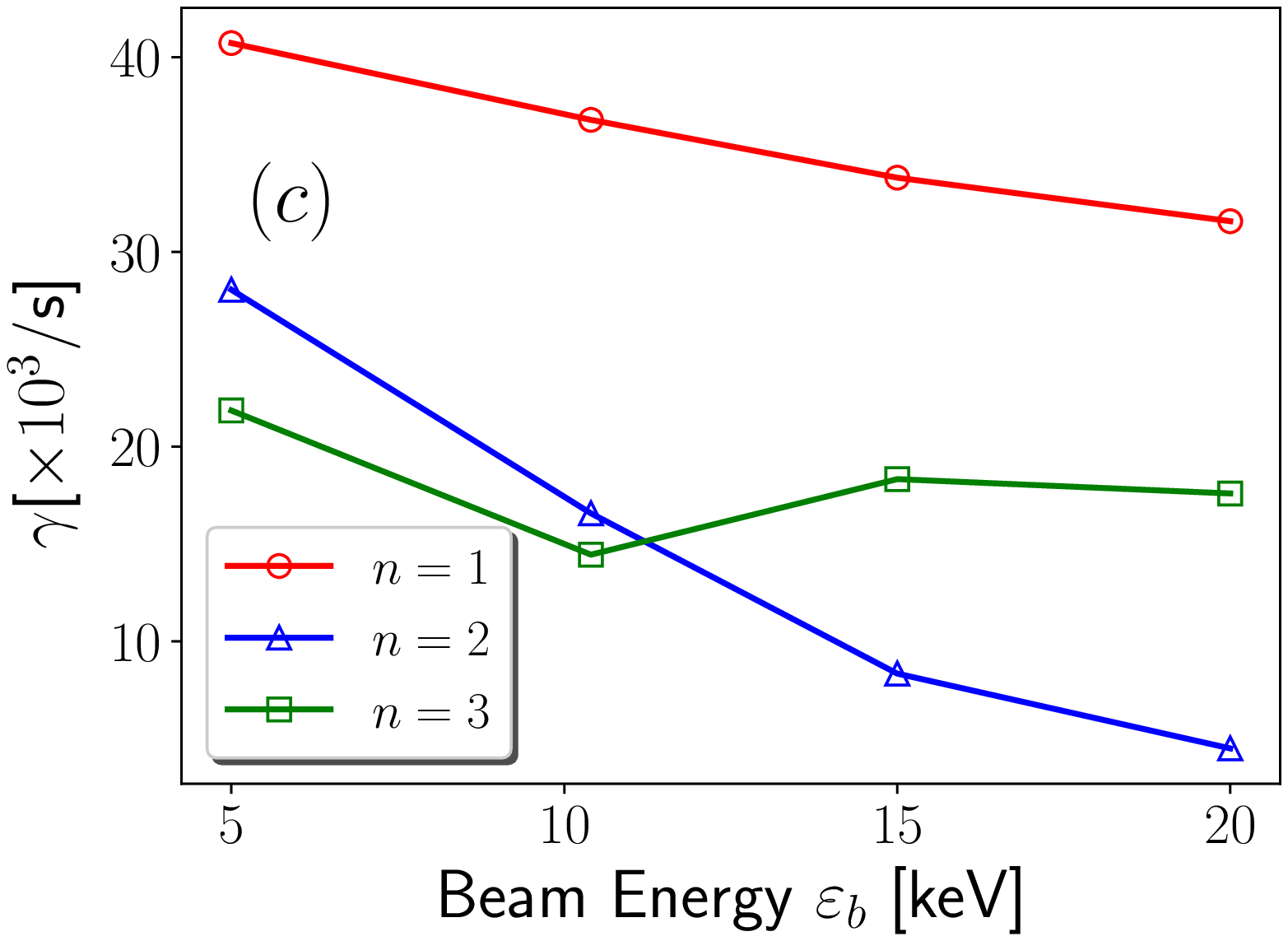}
  \includegraphics[width=8.0cm]{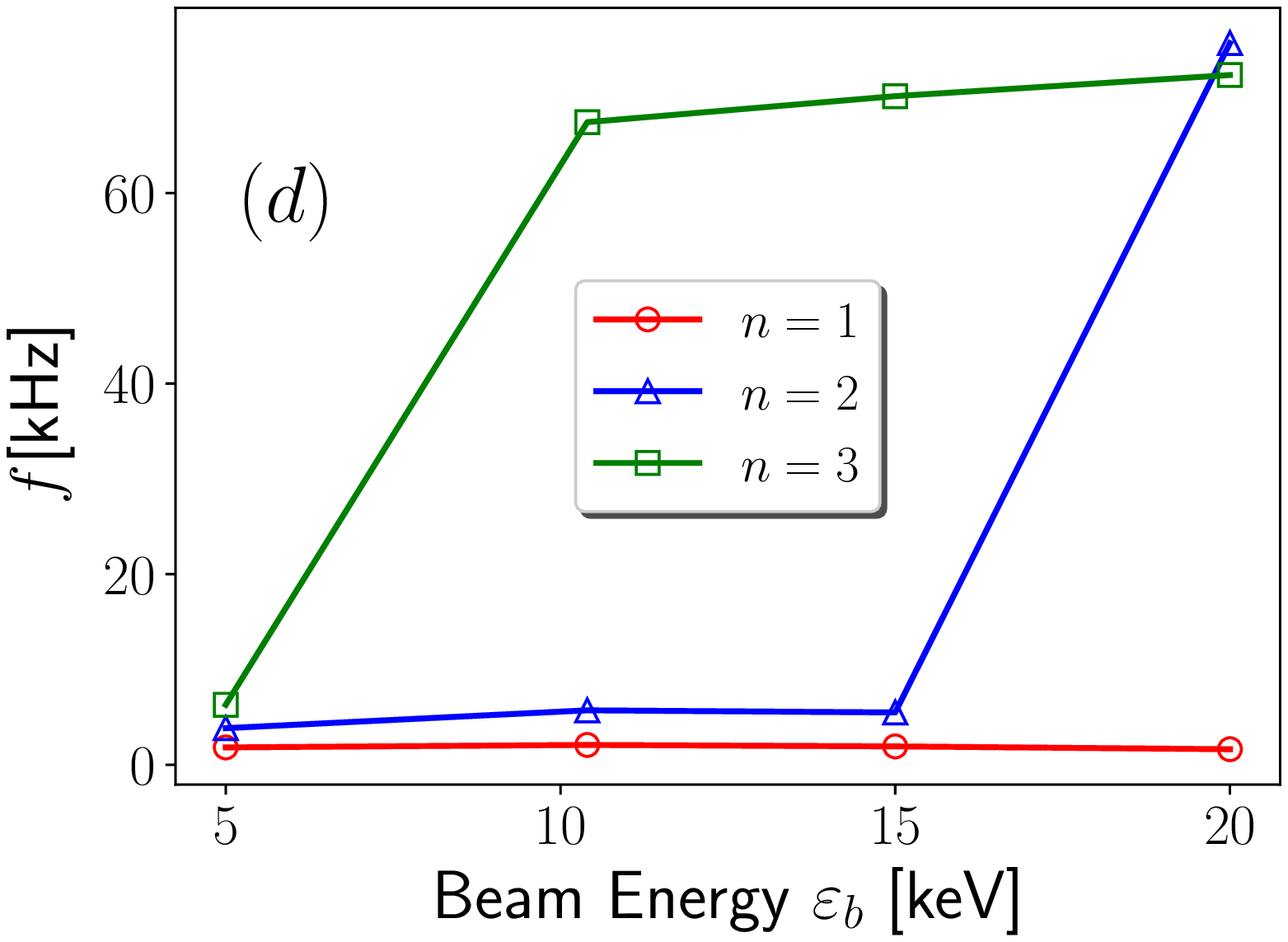}
  \caption{\label{fig:fig12} Growth rates [(a) and (c)] and
    mode frequencies [(b) and (d)] as functions of beam energy $\varepsilon_b$
    for the $1/1$, $2/2$ and $3/3$ modes, where $q_0=0.9$ and $\beta_f=0.1$.
    (a) and (b) are based on the M420 equilibrium,
    (c) and (d) are based on the M452 equilibrium.}
\end{figure}

Now we scan beam energy $\varepsilon_b$ to
study its effects on FM, and we set $q_0=0.9$ and $\beta_f=0.1$.
For the M420 case, the growth rate decreases with the beam energy
$\varepsilon_b$ when the mode frequency is low ($\sim 10\,\mkHz$)
[figure \ref{fig:fig12}(a)].
When beam energy $\varepsilon_b$ is less than $15\,\mkeV$,
the mode frequency is proportional to $n$
approximately. However, for the $3/3$ mode,
as the beam energy $\varepsilon_b$ increases
from $15\,\mkeV$ to $20\,\mkeV$, the frequency jumps from
$10\,\mkHz$ to $86\,\mkHz$, and the growth rate increases as well
suggesting the onset of another branch of mode.
For the M452 case,  the dependence of growth rate decreases
on the beam energy is similar [figure \ref{fig:fig12}(c)].
The FM is broken with $5\,\mkeV <\varepsilon_b<10\,\mkeV$ for $n=3$, and
$15\,\mkeV <\varepsilon_b<20\,\mkeV$ for $n=2$. Both $2/2$ and $3/3$ mode
frequencies jump to a higher mode branch above certain but
different beam energy level. \par
Comparing the results of the M420 and M452 cases,
we can see that, FM can be more easily lost with weaker background
plasma pressure gradient
[figure \ref{fig:fig12}(b) and figure \ref{fig:fig12}(d)],
which is consistent with the results of section \ref{subsec:qep}
and \ref{subsec:bfrac}. \par
\subsection{Effects of EP pressure gradient}\label{subsec:epg}
\begin{figure}[ht]
  \centering
  \includegraphics[width=8.0cm]{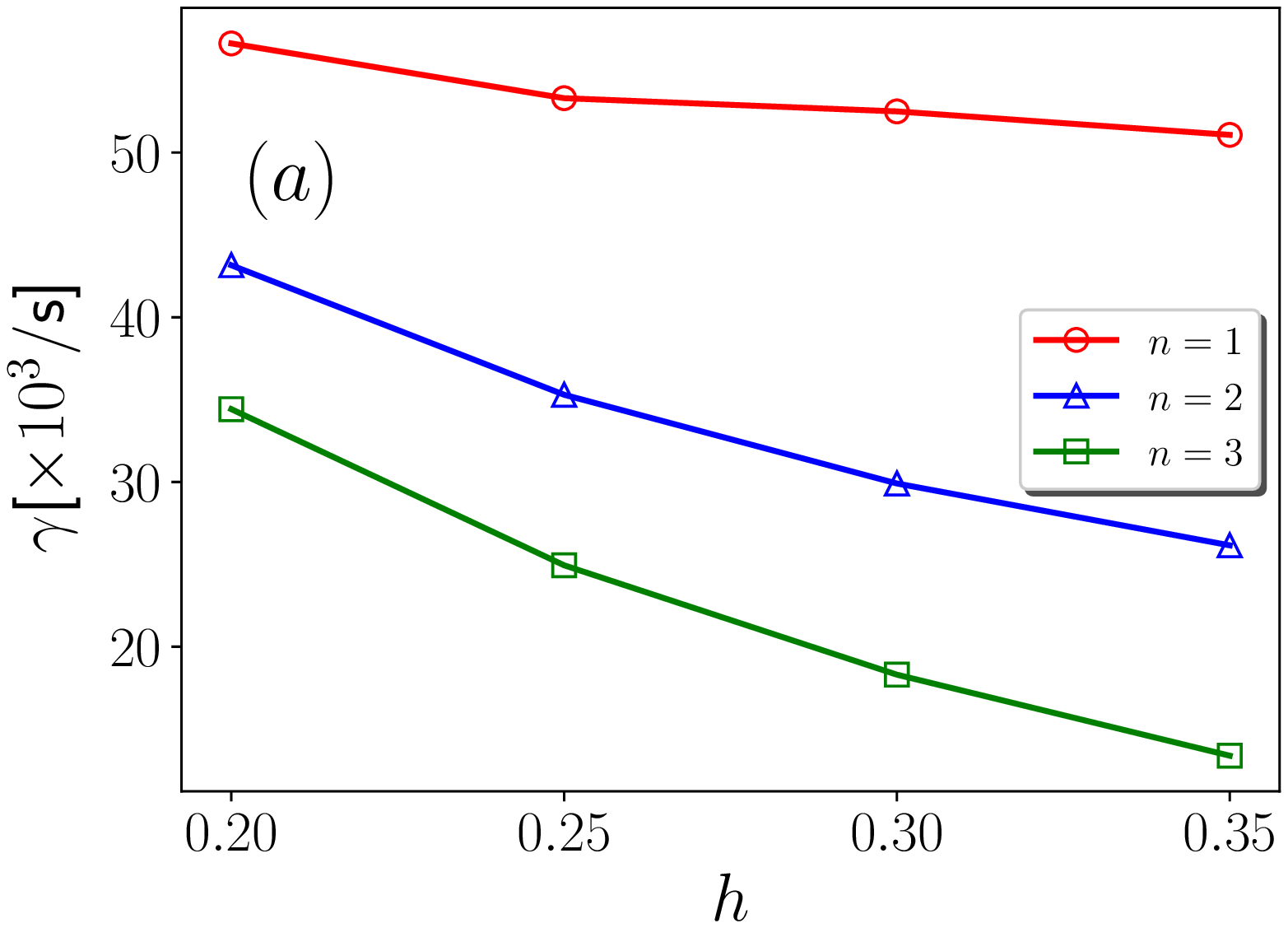}
  \includegraphics[width=8.0cm]{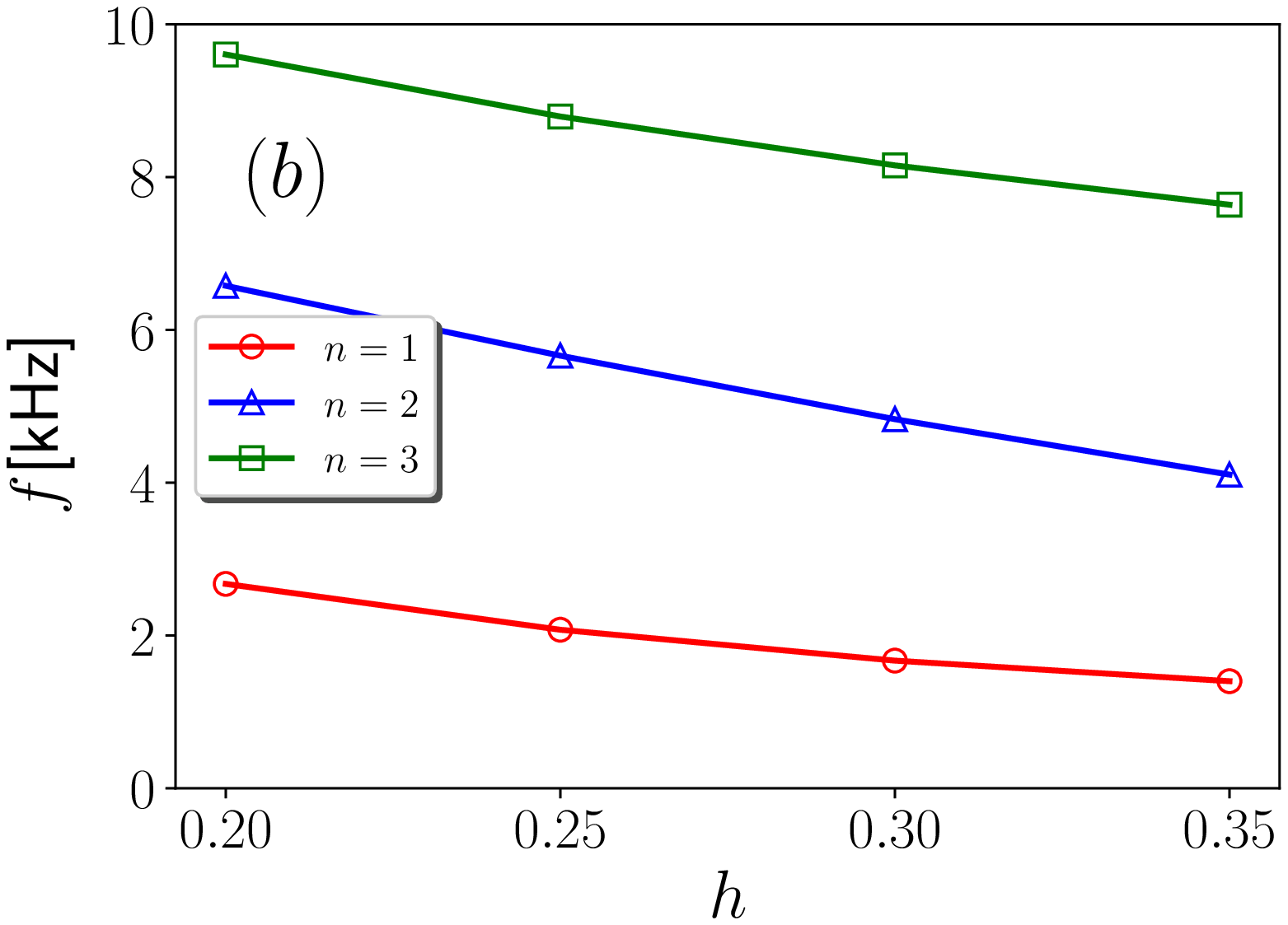}
  \includegraphics[width=8.0cm]{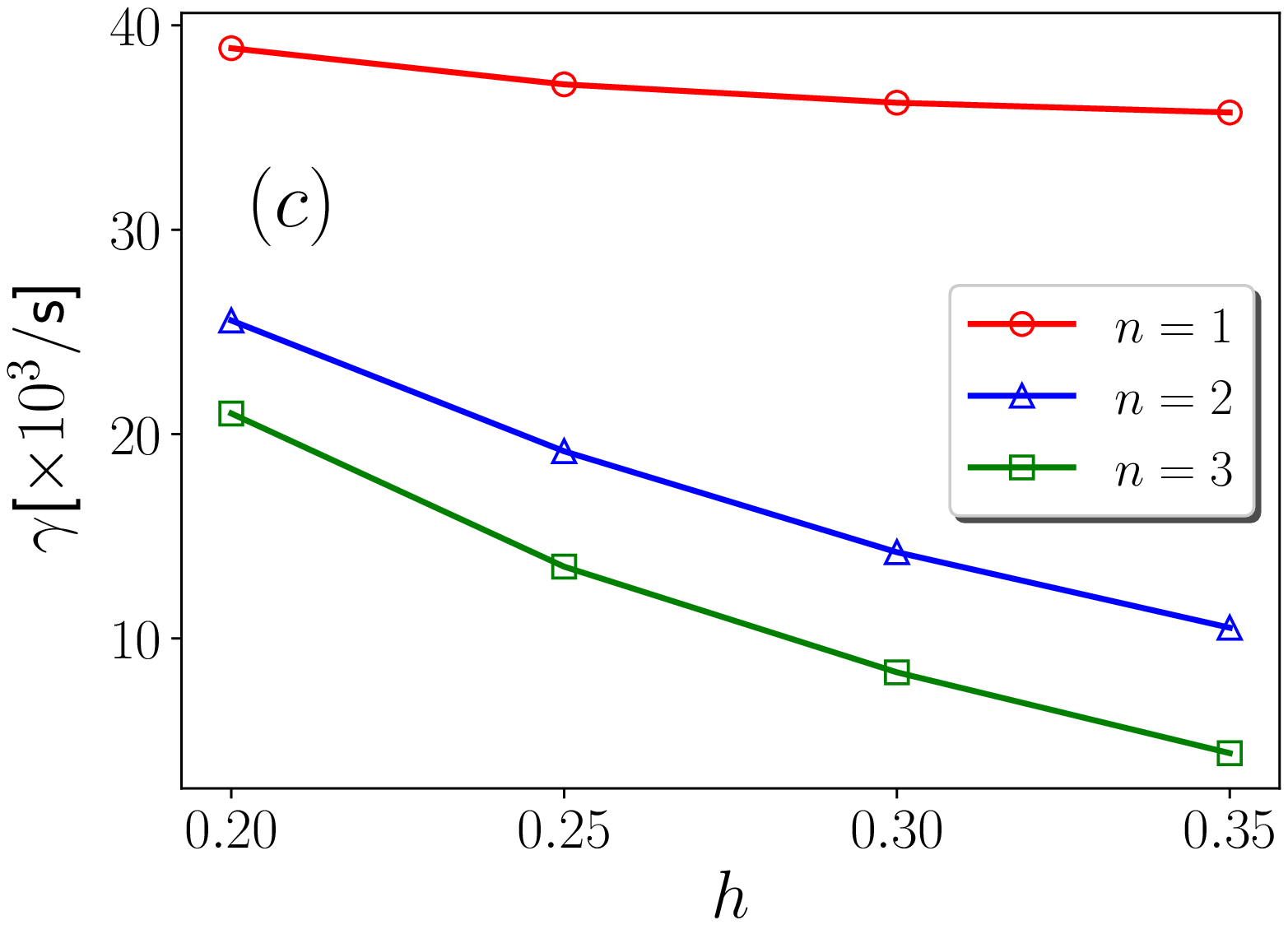}
  \includegraphics[width=8.0cm]{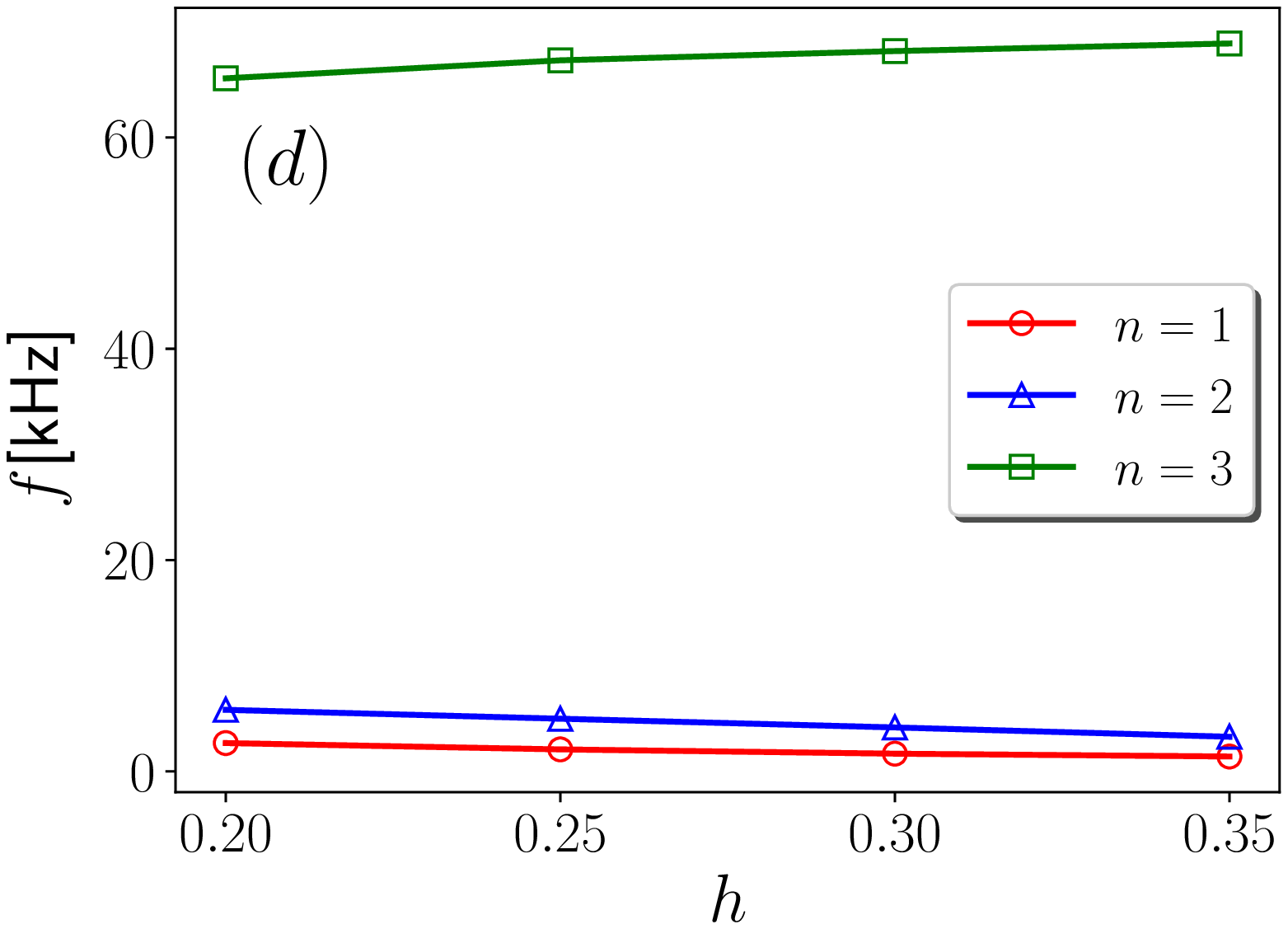}
  \caption{\label{fig:fig13} Growth rates [(a) and (c)] and
    mode frequencies [(b) and (d)] as functions of
    EP pressure gradient coefficient $h$
    for the $1/1$, $2/2$ and $3/3$ modes, where $q_0=0.9$ and $\beta_f=0.1$.
    (a) and (b) are based on the M420 equilibrium,
    (c) and (d) are based on the M452 equilibrium.}
\end{figure}

In the end, we study the effects of EP pressure gradient on FM.
For the M420 case, as the EP pressure gradient coefficient $h$ increases,
the radial profile of EPs becomes more flat with
smaller EP pressure gradient and the growth rate decreases
[figure \ref{fig:fig13}(a)].
The mode frequency decreases slightly with $h$, and remains to be
proportional to $n$ clearly [figure \ref{fig:fig13}(b)].
For the M452 case, the effects of $h$ on the growth
rate are similar, but is much weaker on the mode frequency.
The mode frequency keeps almost constant as $h$ increases.
For $n=1, 2$, the mode frequencies are low ($\sim 10\,\mkHz$),
and proportional to $n$ approximately. For $n=3$, the mode
frequency ($\sim 60\,\mkHz$) can stay on the higher branch for a wide
range of EP pressure gradient [figure \ref{fig:fig13}(d)].
Based on the above results, the pressure gradient driven nature
of FM for LLMs has been further confirmed and the FM
is not much influenced by the EP pressure gradient, which is
probably due to the fact that EP pressure is relatively small
comparing to background plasma pressure. \par
\section{Conclusion}
The ($n=1, 2, 3$) kink/fishbone mode driven by EPs on HL-2A tokamak
is investigated through kinetic-MHD simulations.
The mode frequency is found to be proportional to toroidal mode
number $n$ (frequency multiplication, FM)
even in the absence of equilibrium flow,
when the background plasma pressure gradient is strong,
and neither the beam energy  or the  EP $\beta$ fraction is too high.
Above certain threshold for the EP beam energy or
the EP $\beta$ fraction, the FM becomes
broken, and the higher-$n$ modes can transform to the AE branch,
and such a transition tends to be facilitated by weaker background
plasma pressure. The $q$ profile with varied
$q_0$ ($0.85<q_0<0.95$) have weak influence on the FM.
Although in the absence of EPs, the growth rate of the $1/1$ mode is
greater than that of the higher-$n$ mode, the growth rate of higher-$n$
mode increases more rapidly with $\beta_f$ than that of
$1/1$ mode, suggesting the dominance of higher-$n$ modes in experiments
with  higher EP fraction. \par
The frequency range we calculate does not match that from
experiments for the lacking of plasma rotation in our simulation model.
After adding the frequency of plasma rotation, we get frequencies close
to experimental measurements. Although
we perform the simulations in the circular cross-section tokamak,
our results may also apply to the non-circular cross-section tokamaks,
such as DIIID and EAST, because the LLMs are localized in the
core region, where magnetic flux surface is near circular
for non-circular cross-section tokamaks. In addition, this work is
based on the linear simulations
with simplified EP model. In future, we plan on performing
nonlinear simulations in order to reveal more physical details of LLMs
(such the long-lasting feature)
in the advanced tokamaks with weak or reversed central magnetic shear.
\section*{Acknowledgments}
We thank Prof. Lu Wang and Dr. Da Li for their helpful discussions
and suggestions. We appreciate the assistance from Dr. Haolong Li.
We are grateful for the support from the NIMROD team.
This work was supported by the National Magnetic Confinement Fusion
Program of China (No. 2019YFE03050004),
National Natural Science Foundation of China (Nos. 11875253,
11775221, 51821005, 11875018), the Fundamental Research Funds for the Central
Universities (Nos. WK3420000004 and 2019kfyXJJS193),
the Collaborative Innovation Program of Hefei Science Center, CAS
(No. 2019HSC-CIP015), the U.S. Department of Energy
(Nos. DE-FG02-86ER53218 and DE-SC0018001).
This research used the computing resources from the Supercomputing
Center of University of Science and Technology of China.

%\printfigures
\end{document}